\documentclass[twocolumn,
preprintnumbers,prx,nofootinbib]{revtex4-1}

\usepackage{amsmath,amssymb}
\usepackage{hyperref}
\usepackage{graphicx}
\usepackage{times}
\usepackage{dcolumn}

\def\tr{\mathop{\mathrm{tr}}}

\begin{document}

\title{M-theoretic Genesis of Topological Phases}

\author{Gil Young Cho}
\affiliation{Department of Physics, Pohang University of Science and Technology (POSTECH), Pohang 37673, Republic of Korea}
\affiliation{Asia Pacific Center for Theoretical Physics (APCTP), Pohang 790-784, Korea}
\author{Dongmin Gang}
\affiliation{Department of Physics, Pohang University of Science and Technology (POSTECH), Pohang 37673, Republic of Korea}
\affiliation{Asia Pacific Center for Theoretical Physics (APCTP), Pohang 790-784, Korea}
\author{Hee-Cheol Kim}
\affiliation{Department of Physics, Pohang University of Science and Technology (POSTECH), Pohang 37673, Republic of Korea}
\affiliation{Asia Pacific Center for Theoretical Physics (APCTP), Pohang 790-784, Korea}

\date{\today}

\begin{abstract}
We present a novel M-theoretic approach of constructing and classifying anyonic topological phases of matter, by establishing a correspondence between (2+1)d topological field theories and non-hyperbolic 3-manifolds. In this construction, the topological phases emerge as macroscopic world-volume theories of M5-branes wrapped around certain types of non-hyperbolic 3-manifolds. We devise a systematic algorithm for identifying the emergent topological phases from topological data of the internal wrapped 3-manifolds. As a benchmark of our approach, we reproduce all the known unitary bosonic topological orders up to rank 4. Remarkably, our construction is not restricted to an unitary bosonic theory but it can also generate fermionic and/or non-unitary topological phases in an equivalent fashion. Hence, we pave a new route toward the classification of topological phases of matter. 
\end{abstract}

\maketitle
\tableofcontents

\newpage
\section{Introduction}

Compactification from higher dimensional field theories is a powerful tool for engineering consistent lower dimensional theories.
One of well-studied such examples in M-theory context is compactifications of the six-dimensional $\mathcal{N}=(2,0)$ superconformal field theory (SCFT) living on a stack of M5-branes. 
A large number of (5+1)d and (4+1)d SCFTs have been realized in String theory/M-theory \cite{Witten:1995zh,Seiberg:1996bd,Heckman:2013pva,Heckman:2015bfa}, and compactifications of these higher dimensional theories give rise to a rich class of supersymmetric theories in lower dimensions. 

For example, compactifications of the (5+1)d SCFTs on a circle realizes (4+1)d Kaluza-Klein theories that are conjectured to be the progenitors of all (4+1)d SCFTs via renormalization group (RG) flows \cite{Jefferson:2017ahm,Jefferson:2018irk}, and also families of interesting $\mathcal{N} = 1,2$ supersymmetric quantum field theories in four dimensions have been constructed in compactifications from six dimensions on 2d Riemann surfaces \cite{Gaiotto:2009we,Gaiotto:2015usa,Kim:2017toz}.
Similarly, many examples of (2+1)d superconformal field theories have been geometrically realized in M-theory using compactifications of the (5+1)d $\mathcal{N}=(2,0)$ SCFT on  three-manifolds \cite{Dimofte:2010tz,Terashima:2011qi,Dimofte:2011ju}.
These constructions elucidate deep connections between the (2+1)d physics and geometric properties of 3-manifolds in the compactifications.

The aim of this paper is to construct a new classification scheme for topological field theories in 2+1 dimensions based on the geometric properties of 3-manifolds in the compactifications of the six-dimensional $\mathcal{N}=(2,0)$ theory. In condensed matter physics text, this corresponds to building a new approach for generating and thus classifying a series of different (2+1)d topological phases supporting anyons, which is currently an active research topic. The anyons are the particle-like excitations, whose braiding statistics is different from those of conventional fermions and bosons. They hold the key to the decoherence-free quantum computation, and thus have been pursued extensively both in theory and experiment \cite{tqc}. Theoretically, the mathematical framework for bosonic anyon theories, i.e., ``unitary modular tensor category (UMTC)", has been identified and scrutinized. Essentially, a UMTC consists of a few defining data of anyons (such as fusion rule) and their algebraic relations. See Appendix \ref{UMTC-Suppl} and \cite{Kitaev, Bonderson} for a review. Because different solutions of the algebraic equations correspond to different anyon theories, we can in principle obtain a complete classification of topological phases by generating all the possible solutions of UMTCs. Unfortunately, this has not been completed despite of significant amount of efforts, see for example \cite{Wen_2015, Lan, Rowell, MS}. Hence, it is desirable to develop an entirely new, \textit{physics-oriented} approach for generating the consistent anyon theories, which is independent with the previous algebraic approaches. 

We achieve this by establishing a novel  correspondence between (2+1)d topological field theory and geometry of non-hyperbolic 3-manifolds. When the 3-manifold, around which we compactify the 6d $\mathcal{N}=(2,0)$ theory, is a hyperbolic manifold, it is well established that the (2+1)d theory flows to a superconformal fixed point in the low energy limit. These types of (2+1)d SCFTs have been studied extensively in the earlier literature, see for example \cite{Dimofte:2010tz,Terashima:2011qi,Dimofte:2011ju}. In a sharp contrast to this, we consider the cases with \textit{the internal non-hyperbolic 3-manifolds}. Unlike the hyperbolic type, the (2+1)d field theories constructed out of non-hyperbolic manifolds do not, in general, flow to conformal theories. Instead, we show that such theories of non-hyperbolic manifolds (enjoying certain properties) flow to topological quantum field theories (TQFT) with anyons at the infrared (IR) fixed point. In that case, properties of the IR anyon theories are fully controlled by the topology of the non-hyperbolic manifolds. Exploiting this, we explicitly build a map between non-hyperbolic 3-manifolds and (2+1)d TQFTs. Hence, we find a systematic classifying algorithm for topological phases from the topology of non-hyperbolic manifolds.

As a beginning of our classification program using 3-manifolds, we provide a  list of non-hyperbolic 3-manifolds and flat connections that realize all known UMTCs of rank $\le 4$. See Table \ref{Table : full list}.
This includes famous TQFTs like Fibonacci, Toric code, and Ising. For these models, modular structures of TQFTs, i.e., $S$- and $T$-matrices, are directly obtained from the partition functions of the associated 3-manifolds. As a bonus, we also obtain the chiral central charges $c_{2d}$ (mod $\frac{1}{2}$) for our particular geometric realizations of TQFTs. Although we primarily focus on bosonic anyon theories at lower ranks $\le 4$, our geometric approach is more general and hence can capture higher rank ones and fermionic anyon theories as well. Indeed, we provide explicit constructions of some higher rank UMTCs and a unitary fermionic MTC via non-hyperbolic 3-manifolds, which exhibits the power of our approach.

One advantage of our approach is that it can immediately generate a UV-complete, consistent field theory description for the anyon theories. Previously, they were described only by abstract modular data (and $c_{2d}$) or lattice constructions for limited cases. This advantage is due to the dictionary between (2+1)d supersymmetric field theories and the 3-manifolds \cite{Dimofte:2010tz,Dimofte:2011ju}. Any closed 3-manifold can be constructed by gluing a number of minimal building blocks, called ideal tetrahedron and solid-torus, together. In (2+1)d field theory context, this construction amounts to gauging some flavor symmetries of elementary chiral multiplets with particular choice of Chern-Simons levels as well as superpotential couplings. The resulting continuum field theory is the UV avatar for a desired TQFT. Its supersymmetric nature allows us to compute various observables of the IR TQFT. For example, the ground state degeneracy on a genus-$g$ Riemann surface and the modular structures in the IR TQFTs can be obtained using localization technique.

On top of these, our approach can also be used to produce natural unitary embeddings of non-unitary TQFTs such as Lee-Yang model, from which we can initiate the classification of the non-unitary MTCs. We will find that some of non-hyperbolic manifolds exhibit modular structure of the non-unitary TQFTs. This sounds puzzling at first sight: since we compactify a unitary (5+1)d theory on a Riemannian manifold, the low energy theory is also expected to be unitary. As we demonstrate through several examples in this paper, it turns out that the (2+1)d field theories from this type of non-hyperbolic manifolds enjoy emergent global symmetry in IR. Taking into account the IR R-symmetry correctly, we find that these theories in fact flow to a unitary superconformal field theory. The seemingly non-unitary TQFT structure arises just from a particular sub-sector of the unitary SCFT, whose correlation functions can be obtained by a non-unitary mass deformation in the path integral. Similar non-unitary sub-sectors appear in various supersymmetric models, e.g.,  the Schur operators in 4d $\mathcal{N}=2$ SCFTs equipped with 2d chiral algebra structure \cite{Beem:2013sza}. Keeping this in mind, in this paper we build unitary embeddings and classification of the non-unitary MTCs up to rank $\le 4$ as a concrete demonstration of our proposal.

The rest of the paper is organized as follows. In section \ref{sec : TQFT from M5s}, we introduce the recipes for geometric engineering of (2+1)d TQFT, notated as $\textrm{TFT}[M]$, from wrapped M5-branes and clarify the physical meaning of emergent modular structures at IR. In section \ref{sec : full modular structure}, we next give a systematic algorithm for computing modular structure of bosonic TQFTs for the case of trivial $H_1(M, \mathbb{Z}_2)$. In section \ref{sec : effecitve field theory}, we give the construction of an effective continuum field theory for $\textrm{TFT}[M]$. In section \ref{sec : Examples}, we present the concrete examples in section \ref{sec : TQFT from M5s}.  In section \ref{sec : non-trivial Z2 homology}, we extend our analysis to the cases with non-trivial $H_1(M, \mathbb{Z}_2)$. In Appendices, we collect the technical details.

\begin{table}[t]
	\begin{center}
		\begin{tabular}{|c|c|l|c|l|}
			\hline
	        $N_c^B$ & 3-manifold & $h_1,h_2,\cdots,h_N$ & comment & Sec.
			\\
			\hline
			$1_1^B$  &$S^2 (2,3,3)$ & $0$ & & \eqref{para:trivial}
			\\
			\hline 
			$2^B_{\pm 14/5}$ & $S^2 (2,3,5)$ & $0,\pm \frac{2}{5}$ &Fibonacci &\eqref{para:Fibonacci}
			\\
			 $2_{\pm 1}^B$ & $S^2(3,3,3)$ & $0,\pm\frac{1}{4}$ & Semion &\eqref{para:Semion}
			\\
			\hline
			 $3^B_{\pm 8/7}$ & $S^2 (2,3,7)$ & $0,\mp \frac{1}{7}, \pm \frac{2}{7}$ & $(A_1,5)_{\frac{1}{2}}$ & \eqref{para:A1-5}
			\\
			  $3^B_{\pm 1/2}$ & $S^2 (3,\frac{3}{2},4)_{\rm I}$ & $0,\frac{1}{2},\pm\frac{1}{16}$ & Ising & \eqref{para:Ising}
			 \\
			 $3^B_{\pm 7/2}$ & $S^2 (3,\frac{3}{2},4)_{\rm II}$ & $0,\frac{1}{2},\pm\frac{7}{16}$ & $SO(7)_1$ &\eqref{para:Ising}
			\\
			$3^B_{\pm 3/2}$ & $S^2 (3,3,4)_{\rm I}$ & $0,\frac{1}{2},\pm\frac{3}{16}$ & $(A_1,2)$ & \eqref{para:A1-2}
			\\
			$3^B_{\pm 5/2}$ & $S^2 (3,3,4)_{\rm II}$ & $0,\frac{1}{2},\pm\frac{5}{16}$ &$SO(5)_1$ &\eqref{para:A1-2}
			\\
			 $3_{\pm 2}^{B}$ & $S^2(2,3,6)$ & $0,\pm\frac{1}{3},\pm\frac{1}{3}$ & $(A_2,1)$ & \eqref{App : Z3}
			\\
			\hline
			  $4^B_{\pm 10/3}$ & $S^2 (2,3,9) $ & $0,\pm\frac{1}{3},\pm\frac{2}{9},\mp\frac{1}{3}$ & $(A_1,7)_{\frac{1}{2}}$ & \eqref{para:239}
			\\
			 $ 4^B_{\pm 19/5}$ & $S^2 (3,3,5)_{\rm I}$ & $0,\pm\frac{1}{4},\mp\frac{7}{20},\pm\frac{2}{5}$ &  & \eqref{para:335}
			\\
			  $ 4^B_{\pm 9/5}$ &  $S^2 (3,3,5)_{\rm II}$ & $0,\mp\frac{1}{4},\pm\frac{3}{20},\pm\frac{2}{5}$ & &\eqref{para:335}
			\\
			 $4^B_{\pm 12/5}$ & $S^2 (2,5,5)$ & $0,\mp\frac{2}{5},\mp\frac{2}{5},\pm\frac{1}{5}$ & & \eqref{para:255}
			\\
			 $4^{B,c}_{0}$ & $ S^2 (2,4,\frac{5}4)$ & $0,0,\frac{2}{5},-\frac{2}{5}$ & & \eqref{para:245/4}
			\\
			 $4_0^{B,a}$ & $S^2(4,\frac{4}7, \frac{3}2)_{\rm I}$ & $0,0,0,\frac{1}{2}$ & Toric code & \eqref{App : Z4}
			 \\
			 $4_{ 4}^{B}$ & $S^2(4,\frac{4}7, \frac{3}2)_{\rm II}$ & $0,\frac{1}{2},\frac{1}{2},\frac{1}{2}$ & $(D_4,1)$ & \eqref{App : Z4}
			\\
			 $4_{\pm 1}^{B}$ & $S^2(4,4, \frac{3}2)_{\rm I}$ & $0,\pm\frac{1}{8},\pm\frac{1}{8},\frac{1}{2}$ & $(A_3,1)$ & \eqref{App : Z4}
			 \\
			 $4_{\pm 3}^{B}$ & $S^2(4,4, \frac{3}2)_{\rm II}$& $0,\pm\frac{3}{8},\pm\frac{3}{8},\frac{1}{2}$ & & \eqref{App : Z4}
			 \\
			 $4_{\pm 2}^{B}$ & $S^2(4,\frac{4}3, \frac{3}2)_{\rm I}$& $0,\pm\frac{1}{4},\pm\frac{1}{4},\frac{1}{2}$ & & \eqref{App : Z4}
			 \\
			 $4_{0}^{B,b}$ & $S^2(4,\frac{4}3, \frac{3}2)_{\rm II}$& $0,0,\frac{1}{4},-\frac{1}{4}$ & Double semion & \eqref{App : Z4}
			\\
			\hline
		\end{tabular}
	\end{center}
	\caption{Geometric realizations of topological phases $N^B_c$. In this table, we used the nomenclature in \cite{Wen_2015} to denote unitary bosonic topological phases with rank $N$ and central charge $c$. The geometry denoted by $S^2(r_i)_a$ with three rational numbers $r_{i=1,2,3}$ is a Seifert fiber 3-manifold and the subscript $a$ labels the maps between anyons and flat connections. $N$ anyons have spins $h_1,h_2,\cdots,h_N$. The table contains all bosonic topological phases with $N\le 4$. }
	\label{Table : full list}
\end{table}

\section{TQFT from wrapped M5-branes} \label{sec : TQFT from M5s}

In this section, we present a brief introduction to the construction of (2+1)d quantum field theories by compactifying (5+1)d $\mathcal{N}=(2,0)$ SCFTs on a compact 3-manifold $M$. Such construction has been mainly focused on the hyperbolic $M$. Basing on this, we extend it to the non-hyperbolic $M$ and the conditions for $M$ to generate a (2+1)d topological theory at IR.

There are two fundamental objects in M-theory: M2-brane and M5-brane. A variety of quantum systems can be engineered by embedding the fundamental objects into subspaces of various (10+1) dimensional  space-time.
In this paper, we use M5-branes to geometrically engineer (2+1)d topological phases. The concrete set-up is as follows:
\begin{align}
\begin{split}
&\textrm{11-dimensional space-time : }  \mathbb{R}^{1,2} \times (T^*M)\times \mathbb{R}^2\;,
\\
&\textrm{Two coincident M5-branes on $\mathbb{R}^{1,2} \times M$}\;. \label{wrapped M5-branes}
\end{split}
\end{align}
$M$ is a compact 3-dimensional manifold and $T^*M$ is its cotangent bundle. The cotangent bundle is a local Calabi-Yau manifold and the configuration is stable thanks to supersymmetry. 
At  length scale above the  characteristic size of the compact internal 3-manifold,  the world-volume theory of M5-branes is described by a quantum  theory on  $\mathbb{R}^{1,2}$. The lower dimensional (2+1)d quantum  theory  depends only on the topology of the internal manifold $M$ (modulo some  discrete choices explained in Appendix \ref{appendix : T[M]}). This fact allows us to denote the (2+1)d theory  by 
\begin{align}
\mathcal{T}[M]:=\textrm{(2+1)d quantum  theory determined by $M$}\;. \label{6d def of T[M]}
\end{align}
From the geometrical construction of $\mathcal{T}[M]$ in  \eqref{6d def of T[M]},  we expect that the 3d   theory  enjoys supersymmetry with 4 supercharges as well as an  $U(1)$ R-symmetry. The $U(1)$ R-symmetry  comes from  rotational symmetry  acting on the  transverse $\mathbb{R}^2$ in the above set-up.  We call the $U(1)$ R-symmetry as {\it UV R-symmetry} since it originates from the symmetry of the UV M-theoretical set-up. 
During the last decade, there have been remarkable progresses in understanding this type of theories \cite{Dimofte:2010tz,Terashima:2011qi,Dimofte:2011ju}.
For example, a systematic algorithm of finding purely gauge theoretical  descriptions for $\mathcal{T}[M]$ was introduced in \cite{Dimofte:2011ju,Gang:2018wek}. 
Most studies on the subject have focused on the theories generated from hyperbolic manifolds.  For a hyperbolic $M$, the corresponding gauge theory  flows to a non-trivial superconformal theory at the end of a RG-flow. However, this is in general not the case for non-hyperbolic manifolds. 

This article is devoted to understanding quantum field theories associated with non-hyperbolic 3-manifolds.
In the followings, we will argue that there exists a large class of non-hyperbolic 3-manifold $M$ hosting modular structures and interpret the modular structure of $M$ as the modular structure of the (2+1)d topological theory  obtained from  M5-branes wrapped on $M$. More precisely, we clarify \textbf{a)} what kind of  3-manifolds  gives such topological  phases (in equation \eqref{3-manifolds for top}),  \textbf{b)} how to obtain the full $S$-matrix using a flat-connection-to-loop-operator map (in section \ref{sec : flat-to-loop}),  \textbf{c)} supersymmetric gauge theories $\mathcal{T}[M]$ for the topological phases (in section  \ref{sec : effecitve field theory}), and \textbf{d)} topological phases with non-trivial $H_1(M, \mathbb{Z}_2)$ (in section \ref{sec : non-trivial Z2 homology}). 

By passing, we note that previously curious modular structures for some non-hyperbolic 3-manifolds were  reported in   mathematics literatures  \cite{lawrence1999modular,Hikami_2003,HIKAMI_2005} from the study of Witten-Reshetikhin-Turaev invariant \cite{witten1989quantum,reshetikhin1991invariants}, but those references missed the connection of the modular structure with the physics of (2+1)d TQFTs.

\subsection{Topological field theories}
We first claim that for a non-hyperbolic 3-manifold $M$ satisfying the following conditions, the corresponding (2+1)d theory  $\mathcal{T}[M]$ has a modular structure:
\begin{align}
\begin{split}
&i) \;\textrm{There are  (non-empty) finitely many }
\\
& \quad \textrm{ irreducible $SL(2,\mathbb{C})$ flat connections on $M$ },
\\
&
ii)\; \textrm{All of them are  gauge equivalent to  either} 
\\
&\quad \textrm{ $SU(2)$ or $SL(2,\mathbb{R})$ flat connections,} \label{3-manifolds for top}
\\
\end{split}
\end{align}
There are infinitely many examples of 3-manifolds satisfying the above conditions \cite{kitano2016sl2rrepresentations}.  We will present a systematic algorithm to compute the modular data of $\mathcal{T}[M]$ from the non-hyperbolic manifold $M$ with \eqref{3-manifolds for top} in the next section. Before presenting the algorithm, let us  first clarify the physical meaning of the modular structure. It has distinct  physical interpretations depending on whether the modular structure is  unitary or not.

We first consider the unitary case. We conjecture that
\begin{align}
&\textbf{Conjecture: }\mathcal{T}[M] \textit{ for $M$ obeying  \eqref{3-manifolds for top} flows to a topological} \nonumber
\\
&\textit{field theory described by a modular structure at IR if the }\nonumber
\\
&\textit{associated modular structure is unitary.} \label{main conjecture}
\end{align} 
Here, the modular structure being unitary means that the first column of the $S$-matrix satisfies $|S_{00}| 
\leq  |S_{0\alpha}|$ for all $\alpha$. 
We denote the resulting (2+1)d topological phase  by
\begin{align}
\begin{split}
\textrm{TFT}[M] :=& \textrm{TQFT from  non-hyperbolic $M$ obeying  \eqref{3-manifolds for top}} . \label{TFT[M]}
\end{split}
\end{align}

Explicit checks  of the conjecture with several choices of $M$ will be given in the section \ref{sec : effecitve field theory} and \ref{sec : Examples}.

Let us sketch our reasoning behind this conjecture, while deferring the details to the appendix \ref{appendix : IR phases}. In the compactification (\ref{wrapped M5-branes}), there exists a natural map between the flat connections on $M$ and the classical vacua in the (2+1)d supersymmetric theory \cite{Dimofte:2010tz}. The vacua are sometimes called Bethe-vacua in the recent literature and they play important roles in the study of IR dynamics of (2+1)d supersymmetric gauge theories \cite{Beem:2012mb,Nekrasov:2014xaa,Gukov:2015sna,Closset:2017zgf,Closset:2018ghr, Gang:2019jut}.  Various supersymmetric partition functions can be written in terms of the Bethe-vacua.  For example, 
\begin{align}
\begin{split}
&\mathcal{I}_{\rm sci}(x) = \sum_{\alpha : \textrm{Bethe-vacua}} \mathbb{B}^\alpha (x) \left( \mathbb{B}^\alpha (x^{-1})  \right)^*\;,
\\
&\mathcal{I}_{\rm top}(x) =\sum_{\alpha : \textrm{Bethe-vacua}} \mathbb{B}^\alpha (x)  \mathbb{B}^\alpha (x^{-1}) \;. \label{factorization}
\end{split}
\end{align} 
Here, the superconformal index $\mathcal{I}_{\rm sci} (x)$ and the (topologically) twisted refined index $\mathcal{I}_{\rm top} (x)$ are partition functions of a (2+1)d supersymmetric  theory on two distinct supersymmetric backgrounds with topology $S^2 \times S^1$  \cite{Kim:2009wb,Imamura:2011su,Benini:2015noa}.
$\mathbb{B}^\alpha $(x) is the so-called {\it holomorphic block}, which computes the partition function on $\mathbb{R}^2\times S^1$ with an asymptotic boundary condition determined by the choice of a Bethe-vacuum $\alpha$. The fugacity $x$ is related to the circumference  $\beta$ of Euclidean time circle $S^1$ as $x =e^{-\beta}$.

In the language of canonical quantization, the superconformal index  counts supersymmetric (so-called BPS) local operators (i.e. supersymmetric states in the radially quantized Hilbert-space) preserving 2 supercharges, whereas the twisted refined index counts the supersymmetric ground states on a topologically twisted $S^2$ with unit background magnetic flux coupled to the $U(1)$ R-symmetry.
\begin{align}
\begin{split}
&\mathcal{I}_{\rm sci} (x) = \textrm{Tr}_{\mathcal{H}_{\rm rad}(S^2)}  (-1)^{R_{}} x^{\frac{R_{}}2+ j_3}\;,
\\
&\mathcal{I}_{\rm top} (x) = \textrm{Tr}_{\mathcal{H}_{\rm top}(S^2)} (-1)^{R_{}} x^{ j_3}\;,	\label{eq:indices}
\end{split}
\end{align}
where $\mathcal{H}_{\rm rad}(S^2)$ stands for the radially quantized Hilbert-space of $\mathcal{T}[M]$ and $\mathcal{H}_{\rm top}(S^2)$ stands for the ground states Hilbert-space of $\mathcal{T}[M]$ on the topologically twisted $S^2$.   $R$ denotes the charge of the UV $U(1)$ R-symmetry and $j_3 \in \mathbb{Z}/2$ is the Cartan charge of the $SO(3)$ isometry of $S^2$.  The UV R-charge is always quantized as $R\in \mathbb{Z}$.

Now here comes our key observation. The conditions in \eqref{3-manifolds for top} imply that $(B^{\alpha}(x))^* = B^\alpha (x)$ for all $\alpha$ (see Appendix \ref{appendix : IR phases} for details). Therefore, for the $\mathcal{T}[M]$ from the 3-manifold $M$ satisfying \eqref{3-manifolds for top}, we find
\begin{align}\label{eq:idx=top}
\mathcal{I}_{\rm sci} (x)  = \mathcal{I}_{\rm top} (x) \quad   \textrm{ of } \mathcal{T}[M]  \;. 
\end{align}
The only possible explanation on the  match of two indices is that the theory $\mathcal{T}[M]$ flows to a topological theory so that the  partition functions only depend on the topology of the background Eucledian space-time. Let us explain this below. 

The superconformal index is in general an infinite power series in the fugacity $x^{1/2}$ for $R+2j_3$ charge. Note that the BPS operators captured in the superconformal index are labelled by their conformal dimensions $\Delta$ and spins via the BPS relation $R+2j_3=\Delta + j_3$. Generic (2+1)d SCFTs has no bound on $\Delta+j_3$ of the supersymmetric operators.
On the other hand, the refined index is a (finite) Laurent polynomial in $x^{1/2}$ since $\mathcal{H}_{\rm top}(S^2)$ is finite dimensional. This strongly suggests that the theories with \eqref{eq:idx=top} are topological. For topological theories, the superconformal index is  just 1, i.e. $\mathcal{I}_{\rm sci}(x)=1$, since the theory has no non-trivial local operator, while the refined twisted index is also  given by 1 since there is only one ground states on $S^2$.  
Independence of superconformal index on the fugacity $x$ is a characteristic property of topological  field theory. One can actually compute the superconformal index $\mathcal{I}_{\rm sci}(x)$ of $\mathcal{T}[M]$ using localization and show that it becomes 1 for various examples of $M$ satisfying \eqref{3-manifolds for top}. See \cite{Gang:2018gyt} for some concrete examples.

With this in mind, let us give some technical and general comments on flat connections on 3-manifolds. Conventionally, a $G$ flat connection for a Lie group $G$ is described by a gauge field configuration $\mathcal{A}_\alpha$ satisfying the flatness equation, $d\mathcal{A}_\alpha+\mathcal{A}_\alpha\wedge \mathcal{A}_\alpha=0$, modulo gauge transformations. A $G$ flat connection is fully characterized by its holonomy matrices $\rho_\alpha (a)$ along closed loops $\{a \}$ in the fundamental group,
\begin{align}
\rho_\alpha (a) = P \exp \left( \oint_a \mathcal{A} \right) \in G\;, \quad a \in \pi_1 M\;.
\end{align}
Here $P$ represents the path-ordered integral. Gauge transformations act on the holonomy matrices as simultaneous conjugation of $G$. The holonomy matrices of the flat connections  satisfy the fundamental group relations. Thus a $G$ flat connection can be alternatively described by a homomorphism $\rho_\alpha \in \textrm{Hom}[\pi_1 M \rightarrow G]$ up to conjugation of $G$. 
\begin{align}
\begin{split}
&\textrm{$G$ flat connection $\mathcal{A}_\alpha$ on $M$ }
\\
&\leftrightarrow  \frac{\textrm{Group homomorphsim $\rho_\alpha$ from $\pi_1 (M)$ to $G$}}{\textrm{Conjugation of $G$}}\;.
\end{split} 
\end{align}
Using the above  identification, the symbols $\mathcal{A}_\alpha$ and  $\rho_\alpha$ are interchangeably used throughout this paper.  The homomorphism description is much easier to handle in the actual computation and  distinguishes differences in the global structure of $G$, such as the difference between $SL(2,\mathbb{C})$ and $PSL(2,\mathbb{C})$.  An $SL(2,\mathbb{C})$ flat connection $\rho_\alpha$ is called {\it reducible} if their holonomy matrices are all mutually commuting and called {\it irreducible} otherwise. 

\subsection{Non-unitary modular structures}

We next discuss the $\mathcal{T}[M]$ theories with non-unitary modular structures.

In the above discussions, we assumed that there is no accidental $U(1)$ flavor symmetry in the IR limit of $\mathcal{T}[M]$.  However, there could be an additional accidental Abelian symmetry  in $\mathcal{T}[M]$.\footnote{When we say {\it accidental} $U(1)$ symmetry, it is a symmetry not manifest in the UV M-theoretical set-up  \eqref{6d def of T[M]}. The accidental symmetry is not necessarily accidental in the effective (2+1)d gauge theory $\mathcal{T}[M]$ proposed in \cite{Dimofte:2011ju,Gang:2018wek}. For most cases, the accidental symmetry is manifest in the effective 3d gauge theory. } 
If a new $U(1)$ flavor symmetry emerges in the lower dimensional theory, then the UV $U(1)$ R-symmetry of $\mathcal{T}[M]$ theory may differ from the proper IR R-symmetry  because  the UV R-symmetry can mix with the emergent $U(1)$ symmetry in the IR fixed point. In this case the superconformal R-symmetry of the IR CFT can be computed using F-extremization given in \cite{Jafferis:2010un}.  
Then the index $\mathcal{I}^{UV}_{\rm sci}(x)$ in \eqref{eq:indices}  computed  using the UV R-charge does not agree with the usual superconformal index of the IR CFT computed using IR R-charge. Namely, $\mathcal{I}^{UV}_{\rm sci}(x)\neq \mathcal{I}^{IR}_{\rm sci}(x)$ when the UV R-symmetry is not equal to the IR R-symmetry.
Our geometric conditions in \eqref{3-manifolds for top} only guarantees that $\mathcal{I}^{UV}_{\rm sci}(x)=1$, but not $\mathcal{I}^{IR}_{\rm sci}(x)=1$ when $\mathcal{I}^{UV}_{\rm sci}(x)\neq \mathcal{I}^{IR}_{\rm sci}(x)$. So we cannot say that the IR theory is topological. 

We suggest that the $\mathcal{T}[M]$ theory from a non-hyperbolic manifold $M$ of \eqref{3-manifolds for top} equipped with a non-unitary modular structure has such an emergent Abelian symmetry at low energy.
In addition, the IR theory at the fixed point is a SCFT instead of being topological. 
We again emphasize that the IR SCFT is a unitary theory as expected in the compactification of the unitary (5+1)d SCFT.
The non-unitary modular structure just resides in a special subset of the spectrum in the IR SCFT. We call this subset as the {\it non-unitary sector}.  The UV index $\mathcal{I}^{UV}_{\rm sci}(x)$ for such theory can be interpreted as a certain degeneration limit (by specializing the fugacity for the emergent $U(1)$ symmetry) of the full superconformal index $\mathcal{I}^{IR}_{\rm sci}(x)$ in the IR CFT. In this degeneration limit, the path-integral receives contributions only from the states in the non-unitary sector. One finds that $\mathcal{I}^{UV}_{\rm sci}(x)=1$ for the $\mathcal{T}[M]$ theories with non-unitary modular structures. This implies that the non-unitary sector has no local excitations and thus topological. We expect that the non-unitary sector of the IR CFT is described by the non-unitary modular tensor category. To ease our convention, we  will again refer to the non-unitary sector as $\textrm{TFT}[M]$. 
\begin{align}
\textrm{TFT}[M]:=\textrm{non-unitary sub-sector of $\mathcal{T}[M]$}\;.
\end{align}
As a consequence, we claim that a unitary theory $\mathcal{T}[M]$ provides a unitary extension of the non-unitary MTC.
Although denoted in the same fashion, one should distinguish  physical meaning of $\textrm{TFT}[M]$ for unitary cases from that for non-unitary cases. $\textrm{TFT}[M]$ describes the full IR dynamics of $\mathcal{T}[M]$ for unitary cases,  while $\textrm{TFT}[M]$ for non-unitary cases describes only the non-unitary sub-sector of $\mathcal{T}[M]$ in IR. 

In summary, we have the following conclusion
\begin{align}
\begin{split}
&\textrm{Modular structure associated to $M$ is non-unitary}
\\
&\Leftrightarrow \textrm{$\mathcal{T}[M]$ flows to a SCFT with  Abelian flavor symmetry}
\\
& \quad \;\;\textrm{containing non-unitary modular structure in its sub-sector}\;.
 \label{non-unitary/accidental symmetry}
\end{split}
\end{align}
Explicit checks  of this claim will be given in section \ref{sec : effecitve field theory} and \ref{sec : Examples}.

\section{Full Modular structure  of TFT$[M]$ with trivial $H_1 (M, \mathbb{Z}_2)$} \label{sec : full modular structure}
A consistent bosonic anyon theory, namely a UMTC, is characterized by the set of defining data associated with fusion and braiding. See the brief review at Appendix \ref{UMTC-Suppl}. The most fundamental data are the fusion rules $N^{\gamma}_{\alpha \beta}$ of anyons, and the two \textit{gauge-dependent} unitaries, i.e., $F$, $R$ symbols. From these, the two \textit{gauge-independent} topological data, which label the topological phases, can be derived: namely, $S$- and $T$-matrix. They contain the most quintessential properties of anyons, i.e., self and mutual statistics, and are called as ``modular structure" of the anyon theory. They can label the UMTC and thus can be used to distinguish the different UMTCs. 

As one of our main results of this paper, we propose a systematic algorithm for determining the modular structure of our geometric realization TFT$[M]$ from topological data of $M$. The algorithm is summarized in Table \ref{Table : MTC[M]}. 
\begin{table}[h]
	\begin{center}
		\begin{tabular}{|c|c|}
			\hline
			$\quad \textrm{TFT}[M]$ \quad &  3-manifold $M$
			\\
			\hline
			& \textrm{ Irreducible $SL(2,\mathbb{C})$ flat connections $\rho_\alpha$ on $M$}
			\\
            \textrm{Anyons } & \textrm{ or }
             \\
               & \textrm{Loop operators $\bigotimes_{\kappa }(a^{(\kappa)}_\alpha, R^{(\kappa)}_\alpha)$ on $M$}
			\\
			\hline
			$\textrm{GSD}_g$ &    $\sum_\delta (2 \textrm{Tor}[\rho_\delta])^{g-1} $
			\\
			\hline
			$T^\alpha_\beta$&  $ \delta^{\alpha}_\beta\exp \big{(}   -2\pi i  CS [\rho_\alpha ]\big{)}$
			\\
			\hline
			$S_{0 0}$&   $\big{|}  \sum_{\delta }  \frac{ \exp \left(- 2\pi i   CS[\rho_\delta]\right) }{2 \textrm{Tor}[\rho_\delta]} \big{ |} =  (2 \textrm{Tor}[\rho_{\alpha=0}])^{-1/2}$
			\\
			\hline
			$\mathcal{W}_\beta(\alpha)$ & $\pm \prod_{\kappa} \textrm{Tr}_{R^{(\kappa)}_\alpha} (\rho_\beta (a^{(\kappa)}_\alpha))$, \;Equation \eqref{Wab}
			\\
			\hline
			\textrm{Unitarity} &  \textrm{Equation }\eqref{unitarity}
			\\
			\hline
		\end{tabular}
	\end{center}
	\caption{Modular structure of TFT$[M]$, the topological phase associated to a non-hyperbolic 3-manifold $M$ satisfying \eqref{3-manifolds for top} with trivial $H_1(M,\mathbb{Z}_2)$. The information of central charge $c_{2d}$ and topological spins $\{h_\alpha\}$ of the   boundary chiral conformal field theory are encoded in the $T$-matrix as given in \eqref{T-matrx}. $\textrm{GSD}_g$ denotes the ground state degeneracy on a Riemann surface $\Sigma_g$ of genus $g$. We define $\mathcal{W}_\beta(\alpha):=\frac{S_{\alpha \beta}}{S_{0 \beta}}$,  from which one can compute S-matrix
	$S_{\alpha \beta} = \mathcal{W}_\beta (\alpha) \mathcal{W}_0 (\beta) S_{00}$ and the fusion coefficients $N_{\alpha \beta \gamma}=\sum_\delta (S_{0 \delta})^{2}\mathcal{W}_\delta (\alpha ) \mathcal{W}_\delta (\beta )\mathcal{W}_\delta (\gamma )  $. The primitive loop operator $(a,R)$ is labelled by the choice of an 1-cycle $a$ on $M$ and an irreducible representation $R$ of $SU(2)$. }
	\label{Table : MTC[M]}
\end{table}
In the table, $CS[\rho_\alpha]$ and $\textrm{Tor}[\rho_\alpha]$ are the Chern-Simons action and the adjoint Reidemeister torsion of an irreducible $SL(2,\mathbb{C})$ flat connection $\rho_\alpha$ on the internal 3-manifold $M$ respectively.  See Appendix \ref{app : CS and Tor} for more explanations on those invariants. These invariants  are complex valued  for general $SL(2,\mathbb{C})$ flat connections. However, for the 3-manifolds $M$ subject to the conditions in \eqref{3-manifolds for top}, the invariants evaluated on the flat connections are real since the flat connections are conjugate to $SU(2)$ or $SL(2,\mathbb{R})$ flat connections. 

In this  and subsequent sections, we mainly focus on $\textrm{TFT}[M]$ for a 3-manifold $M$ with trivial $H_1 (M, \mathbb{Z}_2)$.   From geometrical analysis in \cite{Eckhard:2019jgg}, the corresponding $\textrm{TFT}[M]$ is expected to be  a bosonic (non-spin) self-dual TQFT 
\begin{align}
\begin{split}
&\textrm{TFT}[M] = \textrm{Bosonic self-dual TQFT}\;,
\\
&\textrm{when $H_1 (M, \mathbb{Z}_2)$ is trivial}\;. \label{decoupled invertible spin TQFT}
\end{split}
\end{align}
Here, `self-dual' means that the charge conjugation in the TFT maps an anyon to itself, i.e. $S^2=C={\bf 1}$ (Appendix \ref{UMTC-Suppl}).
Generalization to the cases with non-trivial $H_1(M, \mathbb{Z}_2)$ will be discussed  in section \ref{sec : non-trivial Z2 homology} where we will see the emergence of a richer structures of topological phases.

\subsection{Dictionary for $\textrm{GSD}_g, S_{00}$ and $T^{\alpha}_\beta$} 
Certain supersymmetric observables, such as topologically twisted partition functions, in the UV theory $\mathcal{T}[M]$ can be written in terms of topological invariants on the internal 3-manifold $M$ \cite{Gang:2018hjd,Gang:2019uay,Benini:2019dyp}. In addition, these supersymmetric observables are protected and thus receive no quantum corrections along RG-flows. This enables us to find an explicit map between the UV observables in $\mathcal{T}[M]$ and the modular data in the IR topological theory TFT$[M]$. Exploiting this map, in this section, we clarify the dictionary summarized in Table \ref{Table : MTC[M]}. 

For this, we express some basic modular data encoded in the partition functions of $\textrm{TFT}[M]$ in terms of topological data of $M$. Recently, there have been huge progresses in computing partition functions of supersymmetric gauge theories on curved backgrounds using localization technique. Combining the localization technique with the field theoretic construction of $\mathcal{T}[M]$, we can exactly compute the partition functions of the IR $\textrm{TFT}[M]$ on appropriate curved backgrounds, from which we will extract the topological data.

Let $\mathcal{M}_{g,p}$ be the degree $p$ $S^1$-bundle over a Riemann surface $\Sigma_g$ of genus $g$.  When $g=0$ and $p=1$, for example, the manifold is the 3-sphere $S^3$, while the manifold is just $\Sigma_g \times S^1$ when $p=0$. By properly choosing the metric and turning on the background fields coupled to the $U(1)$ R-symmetry, we can put (2+1)d supersymmetric theories on $\mathcal{M}_{g,p}$  while preserving some supercharges  \cite{Benini:2015noa,Benini:2016hjo,Closset:2016arn,Closset:2017zgf,Closset:2019hyt}. Applying the localization technique to the supersymmetric $\mathcal{T}[M]$ on the $\mathcal{M}_{g,p}$ background, we can find\cite{Gang:2018hjd,Gang:2019uay,Benini:2019dyp}
\begin{align}
&Z\big{[}\mathcal{T}[M] \textrm{ on } \mathcal{M}_{g,p \in 2 \mathbb{Z}} \textrm{ with  fixed metric,\,background fields} \big{]}\nonumber
\\
& = \sum_\alpha \textrm{}(2 \textrm{Tor}[\rho_\alpha])^{g-1}\exp \left(-2\pi i  p CS[\rho_\alpha]\right)\;, \label{3d-3d for twisted parition functions}
\end{align}
which is valid for any compact 3-manifold $M$ with trivial $H_1(M, \mathbb{Z}_2)$. On $\mathcal{M}_{g,p}$ with even $p$, there are two possible spin structures preserving some supercharges depending on whether fermions are periodic or anti-periodic along the fiber $S^1$-direction \cite{Closset:2018ghr}.  The above formula is valid only for the anti-periodic boundary condition.  The anti-periodic boundary condition is not compatible with supersymmetry for odd $p$ and thus the above formula does not work for odd $p$. 
The summation is over all  irreducible $SL(2,\mathbb{C})$ flat connections on $M$. Protected by the supersymmetry, this partition function is invariant under the RG-flow.

When $\mathcal{T}[M]$ flows to a bosonic topological theory, i.e. when $M$ satisfies the conditions in \eqref{3-manifolds for top} and  has  trivial $H_1(M, \mathbb{Z}_2)$ , the partition functions on  $\mathcal{M}_{g,p}$ are independent of the choice of the metric and the background fields as well as the spin structure. So, in the case at hand, the above quantity becomes
\begin{align}
\begin{split}
&Z\big{[}\textrm{TFT}[M] \textrm{ on } \mathcal{M}_{g,p}  \big{]}
\\
& = \sum_\alpha \textrm{}(2 \textrm{Tor}[\rho_\alpha])^{g-1}\exp \left(-2\pi i  p CS[\rho_\alpha]\right)\;.
\end{split} \label{Z[M-pg]}
\end{align}
Note that we did not need to specify the choices, e.g., those of the metric, the background fields or the spin structure. Moreover, the formula is  valid even for odd $p$. 

This formula with $p=0$ reproduces the dictionary for the ground state degeneracy, $\textrm{GSD}_g$, in Table \ref{Table : MTC[M]}:
\begin{equation}
\textrm{GSD}_g= Z[\textrm{TFT}[M] \textrm{ on }  \Sigma_g \!\times\! S^1]=\sum_\delta (2 \textrm{Tor}[\rho_\delta])^{g-1}\ .
\end{equation}
From this result, one can also deduce that the flat connections on $M$ are mapped to the line operators, which represent anyons in the topological theory.
Furthermore, by comparing the above expression with the following general result of topological theories \cite{witten1989quantum,Blau:2006gh}
\begin{align}
\begin{split}
&Z\big{[}\textrm{TFT} \textrm{ on }\mathcal{M}_{g,p}  \big{]}= \sum_\alpha \textrm{}(S_{0\alpha})^{2-2g}(T^\alpha_\alpha)^p \ ,
\end{split}
\end{align}
we find
\begin{align}
\begin{split}
&(S_{0\alpha})^2 = \frac{1}{2\textrm{Tor}[\rho_\alpha]}\;, \quad T^\alpha_\alpha \sim \exp (-2\pi i CS[\rho_\alpha])\;.
\end{split} \label{S0a,T}
\end{align}
We note that the second relation is not the exact equality, but only the proportionality $\sim$. This is because the partition function \eqref{Z[M-pg]} suffers from an overall phase ambiguity. One can freely add the gravitational Chern-Simons counterterm to the UV Lagrangian and shift the overall phase of the partition function \cite{Closset:2012vp}. The partition function \eqref{Z[M-pg]} is essentially the partition function of the theory $\mathcal{T}[M]$ put on spin manifolds $\mathcal{M}_{g,p}$. So for the spin $\mathcal{M}_{g,p}$, the physically meaningful factor in the overall phase $e^{\frac{2\pi i k_g}{24}}$ is $k_g$ mod $\frac{1}{2}$. Here, $k_g$ is the gravitational Chern-Simons coefficient in the (2+1)d theory.

The $T$-matrix in a MTC is related to the basic data of the corresponding 2d boundary chiral CFT as follows \cite{witten1989quantum}
\begin{align}
T^{\alpha}_\beta = \delta^\alpha_\beta \exp \left(  \pm  2\pi i (h_\alpha - \frac{c_{2d}}{24})\right)\;. \label{T-matrx}
\end{align}
Here $h_\alpha$ (mod 1) is the conformal weight of the 2d primary field on which $\alpha$-th anyon ends, and $c_{2d}$ (mod 24) is the chiral central charge of the 2d CFT.
The conformal weight $h_\alpha$ is identified with the topological spin of the anyon. We have $h_{\alpha} =0$ for the vacuum state $\alpha=0$. The second relation in \eqref{S0a,T} allows us to compute $h_\alpha$ precisely from $CS[\rho_\alpha]$. However, the central charge $c_{2d}$ that is now identified with $k_g$ in the above relations is determined only  by $c_{2d}$ mod $\frac{1}{2}$, due to the overall phase ambiguity.

Finally, the sign $\pm$ choice \eqref{T-matrx} depends on the (2+1)d space-time orientation.  This orientation choice is also correlated with the orientation of the associated internal 3-manifold $M$. The Chern-Simons action $CS[\rho_\alpha]$ flips its overall sign under the orientation-reversal of either (2+1)d space-time or the internal $M$.

Let us now determine $S_{00}$ using the first equation of \eqref{S0a,T}. We first need to figure out which flat connection $\rho_{\alpha}$ corresponds to the trivial anyon ($\alpha =0$).  For this, we use the following   universal property of TQFTs
\begin{align}
\begin{split}
&S_{00} = |Z(\textrm{TFT on $S^3$})| = \big{|}\sum_\alpha (S_{0\alpha})^2 T^{\alpha}_\alpha\big{|}
\\
&\Rightarrow \frac{1}{\sqrt{2 \textrm{Tor}[\rho_{\alpha =0}]}} = \bigg{|}\sum_\alpha \frac{\exp (-2\pi i CS[\rho_\alpha])}{2 \textrm{Tor}[\rho_\alpha]} \bigg{|}\;. \label{trivial vacuum}
\end{split}
\end{align}
The first line follows from the $SL(2,\mathbb{Z})$ relation, $S_{00} = (T^{-1}ST^{-1}ST^{-1})_{00}$, combined with diagonality and unitarity of $T$. The  $SL(2,\mathbb{Z})$ structure  emerges only when $\mathcal{T}[M]$ flows to a topological theory. 
On the other hand, the summation on RHS can be defined for any 3-manifold $M$. For a general $M$, however, we do not expect that there is a special flat connection $\rho_{\alpha=0}$  satisfying the above relation. Only for $M$ with trivial $H_1 (M,\mathbb{Z}_2)$ with the condition in \eqref{3-manifolds for top}, we expect that there exists a special flat connection $\rho_{\alpha=0}$ satisfying the above relation, which we identify with the trivial anyon. 
From this analysis, we now have the following topological criterion on $M$ for the associated modular structure to be unitary 
\begin{align}
\begin{split}
&\textrm{Modular structure of $[M]$  is } 
\\
&\begin{cases}
&\textrm{unitary,} \;\textrm{if $ \big{|}  \sum_{\alpha }  \frac{ \exp \left(- 2\pi i CS[\rho_\alpha]\right) }{2 \textrm{Tor}[\rho_\alpha]} \big{ |} \leq   |2 \textrm{Tor}[\rho]|^{-1/2}$  \; $\forall \rho $ }
\\
&\textrm{non-unitary,} \quad  \textrm{otherwise}\;.
\end{cases}
\end{split} \label{unitarity}
\end{align}
This inequality simply means $|S_{00}|\leq |S_{0\alpha}|$ for $\alpha \neq 0$, which is the unitarity condition in modular tensor category. 

For some cases, there are multiple flat connections $\{\rho_{I},\rho_{II},\ldots \}$ which satisfy the condition for $\rho_{\alpha =0}$ in \eqref{trivial vacuum}.
We label such flat connections by Roman capital letters, $I,II,\cdots$.
In the case, we have a freedom to choose the true vacuum (or the trivial anyon) $\rho_{\alpha=0}$. Choosing any of these flat connections as the vacuum we can attempt to construct a modular structure. If a vacuum choice, say $\rho_{\alpha=o}=\rho_I$, leads to a consistent unitary modular structure, then we claim that there exists a (2+1)d topological theory described by the modular structure.
\begin{align}
\textrm{TFT}[M_{I}] := \textrm{TFT$[M]$ at the choice $\rho_{\alpha =0} = \rho_{I}$}\;.\label{True vacuum choice}
\end{align}
The subscript in $M_I$ denotes our vacuum choice.
Different vacuum choices among $\{\rho_{I},\rho_{II},\ldots \}$ for a given 3-manifold $M$ can give rise to distinct (2+1)d topological phases.
Quite interestingly, we observe that these topological phases are all related to each other by the Lorentz symmetry fractionalization (up to time reversal) studied in \cite{Hsin:2019gvb}, which may imply that the vacuum choice in the flat connections of $M$ is correlated with activating a non-trivial background for an anomalous $\mathbb{Z}_2$ 1-form symmetry in the (2+1)d TQFT using the Lorentz group background fields.

\subsection{Flat-connection-to-loop-operator map } \label{sec : flat-to-loop}
One characteristic property of a topological phase is that ground states on a two-torus are indexed by loop operators, or equivalently, anyons. For a general $\mathcal{T}[M]$ in \eqref{6d def of T[M]}, it is known that the ground states correspond to the set of irreducible flat connections on $M$ while the loop operators are related to those of $SL(2,\mathbb{C})$ Chern-Simons theory on $M$.  Therefore  when the $\mathcal{T}[M]$  flows to a topological theory, we expect that there exists a natural map between the two, i.e., flat connections  and loop operators on $M$. In this map, one can interpret  anyons in $\textrm{TFT}[M]$ as  M2-branes wrapping an 1-cycle inside $M$. 
The flat-connection-to-loop-operator map plays an important role below in calculating $S$-matrix. 

To proceed, let us remind a few basic properties of TQFTs. Generally, anyons in bosonic topological phases have two seemingly different physical interpretations. One is to consider anyons as quasiparticles. In the language of abstract quantum field theory,  they correspond to different types of loop operators.  The one-dimensional curve on which a loop operator is  supported  can be thought of as the trajectory of the corresponding quasiparticle. On the other hand,  anyons  can be considered as the label of the ground states on a 2-torus. In a general quantum field theory, two sets (the set of loop operators and of ground states on a 2-torus) are different. In bosonic topological phases, two sets are equivalent since all ground states on a 2-torus can be generated by acting line operators on the trivial vacuum $|0\rangle$
\begin{align}
|\alpha \rangle = \mathcal{O}^B_{\alpha} |0\rangle\;.
\end{align}
Here the quantum  operator $\mathcal{O}^B_\alpha$ denotes the loop operator of type $\alpha$ along a primitive one-cycle $B$ on a 2-torus. The quantum  operators  $\{\mathcal{O}^A_\alpha \}$ along the other primitive one-cycle $A$ are canonically conjugate to $\{\mathcal{O}^B_\alpha \}$. Without losing generality, we choose the ground state basis $|\alpha\rangle$ as simultaneous eigenstates of $\{\mathcal{O}^A_\alpha\}$. The loop operators around the two cycles are related to each other by the conjugation of $S$-matrix
\begin{align}
\mathcal{O}^A_\alpha = S^{-1} \mathcal{O}^B_{\alpha} S\;.
\end{align}
This reflects that the two one-cycles  are related to each other by the $S$-transformation in the mapping class group. 
On this basis, the operators $\mathcal{O}^A_\alpha$  and $\mathcal{O}^B_\alpha$ act as
\begin{align}
\begin{split}
&\mathcal{O}^A_\alpha |\beta \rangle =(S^{-1} \mathcal{O}^B_\alpha  S)  |\beta\rangle =  \mathcal{W}_\beta (\alpha) |\beta \rangle\;,
\\
&\mathcal{O}^B_\alpha |\beta \rangle = \mathcal{O}^B_\alpha \mathcal{O}^B_\beta |0 \rangle =\sum_\gamma N_{\alpha \beta}^\gamma |\gamma\rangle \;. \label{Action of 1-form on ground states}
\end{split}
\end{align}
We used the relation $(S^{-1}N_\alpha S)^{\gamma}_{\beta} = \delta^{\gamma}_{\beta} \mathcal{W}_{\beta}(\alpha)$ in the first line and $(N_\alpha)^{\gamma}_\beta :=N_{\alpha \beta}^\gamma$ is the fusion coefficient. Here $\mathcal{W}_\beta (\alpha) = \frac{S_{\alpha \beta}}{S_{0 \beta}}$. 

Now we use the above two different interpretations to identify the set of anyons in $\textrm{TFT}[M]$ from the topological data of $M$. First, by interpreting anyons as ground states on a 2-torus, one can identify them as a label of vacua on $M$. More precisely, the vacua are turned out to be the set of irreducible $SL(2,\mathbb{C})$ flat connections \cite{Dimofte:2010tz,Dimofte:2011ju}.  
\begin{align}
\begin{split}
&\textrm{Anyon of type $\alpha$ in TFT$[M]$}
\\
&\longleftrightarrow \;  \textrm{Irreducible $SL(2,\mathbb{C})$ flat connection $\rho_\alpha$ on $M$} \label{anyons on M-1}
\end{split}
\end{align}  
The above identification is also expected  from \eqref{3d-3d for twisted parition functions}.

On the other hand, by interpreting anyons as different types of loop operators, we can identify them as  loop operators of $SL(2,\mathbb{C})$ Chern-Simons theory on $M$ \cite{Gang:2015wya}.
A primitive loop operator on $M$ is specified by the choice of  an 1-cycle $a \in (\textrm{conjugacy classes in }\pi_1 (M))$ and an irreducible unitary representation  $R$ of $SU(2)$:
\begin{align}
\textrm{Loop operator } (a, R)  \; : \; \textrm{Tr}_R P \exp (\oint_a \mathcal{A})\;. \label{loop in SL(2) CS}
\end{align} 
Here  $\mathcal{A}$ is the $SL(2,\mathbb{C})$ gauge field.  $\textrm{Tr}_{R}(g)$ is the trace (character) of $g\in SL(2,\mathbb{C})$ taken in the representation $R$. More explicitly,
\begin{align}
&\textrm{Tr}_{\textrm{Sym}^{n=0}  \Box } (g)=1\;, \; \; \textrm{Tr}_{R = \textrm{Sym}^{n=1}  \Box} (g)=\textrm{Tr}(g)\;, \nonumber
\\
& \textrm{Tr}_{ \textrm{Sym}^{n+2}  \Box} (g)=\textrm{Tr}(g)  \textrm{Tr}_{\textrm{Sym}^{n+1}  \Box} (g)-\textrm{Tr}_{\textrm{Sym}^{n}  \Box} (g)\;.
\nonumber
\end{align}
We denote the $(n+1)$-dimensional unitary irreducible representation of $SU(2)$ by $R=\textrm{Sym}^n \Box$, $n$-th symmetric power of the fundamental representation $\Box$. 

The loop operators originate from M2-branes intersecting with the two coincident M5-branes wrapped on $M$ \cite{Alday:2009fs,Gaiotto:2009fs}. 
 To be supersymmetric (thus stable), the M2-branes should stretch along the time direction in (2+1)d space-time and wrap around an one-cycles in $M$. This thus gives line operators or equivalently anyons in $\textrm{TFT}[M]$. The type of the corresponding anyon depends on the choice of $a$, the one-cycle on internal 3-manifold $M$ along which the M2-branes are wrapping, and an irreducible representation $R= \textrm{Sym}^{n}\Box$, where the $n$ is the number of the M2-brane defects. General loop operators on $M$ can be obtained by taking direct products of the primitive ones,
\begin{align}
\begin{split}
&\textrm{Anyon of type $\alpha$ in TFT$[M]$}
\\
&\longrightarrow \; \textrm{Loop operator $\bigotimes_{\kappa } \left(a^{(\kappa)}_\alpha, R^{(\kappa)}_\alpha \right)$ on $M$} \label{anyons on M-2}\;,
\end{split}
\end{align}
where $\kappa$ runs over the primitive loop operators for $\alpha$-th anyon.  
The two relations \eqref{anyons on M-1} and \eqref{anyons on M-2} then naturally suggest
\begin{align}
\begin{split}
&\textbf{flat-connection-to-loop-operator map : }
\\
&\rho_\alpha \quad \longrightarrow  \quad \bigotimes_{\kappa } (a^{(\kappa)}_\alpha, R^{(\kappa)}_\alpha)\;,
\end{split} \label{flat-loop map}
\end{align}
for non-hyperbolic 3-manifolds satisfying \eqref{3-manifolds for top}.

The above two alternative viewpoints on anyons also endow the coefficient $\mathcal{W}_\beta (\alpha) =\frac{S_{\alpha \beta}}{S_{0\beta}}$ with a novel interpretation
\begin{align}
&\mathcal{W}_{\beta} (\alpha )= \langle \beta | \mathcal{O}^A_\alpha | \beta\rangle
\\ 
 & = \textrm{Expectation value of loop operator $\mathcal{O}_\alpha^{A}$ at the vacuum $|\beta\rangle$}\;. \nonumber
\end{align}
Consequently, we have the following dictionary 
\begin{align}
&\mathcal{W}_{\beta} (\alpha )=\textrm{VEV  of loop operator $\bigotimes_{\kappa } (a^{(\kappa)}_\alpha, R^{(\kappa)}_\alpha )$ }\nonumber
\\
& \qquad \qquad  \; \; \textrm{at the irreducible flat connection $ \rho_\beta $  }\;,
\\
&= 
\begin{cases}
 \prod_{\kappa} \textrm{Tr}_{R^{(\kappa)}_\alpha} \left(\rho_\beta (a^{(\kappa)}_\alpha)\right)\;, \quad \textrm{if TFT$[M]$ is unitary}\nonumber
 \\
 \prod_{\kappa} \textrm{Tr}_{R^{(\kappa)}_\alpha} \left(-\rho_\beta (a^{(\kappa)}_\alpha)\right)\;, \quad \textrm{otherwise}\;.\\
\end{cases} \label{Wab}
\end{align}
The  $\textrm{Tr}_R (\rho_\alpha (a))$ is nothing but the classical value of the loop operator $(a, R)$ in \eqref{loop in SL(2) CS} evaluated at the saddle point (flat connection) $\mathcal{A}_\alpha$. For non-unitary cases, the extra sign factor is necessary to obtain the correct $S$-matrix from  $S_{\alpha \beta} = \mathcal{W}_\beta (\alpha) \mathcal{W}_0 (\beta) S_{00}$, which gives a projective representation of  $SL(2,\mathbb{Z})$  combined with the $T$-matrix in \eqref{S0a,T}. 

Now, the remaining non-trivial task  is to  determine the flat-connection-to-loop-operator map in \eqref{flat-loop map}.
First, the flat connection $\rho_{\alpha=0}$ for the trivial anyon in \eqref{trivial vacuum} can naturally be mapped to the trivial loop:
\begin{align}
&\bigotimes_{\kappa } (a^{(\kappa)}_\alpha, R^{(\kappa)}_\alpha )\big{|}_{\alpha =0} = \left(\textrm{trivial loop with $R_\alpha^{(\kappa)} = \mathrm{Sym}^{n=0}\Box$}\right) \nonumber 
\\
& \Rightarrow  \mathcal{W}_\beta (\alpha =0)=1\;.
\end{align}
The non-trivial anyons can be identified as follows. We start with the fact that there are two ways of computing $S_{0\alpha}$: one is using the relation in \eqref{S0a,T} and the other one is using the relation $S_{0\alpha} = \mathcal{W}_{0}(\alpha)S_{00}$ from the dictionary in Table \ref{Table : MTC[M]}. By equating two results as
\begin{align}
\bigg{|}\sum_\beta \frac{\exp (-2\pi i CS[\rho_\beta])}{2 \textrm{Tor}[\rho_\beta]} \bigg{|} |\mathcal{W}_0 (\alpha) |=    \frac{ 1}{\sqrt{2\textrm{Tor}(\rho_\alpha)}} \,, \label{W0a}
\end{align}
one can determine $\mathcal{W}_{0}(\alpha)$ up to sign. The $\mathcal{W}_0(\alpha)$ is the classical value of the loop operator $\bigotimes_{\kappa} (a^{(\kappa)}_\alpha,R_\alpha^{(\kappa)})$ evaluated at the vacuum $\rho_{\alpha =0}$ as presented in \eqref{trivial vacuum}. The classical value severely restricts  possible candidates for the flat-connection-to-loop-operator map up to the gauge equivalence.\footnote{We regard two loop operators $(a,R)$ and $(a',R')$ as equivalent  at infrared if their expectation values $\textrm{Tr}_{R} (\rho_\alpha (a))$ and $\textrm{Tr}_{R'} (\rho_\alpha (a'))$ are identical for all irreducible flat connections $\rho_\alpha$.} 

Once a trial map is prepared, it is straightforward to construct the modular structure $S$- and $T$-matrices as well as the fusion coefficients $N_{\alpha \beta \gamma}$. One then needs to check if the result satisfies the following consistency conditions
\begin{align}
\begin{split}
&N_{\alpha \beta \gamma} \in \mathbb{Z}_{\geq 0}\;, \; N_{\alpha \beta 0} = \delta_{\alpha \beta} \;, \; S_{\alpha \beta} = S_{\beta \alpha}\;, \; 
\\
&S^2 = {\bf1}\;,\quad (ST)^3 = (\textrm{a phase factor})\times{\bf1}\;,
\end{split}
\end{align}
for being a UMTC.
The phase factor in $(ST)^3$ is of the form $e^{ i\frac{\pi n}{8}}|_{n\in \mathbb{Z}}$ and is not physically meaningful in our construction as it can be cancelled by the gravitational counterterms. We find that these  conditions are strong enough to uniquely fix the flat-connection-to-loop-operator map for a bosonic TFT$[M]$.

In the above, we assumed the self-dual property $S^2 = C=1$ for $\textrm{TFT}[M]$ from $M$ with trivial $H_1 (M, \mathbb{Z}_2)$. The self-duality is guaranteed for this case since all loop operators (or equivalently anyons) take values in unitary representations $R$ of $SU(2)$, which are real. To engineer non-self dual anyonic models in our geometric setup, one needs to consider another class of $M$ with non-trivial $H_1 (M, \mathbb{Z}_2)$ or increase the number of M5-branes. We will come back to these generalizations in the  section \ref{sec : non-trivial Z2 homology}.

\section{Supersymmetric gauge theories} \label{sec : effecitve field theory}
One nice feature of our construction of $\textrm{TFT}[M]$ is that it provides a concrete supersymmetric gauge theory  $\mathcal{T}[M]$  \cite{Dimofte:2011ju,Gang:2018wek}. In this section, we will review a general prescription to obtain a supersymmetric gauge theory  $\mathcal{T}[M]$ from a closed 3-manifold $M$ and provide several examples of its applications for non-hyperbolic manifolds.

\subsection{Dehn surgery description of 3-manifold : $M=( S^3\backslash \mathcal{K})_{p/q}$}One well-known way of constructing  closed 3-manifolds is using Dehn surgery along a knot $\mathcal{K}$ on a 3-sphere $S^3$. We denote the 3-manifold obtained by Dehn surgery along a knot $\mathcal{K}$ with slope $p/q \in \mathbb{Q}\cup \{\infty = 1/0\}$ by $(S^3\backslash \mathcal{K})_{p/q}$.

The Dehn surgery procedure is carried out in two steps: drilling and  Dehn filling. First, drilling is a procedure of  removing the tubular neighborhood of a knot $\mathcal{K}$ from a 3-sphere $S^3$. This creates a 3-manifold $S^3\backslash \mathcal{K}$ called the knot complement of $\mathcal{K}$:
\begin{align}
S^3 \backslash \mathcal{K} := S^3 - (\textrm{tubular neighborhood of a  knot $\mathcal{K}$})\;.
\end{align}
The knot complement has a single two-torus ($\mathbb{T}^2$) boundary surrounding  the removed tubular neighborhood.  There is a canonical basis choice for  1-cycles on the boundary called the meridian ($\mu$) and the longitude ($\lambda$) defined as
\begin{align}
H_1 \left(\partial (S^3\backslash \mathcal{K} ),\mathbb{Z}\right) = H_1 (\mathbb{T}^2,\mathbb{Z}) = \mathbb{Z} \times \mathbb{Z} = \langle \mu, \lambda \rangle \ .
\end{align}

Dehn filling is a procedure of gluing  a solid-torus back to the knot complement.  The way of  gluing back is not unique but depends on the choice of a boundary 1-cycle $p\mu + q\lambda$ that  will be glued to the shrinkable boundary cycle of the solid-torus. The closed 3-manifold after the Dehn filling procedure is denoted as
$ (S^3\backslash \mathcal{K} )_{p /q} $.
\begin{align}
&(S^3\backslash \mathcal{K})_{p/q} = \big{(} (S^3\backslash \mathcal{K}) \cup (\textrm{solid-torus}) \big{)}/\sim \;,
\\
& (p\mu + q\lambda) \sim (\textrm{shrinkable boundary 1-cycle of solid-torus})\;. \nonumber
\end{align}
Obviously, $(S^3\backslash \mathcal{K})_{p/q =1/0}$ is just $S^3$ for any knot $\mathcal{K}$. 

The Dehn surgery construction can be extended to more general cases by replacing a knot $\mathcal{K}$ by a general link $\mathcal{L}$ which consists of several knots
\begin{align}
(S^3\backslash \mathcal{L})_{\{p_i/q_i\}}\;, \quad \mathcal{L} = \bigcup_i \mathcal{K}_i\;.
\end{align}
In the case, we need to specify a slope $p_i/q_i$ for each component of the link. As one of fundamental theorems in 3d topology, it is known that any closed orientable 3-manifold can be constructed from a Dehn surgery on a link on $S^3$ \cite{lickorish1962representation,wallace1960modifications}. 

To construct the (2+1)d gauge theory $\mathcal{T}[M]$ associated to a closed 3-manifold $M$, we first need to choose a Dehn surgery description of the 3-manifold. To avoid clutter, let us assume that the 3-manifold is given by a Dehn surgery along a knot $\mathcal{K}$, i.e. $M=(S^3\backslash \mathcal{K})_{p/q}$.  One can first construct a field theory $\mathcal{T}[S^3\backslash \mathcal{K}]$ associated to the knot complement $S^3\backslash \mathcal{K}$ using an  algorithm proposed in \cite{Dimofte:2011ju}, which is based on an ideal triangulation of the knot complement. It is known that this knot complement theory $\mathcal{T}[S^3\backslash \mathcal{K}]$ has a $SO(3)$ flavor symmetry at IR \cite{Gang:2018wek}. Then, the theory associated to the Dehn filled closed 3-manifold $\mathcal{T}[M=(S^3\backslash \mathcal{K})_{p/q}]$ is given by
\begin{align}
\mathcal{T}[M=(S^3\backslash \mathcal{K})_{p/q}] = \frac{\mathcal{T}[S^3\backslash \mathcal{K}]}{SO(3)_{p/q}}\;.
\end{align}
Here, $ \mathcal{T}/ SO(3)_{p/q}$ stands for ``gauging the $SO(3)$ symmetry with Chern-Simons level $p/q$" of a (2+1)d gauge theory $ \mathcal{T}$.
The ``gauging'' procedure in (2+1)d gauge theory corresponds to the Dehn filling operation in the internal 3-manifold. 
It is a usual supersymmetric gauging  when $p/q \in 2\mathbb{Z}$. We also refer to \cite{Gang:2018wek} for more explanations on the ``gauging'' operation with general $p/q \in \mathbb{Q} \cup \{1/0\}$.

In the subsequent sections,  we give  explicit constructions of   $\mathcal{T}[M]$ for $M=(S^3\backslash \mathbf{4}_1)_{p/q},(S^3\backslash \mathbf{5}_2)_{p/q}$ and $(m007)_{p/q}$ where $ \mathbf{4}_1$ is the figure-eight knot (or 2-twist knot) and $\mathbf{5}_2$ is the 3-twist knot. For naming knots,  we follow the Alexander–Briggs notation.  $\mathbf{4}_1$ is the 1st simplest knot with $4$ crossings  while  $\mathbf{5}_2$ is the 2nd simplest (next to $\mathbf{5}_1$, a torus knot) knot with $5$ crossings. These are the 1st and 2nd simplest hyperbolic knots. A knot $\mathcal{K}$ is called hyperbolic when its complement  $S^3\backslash \mathcal{K}$ is hyperbolic. 

$m007$ is a hyperbolic 3-manifold with a torus boundary. Here we follow the nomenclature of Snap{P}y \cite{SnapPy}.  This manifold cannot be realized as a knot complement on $S^3$. Instead, it can be obtained by performing a Dehn filling on one torus boundary of the Whitehead link complement, $\mathbf{5}^2_1$. The link is the 1st simplest link with 2 components and $5$ crossings. $(m007)_{p/q}$ is the closed 3-manifold obtained from the Dehn filling along a boundary 1-cycle determined by $p/q$. In terms of Dehn surgery along the Whitehead link, $(m007)_{p/q} = (S^3 \backslash \mathbf{5}^2_1)_{-3/2, -p/q-3} $. The Whitehead link is symmetric under the exchange of two components and thus $(S^3 \backslash \mathbf{5}^2_1)_{p_1/q_1, p_2/q_2} = (S^3 \backslash \mathbf{5}^2_1)_{p_2/q_2, p_1/q_1}$

\subsection{$\mathcal{T}[M=(S^3\backslash \mathbf{4}_1)_{p/q}]$ }
The (2+1)d gauge theory from $M= (S^3 \backslash \mathbf{4}_1)_{p/q}$ is  \cite{Dimofte:2011ju,Gang:2018wek}
\begin{align}
\begin{split}
&\mathcal{T}\left[M= (S^3 \backslash \mathbf{4}_1)_{p/q} \right]  
= \frac{\mathcal{T}[S^3 \backslash \mathbf{4}_1]}{SO(3)_{p/q}}\quad \textrm{where}
\\
& \mathcal{T}[S^3\backslash \mathbf{4}_1] = \textrm{$U(1)_0$ coupled to 2 $\Phi$'s of charge $+1$}\;.
\end{split} \label{T[41]}
\end{align}
Here $U(1)_k$ means the $U(1)$ gauge theory at Chern-Simons level $k$ and $\Phi$ is a supersymmetric chiral multiplet which consists of a complex scalar  and a Weyl fermion. The theory  $ \mathcal{T}[S^3\backslash \mathbf{4}_1]$ has the manifest $SU(2)\times U(1)$ flavor symmetry as well as the $U(1)$ R-symmetry. The $SU(2)$ acts on the 2 chiral multiplets while the $U(1)$ comes from so-called topological symmetry associated to the dynamical $U(1)$ gauge field. The charge of the topological symmetry is given by the magnetic flux of the Abelian gauge field. 
Interestingly, the theory $ \mathcal{T}[S^3\backslash \mathbf{4}_1]$  is known to have the flavor symmetry enhancement $SU(2)\times U(1)\rightarrow SU(3)$ at the IR fixed point \cite{Gang:2018wek,Benini:2018umh,Gaiotto:2018yjh}.  After  gauging the $SO(3)$ subgroup of the $SU(3)$, however, this theory does not have any flavor symmetry.

The 3-manifolds $(S^3 \backslash \mathbf{4}_1)_{p/q}$ with $p/q$  other than $p/q \in   \{0, \pm 1,\pm 2,\pm 3,\pm 4, 1/0\}$ are all hyperbolic. For those hyperbolic $M$, the gauge theory $\mathcal{T}[M]$ flows to a non-trivial superconformal field theory. When $p/q \in  \{0, \pm 1, \pm 2, \pm 3\}$, on the other hand, the 3-manifolds $(S^3 \backslash \mathbf{4}_1)_{p/q}$ are non-hyperbolic and satisfy the condition in \eqref{3-manifolds for top}. Also, $(S^3 \backslash \mathbf{4}_1)_{p/q}$ with odd $p$ has trivial $H_1(M, \mathbb{Z}_2)$. For those cases, one can compute the superconfomal index and check that \cite{Gang:2018gyt} (for $p/q \in  \{0, \pm 1, \pm 2, \pm 3\}$) ,
\begin{align}
(\mathcal{I}_{\rm sci}(x) \textrm{ of $\mathcal{T}[(S^3\backslash \mathbf{4}_1)_{p/q}]$}) =1\;.
\end{align}
Note that since the IR theory has no flavor symmetry, the IR R-symmetry will be the same as the UV R-symmetry. So the above index as it is is the superconformal index of the IR fixed point and it counts only a single operator, the vacuum. This is a strong signal that the IR theory has no local excitations.
From this, one can expect that  the $\mathcal{T}\left[M= (S^3 \backslash \mathbf{4}_1)_{p/q} \right]$ flows not to a SCFT with local operators, but instead to a topological theory.   Accordingly, we shall compute the modular structures of $M= (S^3 \backslash \mathbf{4}_1)_{p/q = \pm 1, \pm 3}$ in  section \ref{sec : Examples} and confirm that they are all unitary. Unitarity of the modular structure  plugged in the conjecture in \eqref{main conjecture} implies that the IR limit of $\mathcal{T}\left[M= (S^3 \backslash \mathbf{4}_1)_{p/q} \right]$ is a unitary topological quantum field theory, which is indeed in a good agreement with our expectation from the index computation.

\subsection{$\mathcal{T}[M=(S^3 \backslash \mathbf{5}_2)_{p/q} ]$ : additional $U(1)$} The (2+1)d gauge theory from $M=(S^3 \backslash \mathbf{5}_2)_{p/q}$ is \cite{Gang:2018wek}
\begin{align}
\begin{split}
&\mathcal{T}\left[M= (S^3 \backslash \mathbf{5}_2)_{p/q} \right]   =\frac{\mathcal{T}[S^3 \backslash \mathbf{5}_2]}{SO(3)_{p/q}}  \;, \;\textrm{where}
\\
& \mathcal{T}[S^3\backslash \mathbf{5}_2] = \textrm{$U(1)_{-\frac{1}2}$ coupled to 3 $\Phi$'s of charge $+1$}\;. \label{T[52]}
\end{split}
\end{align}
Note that  the correct quantization for  Chern-Simons level $k$ of the dynamical $U(1)$ gauge field is $k \in \mathbb{Z}+ \frac{\textrm{(the number of $\Phi$s)}}2 = \mathbb{Z}+\frac{1}2$ due to the one-loop shift of Chern-Simons level after integrating out the chiral fermions in 3 $\Phi$'s. 

The  theory $\mathcal{T}[S^3\backslash \mathbf{5}_2]$  has $SU(3)\times U(1)$ flavor symmetry, where the $SU(3)$ rotates 3 $\Phi$'s and $U(1)$ is the topological symmetry. In the above, we ``gauge'' the $SO(3)\subset SU(3)$ flavor symmetry to obtain $\mathcal{T}\left[M= (S^3 \backslash \mathbf{5}_2)_{p/q} \right] $. Taking into account of  the background Chern-Simons level $-4$ of the $SO(3)$ flavor symmetry in $\mathcal{T}[S^3\backslash \mathbf{5}_2]$, the actual ``Chern-Simons level" in the gauging should be $p/q-4$.
 After the gauging, the  theory $\mathcal{T}\left[M= (S^3 \backslash \mathbf{5}_2)_{p/q} \right]  $ still has the $U(1)$ flavor symmetry. 

When $p/q \in  \{ 1,  2,  3\}$,  the 3-manifolds $(S^3 \backslash \mathbf{5}_2)_{p/q}$ are non-hyperbolic and satisfy the condition in \eqref{3-manifolds for top}. Furthermore, the 3-manifold $(S^3 \backslash \mathbf{5}_2)_{p/q}$  has trivial $H_1(M,\mathbb{Z}_2)$ when $p/q \in  \{ 1,    3\}$.  One computes the superconformal indices for these cases using localization technique,
\begin{align}
\begin{split}
&(\mathcal{I}_{\rm sci}(x) \textrm{ of $\mathcal{T}[(S^3\backslash \mathbf{5}_2)_{p/q=1}]$}) 
\\
&= 1+\left(u-1 \right) x+ \left(-2+ u^2 +\frac{1}u\right) x^2 +\ldots\;,
\\
&(\mathcal{I}_{\rm sci}(x) \textrm{ of $\mathcal{T}[(S^3\backslash \mathbf{5}_2)_{p/q=3}]$}) 
\\
&= 1+\left(u-1 \right) x+ \left(-2+ u +\frac{1}u\right) x^2 +\ldots\;.
\end{split}
\end{align}
Here, the superconformal index is computed using the UV R-symmetry. $u$ is the fugacity for the additional $U(1)$ symmetry. The superconformal index with a different R-symmetry can be obtained  by replacing $u$ in the above by $u(-x^{1/2})^\nu$.  $\nu$ is the parameter for the mixing of the R-symmetry with the $U(1)$ flavor symmetry that for the superconformal R-symmetry can be fixed by F-maximization in \cite{Jafferis:2010un}. The above indices are non-trivial and thus these theories will flow to non-trivial SCFTs.

Amazingly, in the limit $u\rightarrow1$, a huge cancellation occurs  and the index reduces to just $1$!
\begin{align}
\begin{split}
&(\mathcal{I}_{\rm sci}(x) \textrm{ of $\mathcal{T}[(S^3\backslash \mathbf{5}_2)_{p/q=1,3}]$})|_{u=1} =1 \ .
\end{split}
\end{align}
This is a quite remarkable result showing that the index in this limit receives no contribution from local operators in the CFT.
This strongly suggests that the IR SCFT may have a topological sub-sector in the limit $u\rightarrow1$. 
In  section \ref{sec : Examples}, we will show that the non-hyperbolic 3-manifold $M= (S^3 \backslash \mathbf{5}_2)_{p/q=1,3} $ actually has a non-unitary modular structure. 
We therefore suggest that the $\mathcal{T}\left[M= (S^3 \backslash \mathbf{5}_2)_{p/q=1,3} \right]$  flows to a superconformal field theory with $U(1)$ flavor symmetry, and includes a topological sub-sector, which can be isolated in the special limit $u\rightarrow1$, described by a non-unitary MTC. Thus the above index computations before/after taking the limit $u\rightarrow1$ give a non-trivial confirmation of the claim in \eqref{non-unitary/accidental symmetry}.

\subsection{$\mathcal{T}[(m007)_{p/q} = (S^3 \backslash \mathbf{5}^2_1)_{-3/2, -p/q-3} ]$} 
For $M=(m007)_{p/q}   $, the corresponding (2+1)d gauge theory is \cite{Gang:2018wek}
\begin{align}
\begin{split}
&\mathcal{T}\left[M= (m007)_{p/q}  \right]   =\frac{\mathcal{T}[m007]}{SO(3)_{p/q}}  \;, \;\textrm{where}
\\
& \mathcal{T}[m007] = \textrm{$U(1)_{\frac{3}2}$ coupled to 4 chirals $(\Phi_1, \Phi_2, \Phi_3, M)$ }\;
\\
&\qquad  \qquad  \;\;\;\; \textrm{of charge $(1,1,1,-2)$ with superpotential}
\\
&\qquad  \qquad  \;\;\;\; W = M(\Phi_1^2 +\Phi_2^2 +\Phi_3^2)\;. \nonumber
\end{split}
\end{align}
The  theory $\mathcal{T}[m007]$  has $SO(3)\times U(1)$ flavor symmetry, where the $SO(3)$ rotates 3 $\Phi$'s and the $U(1)$ is the topological symmetry. Gauging the $SO(3)$ flavor symmetry leads to $\mathcal{T}\left[M= (m007)_{p/q}   \right] $. Taking into account of  the background Chern-Simons level $-1$ of the $SO(3)$ flavor symmetry in $\mathcal{T}[m007]$, the actual ``Chern-Simons level" in the gauging should be $p/q-1$.
The resulting theory $\mathcal{T}\left[M= (m007)_{p/q} \right]  $ has the $U(1)$ flavor symmetry. 

When $p/q \in  \{ -2,  -1,  0,1,2\}$,  the 3-manifolds $(m007)_{p/q}$ are non-hyperbolic and satisfy the condition in \eqref{3-manifolds for top}; the 3-manifold $(m007)_{p/q} $  has trivial $H_1(M,\mathbb{Z}_2)$ when $p/q \in  \{ -2, 0,2\}$.   One computes the superconformal indices with the UV R-symmetry for these cases,
\begin{align}
\begin{split}
&(\mathcal{I}_{\rm sci}(x) \textrm{ of $\mathcal{T}[(m007)_{p/q=-2}]$}) 
\\
&= 1+\left(u-1 \right) x+ \left(-2+ u^2 +\frac{1}u\right) x^2 +\ldots\;,
\\
&(\mathcal{I}_{\rm sci}(x) \textrm{ of $\mathcal{T}[(m007)_{p/q=0,2}]$}) = 1\;,
\end{split}
\end{align}
where $u$ is the fugacity for the additional $U(1)$ symmetry. Interestingly, the $u$-dependence in the index disappears for $p/q=0$ or $2$.

The index for the $\mathcal{T}\left[M= (m007)_{p/q=-2} \right]$ is non-trivial and it depends on the fugacity $u$. Moreover, the index reduces to `1' in the limit $u\rightarrow1$.
As for other examples above, we suggest that the $\mathcal{T}\left[M= (m007)_{p/q=-2} \right]$  flows to a superconformal field theory with $U(1)$ flavor symmetry at low energy that contains a non-unitary topological sub-sector of the limit $u\rightarrow1$.  In  section \ref{sec : Examples}, we will confirm that the modular structure associated to the 3-manifold  $M= (m007)_{p/q=-2} $ is actually non-unitary. This gives another non-trivial confirmation of the claim in \eqref{non-unitary/accidental symmetry}. 

On the other hand, the index for the theory $\mathcal{T}\left[M= (m007)_{p/q=0,2} \right]$ is trivial and independent of $u$. This signals that the theory is topological and the $U(1)$ global symmetry decouples at low energy.
In  section \ref{sec : Examples}, we will also confirm that the modular structure associated to the 3-manifold  $M= (m007)_{p/q=0,2} $ is actually unitary. 
We thus conclude that the $\mathcal{T}\left[M= (m007)_{p/q=0,2} \right]$ flows to a topological theory described by a unitary MTC. 
This agrees with  the conjecture in \eqref{main conjecture}.

\section{Examples} \label{sec : Examples}
\subsection{Seifert fiber manifold $S^2(\frac{p_1}{q_1},\frac{p_2}{q_2},\frac{p_3}{q_3})$}
Let us now introduce a class of non-hyperbolic 3-manifolds defined as
\begin{eqnarray}
&S^2 \left(\frac{p_1}{q_1}, \frac{p_2}{q_2},\frac{p_3}{q_3}\right) := (\textrm{Seifert fiber manifold over $S^2$})\;, \nonumber
\\
& \textrm{$(p_i, q_i)$ are coprimes with  $p_i > 0$ }\;,
\end{eqnarray}
which provides infinitely many examples of 3-manifolds  satisfying the topological conditions in \eqref{3-manifolds for top}.
\begin{figure}[h!]
	\begin{center}
		\includegraphics[width=.22\textwidth]{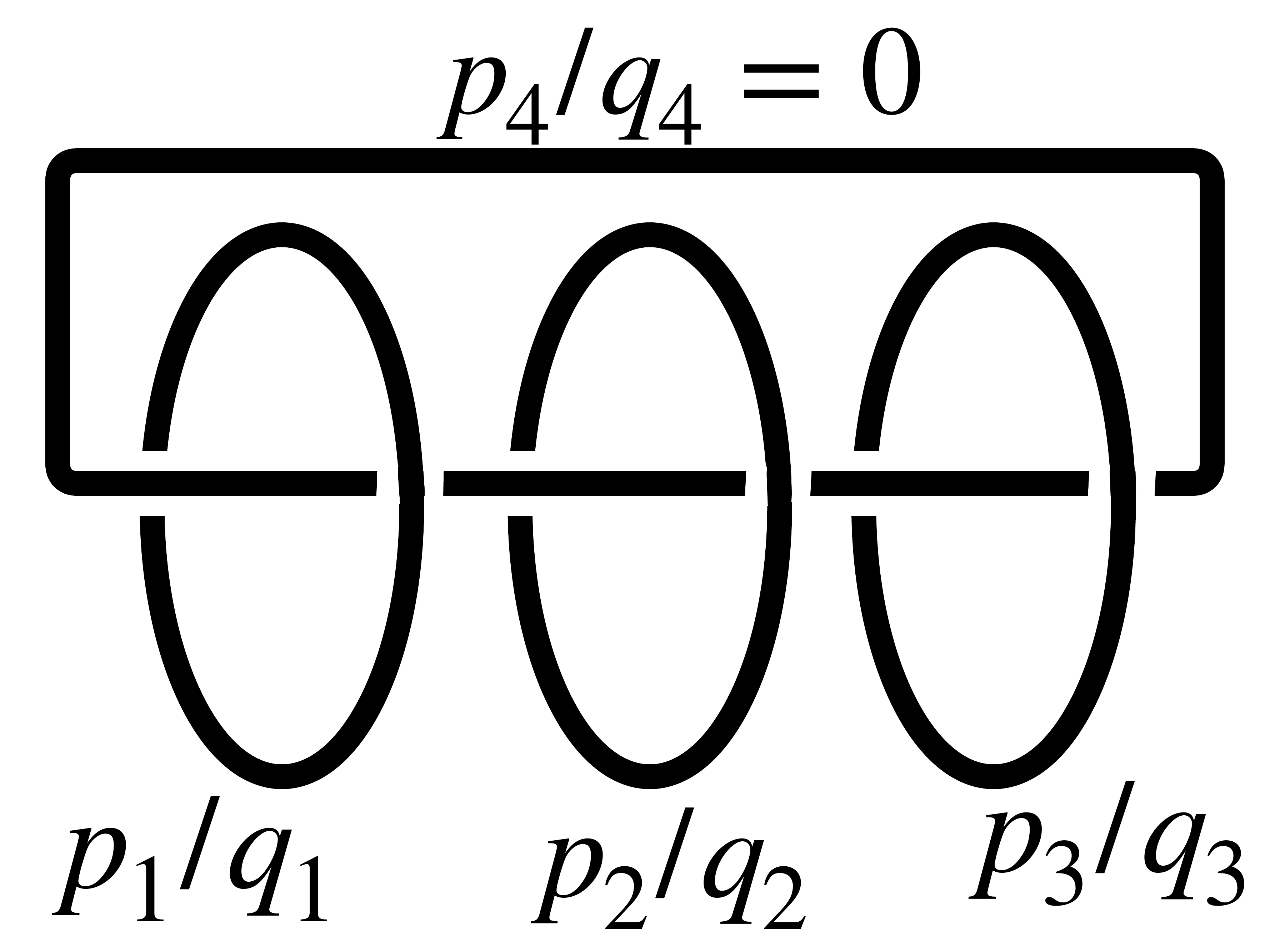}
	\end{center}
	\caption{Dehn surgery presentation of the Seifert fiber manifold $S^2(\frac{p_1}{q_1},\frac{p_2}{q_2},\frac{p_3}{q_3})$ using a link with 4 components.}   
	\label{fig: Seifert-fiber}
\end{figure}
See figure \ref{fig: Seifert-fiber} for the Dehn surgery representation of this 3-manifold.
This manifold is invariant under the permutation of the three rational numbers $\{p_i/q_i\}_{i=1,2,3}$.
The fundamental group of the manifold is
\begin{align}\label{fund group}
&\pi_1 \left(S^2\left(\frac{p_1}{q_1}, \frac{p_2}{q_2},\frac{p_3}{q_3}\right)  \right ) 
\\
&= \bigg{\langle} x_1,x_2,x_3, h \;: \; x_i^{p_i}h^{q_i}=1, \; h x_i = x_i h, \; \prod_{i=1}^3 x_i=1 \bigg{\rangle}\;. \nonumber 
\end{align}
The manifold has trivial $H_1 (M, \mathbb{Z}_2)$ if and only if  
\begin{align}
p_1 p_2 p_3 \left(\frac{q_1}{p_1} + \frac{q_2}{p_2}+ \frac{q_3}{p_3}\right) \in 2\mathbb{Z}+1\;. \nonumber
\end{align}

The Chern-Simons action $CS[\rho]$ and the adjoint Reidemeister torsion $\textrm{Tor}[\rho]$ for irreducible flat connections $\rho$ on the 3-manifold are known in \cite{Rozansky_1995}.
Let  $\rho$ be an irreducible flat connection on $S^2 (\{p_i/q_i\}_{i=1}^3)$ whose holonomy matrices are of the following form, \footnote{To be an irreducible flat connection, $\rho(h)$ should be an element of center $\mathbb{Z}_2 \in SL(2,\mathbb{C})$.  Otherwise, all $\rho(x_i)$ with $i=1,2,3$ as well as $\rho (h)$ belong to the same $GL(1,\mathbb{C})$ subgroup of   $SL(2,\mathbb{C})$ and thus $\rho$ is reducible.  }
\begin{align}\label{n-lambda}
&\rho (h) = \textrm{diag}\{ e^{2\pi i \lambda}, e^{-2\pi i \lambda}\}\; \quad \textrm{with } \lambda_\alpha \in \bigg{\{} 0, \frac{1}2 \bigg{ \}}\;\nonumber
\\
&\textrm{and} 
\\
&\textrm{eigenvalues of $\rho (x_i)$} =\bigg{\{}\! \exp \left(\pm2\pi i \frac{n_{ i }}{p_i}\right)\!\bigg{\}}\;\textrm{with }n_{i} \in \frac{1}2\mathbb{Z}\;. \nonumber
\end{align}
Then we have
\begin{align}
\begin{split}
&CS[\rho] = \sum_{i=1}^3 \left( \frac{r_i}{p_i} n_{ i}^2 - q_i s_i \lambda^2  \right) \quad \textrm{(mod 1)}\;,
\\
& \textrm{Tor}[\rho] = \prod_{i=1}^3 \frac{p_i}{4\sin^2 \big{(}2\pi (\frac{r_i}{p_i} n_{ i} + s_{i} \lambda)\big{)}}\;, \label{CS-and-torsion-Seifert}
\end{split}
\end{align}
where the integers $(r_i,s_i)$ are chosen such that $p_i s_i -q_i r_i =1$. The choice of $(r_i,s_i)$ is not unique, but the above invariants are independent of the choice.

\subsection{$M=S^2(3,3,-\frac{4}3) = (S^3\backslash \mathbf{4}_1)_{p/q=3}$ }  \label{subsec : 3 3 -4/3}
From the  topological fact that $M=S^2(3,3,-4/3) = (S^3\backslash \mathbf{4}_1)_{p/q=3}$  \cite{dunfield2018census,DVN/6WNVG0_2018}, the (2+1)d theory TFT$[M]$ is expected to describe  the  IR physics of  $\mathcal{T}[(S^3\backslash \mathbf{4}_1)_{p/q}]$ in \eqref{T[41]} with $p/q = 3$. In the notation of \cite{dunfield2018census},  $(S^3\backslash \mathbf{4}_1)_{p/q}$ corresponds to $(m004)_{p/q}$.\footnote{More precisely, they denote it by $m004 (p,q)$. The Seifert manifold $S^2(\frac{p_1}{q_1}, \frac{p_2}{q_2},\frac{p_3}{q_3})$ is denoted by $S^2\left((p_1,q_1),(p_2,q_2),(p_3,q_3)\right). $}

 There are 3 irreducible flat $SL(2,\mathbb{C})$ connections $\rho_{\alpha =0,1,2}$ on this 3-manifold with 
\begin{align}
&CS[\rho_{\alpha=0}] = \frac{25}{48} ,\; CS[\rho_{\alpha=1}] = \frac{1}{12},\; CS[\rho_{\alpha=2}] = \frac{1}{48}\; , \nonumber
\\
&\textrm{Tor}[\rho_{\alpha = 0}] = \textrm{Tor}[\rho_{\alpha =2}] =2, \;  \textrm{Tor}[\rho_{\alpha = 1}]=1\;. 
\label{CS-Tor for M[3,3,-4/3]}
\end{align}
Refer to Appendix \ref{app : examples of irreducible flat connections} for detailed computations. According to Table \ref{Table : MTC[M]}, the $\textrm{GSD}_{g}$, $S_{00}$ and $T$-matrix of $M$ are given by
\begin{align}
\begin{split}
&\{ \textrm{GSD}_{g} \}_{g=0,1,2,\ldots}  = \{1,3,10,36,136,528,\ldots \} \;,
\\
&S_{00} = \frac{1}2\;, \quad T = \textrm{diag} \{ e^{-\frac{25 i \pi}{24}}, e^{-\frac{ i  \pi}{6}},e^{-\frac{i \pi}{24}}  \} \ .
\end{split}
\end{align}
The topological spins $h_\alpha  =\pm  ( CS[\rho_{\alpha}] - CS[\rho_{\alpha =0}])$ for three anyons are
\begin{align}
\{h_\alpha \}_{\alpha =0}^{2} = \left\{0,\tfrac{9}{16},\tfrac{1}{2} \right\}\;.
\end{align}
Since  $S_{00}= \big{|}\sum_\delta \frac{\exp(-2\pi i CS[\rho_\delta])}{2 \textrm{Tor}[\rho_\delta]} \big{|} \leq  \frac{1}{\sqrt{2 \textrm{Tor} [\rho_\alpha]}}$ for all $\alpha$, the $\textrm{TFT}[M]$ is a unitary topological theory according to \eqref{unitarity}. The unitarity  is expected since  the effective supersymmetric theory $\mathcal{T}[M]$ in \eqref{T[41]} does not have any Abelian symmetry (see \eqref{non-unitary/accidental symmetry}). In the above, we chose $\rho_{\alpha =0}$ as the true vacuum. The 3-manifold with this vacuum choice will be denoted by $S^2(3,3,-\frac{4}{3})_{\rm I}$. The same manifold with a distinct vacuum choice that we refer to as $S^2(3,3,-\frac{4}{3})_{\rm II}$ will be discussed below.  

Using the formula for the $\mathcal{W}_0 (\alpha)$ in  \eqref{W0a}, we compute 
\begin{align}
|\mathcal{W}_{0} (\alpha =0,2)| =  1\;, \quad |\mathcal{W}_{0} (\alpha =1)| =  \sqrt{2}\;.
\end{align}
From the classical value of loop operator at the flat connection $\rho_{\alpha =0}$, one can easily deduce the flat-connection-to-loop-operator map as follows.
\begin{align}
&  \rho_{\alpha } \rightarrow (a=x_3 , R =\textrm{Sym}^\alpha \Box)\;, \nonumber
\\
&\Rightarrow \{ \mathcal{W}_0 (\alpha ) \}_{\alpha=0}^2= \{1,\sqrt{2},1\}, \quad \{ \mathcal{W}_1 (\alpha ) \}_{\alpha=0}^2 = \{1,0,-1\}, \nonumber
\\
 & \quad \;\; \{ \mathcal{W}_2 (\alpha ) \}_{\alpha=0}^2 = \{1,-\sqrt{2},1\}.
\end{align}
Here we use the explicit holonomy matrices $\rho_{\alpha}(x_3)$ given in \eqref{holonomy matrices} and the formula in \eqref{Wab}. Then, using the relation $S_{\alpha \beta} = \mathcal{W}_\beta (\alpha) \mathcal{W}_0 (\beta) S_{00}$, we find the full $S$-matrix
\begin{align}
S = \left(
\begin{array}{ccc}
\frac{1}{2} & \frac{1}{\sqrt{2}} & \frac{1}{2} \\
\frac{1}{\sqrt{2}} & 0 & -\frac{1}{\sqrt{2}} \\
\frac{1}{2} & -\frac{1}{\sqrt{2}} & \frac{1}{2} \\
\end{array}
\right)\;.
\end{align}
The modular $S$- and $T$- matrices are identical to those of $3^B_{7/2}$ (following notation in \cite{Wen_2015}), up to the fact that in our procedure the 2d central charge $c_{2d}$ can be captured only modulo $\frac{1}{2}$ while that in the usual UMTCs is defined modulo $8$.
From the $T$-matrix and  \eqref{T-matrx}, we compute
\begin{align}
\begin{split}
&\left(c_{2d}\textrm{ of TFT}[M= S^2 (3,3,-4/3)_{\rm I}]\right)  = 0 \;\;  (\textrm{mod }\frac{1}2) \;, \; 
\\
&\textrm{  while }
\\
&\left(c_{2d}\textrm{ of }3^B_{7/2} \right)  =\frac{7}2 = 0 \;\; (\textrm{mod }\frac{1}2) \;.\nonumber
\end{split}
\end{align}
So our result agrees with the 2d central charge of $3^B_{7/2}$ model modulo $\frac{1}{2}$.
We therefore conclude that
\begin{align}
\textrm{TFT}[M= S^2 (3,3,-4/3)_{\rm I }] \sim 3^B_{7/2} \;. \nonumber
\end{align}
In the above, $\textrm{TFT}_1 \sim \textrm{TFT}_2$ means that the two topological theories $\textrm{TFT}_1$ and $\textrm{TFT}_2$ have the same modular structure including $c_{2d}$ modulo $\frac{1}2$. 

Now let us discuss the modular structure of $S^2(3,3,-\frac{4}{3})_{\rm II}$ with the vacuum $\rho_{\alpha=2}$. For convenience, we  first rename flat connections as follows
\begin{align}
\tilde{\rho}_{\alpha =0} := \rho_{\alpha =2}\;,  \quad  \tilde{\rho}_{\alpha =1} := \rho_{\alpha =1}\;,  \quad  \tilde{\rho}_{\alpha =2} := \rho_{\alpha =0}\;,  \nonumber
\end{align}
and choose the $\tilde{\rho}_{\alpha =0} $ as the true vacuum. In the choice, the flat-connection-to-loop-operator map should be modified as follows:
\begin{align}
&  \tilde{\rho}_{\alpha } \rightarrow (a=x_3.h , R =\textrm{Sym}^\alpha \Box)\;.
\end{align}
The $\mathcal{W}_{\alpha} (\beta) $ and the $S$-matrix in this choice are the same as before. However, the $T$-matrix and thus $c_{2d}$ and topological spins change as
\begin{align}
\begin{split}
&\widetilde{T} = \textrm{diag} \{ e^{-\frac{i \pi}{24}} ,e^{-\frac{ i  \pi}{6}}, e^{-\frac{25 i \pi}{24}} \}\;,
\\
&\Rightarrow \tilde{c}_{2d} = 0 \; \left(\textrm{mod } \tfrac{1}2\right) \;, \;\;\{\tilde{h}_\alpha \}_{\alpha =0}^2 = \big{\{} 0, \tfrac{1}{16}, \tfrac{1}2  \big{\}}\;.
\end{split}
\end{align}
From the modular data, we conclude that
\begin{align}
\textrm{TFT}[M= S^2 (3,3,-4/3)_{\rm II }] \sim 3^B_{1/2} \;. \nonumber
\end{align}
Note that the two different choices of the true vacuum lead to the two distinct (2+1)d topological phases described by the UMTCs $3^B_{7/2}$ and $3^B_{1/2}$, respectively, that are related to each other by the Lorentz symmetry fractionalization. 

\subsection{$M=S^2\left(3,3,-\frac{5} 3\right) = (S^3\backslash \mathbf{5}_2)_{p/q=3} $ }  \label{subsec : 3 3 -5/3}
 From the  topological fact that $M=S^2(3,3,-5/3) = (S^3\backslash \mathbf{5}_2)_{p/q=3}$ \cite{dunfield2018census,DVN/6WNVG0_2018}, the TFT$[M]$ arises in the IR limit of  $\mathcal{T}[(S^3\backslash \mathbf{5}_2)_{p/q}]$ in \eqref{T[52]} with $p/q = 3$. In the notation of \cite{dunfield2018census}, $(S^3\backslash \mathbf{5}_2)_{p/q}$ corresponds to $(m015)_{p/q-2}$.
 There are 4 irreducible flat connections $\rho_{\alpha =0,1,2,3}$ on this 3-manifold with
\begin{align}
&\{CS[\rho_{\alpha}] \}_{\alpha=0}^3=\big{\{}  \tfrac{14}{15} , \; \tfrac{11}{60},\; \tfrac{11}{15}, \;  \tfrac{59}{60} \big{\}}\; ,
\\
&\{\textrm{Tor}[\rho_{\alpha}] \}_{\alpha=0}^3= \big{\{} \tfrac{5- \sqrt{5}}2, \;    \tfrac{5- \sqrt{5}}2,\; \tfrac{5+ \sqrt{5}}2,\; \tfrac{5+\sqrt{5}}2\big{\}}\;. \nonumber
\end{align}
According to Table \ref{Table : MTC[M]}, the $\textrm{GSD}_{g}$, $S_{00}$ and $T$-matrix are
\begin{align}
&\{ \textrm{GSD}_{g} \}_{g=0,1,2,\ldots}  = \{1,4,20,120,800,5600,\ldots \} \;,
\\
&S_{00} = \tfrac{1}{\sqrt{5-\sqrt{5}}}\;,  \;T = \textrm{diag} \{ e^{-\frac{28 i \pi}{15}}, e^{-\frac{11 i  \pi}{30}},e^{-\frac{22 i \pi}{15}}, e^{-\frac{59 i \pi}{30}} \}\;, \nonumber
\end{align}
and the topological spins   are
\begin{equation}
\{s_\alpha \}_{\alpha =0}^{3} = \left\{0,\tfrac{1}{4},\tfrac{4}{5},\tfrac{1}{20}\right\}\;.
\end{equation}
Note that  $S_{00} >  \frac{1}{\sqrt{2 \textrm{Tor} [\rho_\alpha]}}$ for  $\alpha=2$ and $\alpha=3$. According to \eqref{unitarity}, this modular data is non-unitary. 

There are two choices of the true vacuum: I.\,$\rho_{\alpha =0}$ and II.\,$\rho_{\alpha=1}$.
Let us first discuss the case with the vacuum I. We then read
\begin{equation}
|\mathcal{W}_{0} (\alpha =0,1) |=  1\;, \quad |\mathcal{W}_{0} (\alpha =2,3) |=  \tfrac{1}{2} \left(1-\sqrt{5}\right) \;.
\end{equation}
This implies the flat-connection-to-loop-operator map
\begin{align}
&  \rho_{\alpha=1 } \rightarrow (a=x_2 , R =\Box)\;, \quad  \rho_{\alpha=2 } \rightarrow (a=x_3^2 , R =\Box)\;,   \nonumber
\\
&\rho_{\alpha=3 } \rightarrow (a=x_2 , R =\Box) \otimes (a=x_3^2 , R =\Box) \nonumber
\\
&\Rightarrow\big{\{}\mathcal{W}_0 (\alpha ) \big{\}}_{\alpha=0}^3=\left\{1,1,\tfrac{1-\sqrt{5}}{2} ,\tfrac{1-\sqrt{5}}{2} \right\} \;, \nonumber
\\
& \quad \;\;  \big{\{}\mathcal{W}_1 (\alpha ) \big{\}}_{\alpha=0}^3=\left\{1,-1,\tfrac{1-\sqrt{5}}{2},\tfrac{\sqrt{5}-1}{2}\right\}\;, \nonumber
\\
& \quad \;\;  \big{\{}\mathcal{W}_2 (\alpha )  \big{\}}_{\alpha=0}^3=  \left\{1,1,\tfrac{1+\sqrt{5}}{2} ,\tfrac{1+\sqrt{5}}{2}\right\}\;, \nonumber
\\
& \quad \;\;  \big{\{}\mathcal{W}_3 (\alpha ) \big{\}}_{\alpha=0}^3  = \left\{1,-1,\tfrac{1+\sqrt{5}}{2},\tfrac{-1-\sqrt{5}}{2}\right\}\;.
\end{align}
Here we use the explicit holonomy matrices  given in \eqref{holonomy matrices2}. Then, using the relation $S_{\alpha \beta} = \mathcal{W}_\beta (\alpha) \mathcal{W}_0 (\beta) S_{00}$, we compute the full $S$-matrix
\begin{align}
S&=\tfrac{1}{\sqrt{5-\sqrt{5}}}\left(
\begin{array}{cccc}
1 & 1 & \frac{ \left(1-\sqrt{5}\right) }{2}& \frac{\left(1-\sqrt{5}\right)}{2}  \\
1 & -1 & \frac{\left(1-\sqrt{5}\right)}{2}  & \frac{ \left(\sqrt{5}-1\right)}{2} \\
\frac{\left(1-\sqrt{5}\right)}{2}  & \frac{\left(1-\sqrt{5}\right) }{2} & -1 & -1 \\
\frac{ \left(1-\sqrt{5}\right)}{2} & \frac{\left(\sqrt{5}-1\right)}{2}  & -1 & 1 \\
\end{array}
\right)\;\nonumber
\\
&= \left(
\begin{array}{cc}
1 &  \frac{(1-\sqrt{5})}2   \\
\frac{(1-\sqrt{5})}2   & -1 \\
\end{array}
\right) \otimes   \left(
\begin{array}{cc}
1 &  1  \\
1 & -1 \\
\end{array}
\right)\;.
\end{align}
The modular $S$ and $T$-matrices are equivalent to those of $(\textrm{Lee-Yang})\otimes (\textrm{Semion})$, the product of the Lee-Yang model and the semion UMTC, up to the central charges
\begin{align}
\begin{split}
&c_{2d}[\textrm{TFT}[S^2(3,3,-5/3)]] =\frac{1}{10}\; \textrm{ (mod $\frac{1}2$)}\;, \textrm{ while}
\\
&c_{2d}[(\textrm{Lee-Yang})\otimes (\textrm{Semion})] 
\\
& =-\frac{22}5+ 1 = -\frac{17}5 = \frac{1}{10} \;\; (\textrm{mod }\frac{1}2) \;.
\end{split}
\end{align} 
As discussed in the previous section, the effective theory $\mathcal{T}[(S^3\backslash \mathbf{5}_2)_{p/q=3}]$ flows to a SCFT with an additional Abelian symmetry. We expect that this (2+1)d SCFT for the vacuum choice I involves a non-unitary sub-sector described by the modular structure of $(\textrm{Lee-Yang})\otimes (\textrm{Semion})$.
\begin{align}
\textrm{TFT}[S^2(3,3,-5/3)_{\rm I}]  \sim (\textrm{Lee-Yang})\otimes (\textrm{Semion})\;. \nonumber
\end{align}

Now let us consider the choice II. For convenience, we  first rename the flat connections as follows:
\begin{align}
\begin{split}
&\tilde{\rho}_{\alpha =0} := \rho_{\alpha =1}\;,  \quad  \tilde{\rho}_{\alpha =1} := \rho_{\alpha =0}\;,  
\\
&\tilde{\rho}_{\alpha =2} := \rho_{\alpha =3}\;,   \quad \tilde{\rho}_{\alpha =3} := \rho_{\alpha =2}\;.
\end{split} \nonumber
\end{align}
In this choice, the flat-connection-to-loop-operator map is given by
\begin{align}
&  \tilde{\rho}_{\alpha=1 } \rightarrow (a=x_2 , R =\textrm{Sym}^3 \Box )\;, \;  \tilde{\rho}_{\alpha=2 } \rightarrow (a=x_3^2 , R =\Box)\;,  \nonumber
\\
&\tilde{\rho}_{\alpha=3 } \rightarrow (a=x_2 , R =\textrm{Sym}^3 \Box ) \otimes (a=x_3^2 , R =\Box)\;.
\end{align}
The $\mathcal{W}_{\alpha} (\beta) $ and $S$-matrix are the same as those for the choice I. The $T$-matrix changes as
\begin{align}
&\tilde{T} = \textrm{diag} \{  e^{-\frac{11 i  \pi}{30}},e^{-\frac{28 i \pi}{15}} , e^{-\frac{59 i \pi}{30}} ,e^{-\frac{22 i \pi}{15}} \}\;. 
\\
&\Rightarrow \tilde{c}_{2d} = \tfrac{1}5  \;\left(\textrm{mod }\tfrac{1}2\right)  \;, \;\;\{\tilde{h}_\alpha \}_{\alpha =0}^3 = \big{\{} 0, \tfrac{3}4, \tfrac{4}5, \tfrac{11}{20}   \big{\}}\;.\nonumber
\end{align}
One notices that the resulting MTC coincides with the non-unitary $(\textrm{Lee-Yang})\otimes (\mathcal{F}_{\rm Lorentz}[\textrm{Semion}])$ MTC where $\mathcal{F}_{\rm Lorentz}[\textrm{Semion}]$ denotes the Lorentz symmetry fractionalization of the semion MTC with respect to the 1-form symmetry generated by the semion with spin $\frac{1}{4}$. We thus conclude that
\begin{align}
\begin{split}
&\textrm{TFT}[S^2 (3,3,-5/3)_{\rm II }] \sim  (\textrm{Lee-Yang})\otimes (\mathcal{F}_{\rm Lorentz}[\textrm{Semion}]) \;. \nonumber
\end{split}
\end{align}

\subsection{More 3-manifolds studied in section \ref{sec : effecitve field theory} } 
In section \ref{sec : effecitve field theory}, we study the supersymmetric gauge theories $\mathcal{T}[M]$ for seven non-hyperbolic 3-manifolds, $(S^3\backslash \mathbf{4}_1)_{p/q=1,3},(S^3\backslash \mathbf{5}_2)_{p/q=1,3}$ and $(m007)_{p/q=-2,0,2}$,  obeying the topological conditions \eqref{3-manifolds for top} with trivial $H_1(M, \mathbb{Z})$. Based on  \eqref{main conjecture} and \eqref{non-unitary/accidental symmetry} combined with the superconformal index computation, we have conjectured that
\begin{align}\label{eq:example-conjecture}
\begin{split}
&\mathcal{T}[M] \textrm{ with } M=(S^3\backslash \mathbf{4}_1)_{p/q=1,3} \textrm{ or } (m007)_{p/q=0,2}
\\
& \textrm{flows to a unitary topological theory},
\\
&\textrm{while}
\\
&\mathcal{T}[M] \textrm{ with } M=(S^3\backslash \mathbf{5}_2)_{p/q=1,3} \textrm{ or } (m007)_{p/q=-2}
\\
& \textrm{flows to a SCFT with $U(1)$ symmetry which contains} 
\\	
& \textrm{a non-unitary topological phase as a sub-sector}.
\end{split}
\end{align}
In the previous subsection, we have shown that the modular structures associated to $M=(S^3\backslash \mathbf{4}_1)_{3}$ are indeed unitary, whereas the modular structures associated to $M=(S^3\backslash \mathbf{5}_2)_{3}$ are non-unitary, which confirms two cases with $p/q=3$ in the conjecture \eqref{eq:example-conjecture}.

Let us now confirm the conjecture \eqref{eq:example-conjecture} for  other 5 manifolds.
We shall basically employ the unitarity criterion in \eqref{unitarity} to show that the associated modular structures are unitary/non-unitary.

\paragraph{$(S^3\backslash \mathbf{4}_1)_{p/q=1} =   S^2 (2,3, -\frac{7}6 )$} There are 3 irreducible $SL(2,\mathbb{C})$ flat connections $\{\rho_\alpha\}_{\alpha=0}^2$ with
\begin{align}
&\left\{ CS[\rho_\alpha]\right\}  =\left\{\tfrac{1}{168},\tfrac{121}{168},\tfrac{25}{168}\right\}\;, \; \nonumber
\\
&\textrm{Tor}[\rho_\alpha] = \tfrac{7}{8} \csc^2\left(\tfrac{(2\alpha+1)\pi }{7}\right)\;. \nonumber
\end{align} 
Thus, according to \eqref{unitarity}, the associated modular structure is unitary. 

\paragraph{$(S^3\backslash \mathbf{5}_2)_{p/q=1} =   S^2 (2,3, -\frac{11}9 )$} There are 5 irreducible $SL(2,\mathbb{C})$ flat connections $\{\rho_\alpha\}_{\alpha=0}^4$ with
\begin{align}
&\left\{ CS[\rho_\alpha]\right\}  =\left\{\tfrac{239}{264},\tfrac{215}{264},\tfrac{167}{264},\tfrac{95}{264},\tfrac{263}{264}\right\} \;, \nonumber
\\
&\textrm{Tor}[\rho_\alpha] = \tfrac{11}{8} \csc^2\left(\tfrac{5(2\alpha+1)\pi }{11}\right)\;. \nonumber
\end{align} 
Thus, according to \eqref{unitarity}, the associated modular structure is non-unitary.

\paragraph{$(m007)_{p/q=-2} =   S^2 (2,3, -\frac{9}7 )$} There are 4 irreducible $SL(2,\mathbb{C})$ flat connections $\{\rho_\alpha\}_{\alpha=0}^3$ with
\begin{align}
&\left\{ CS[\rho_\alpha]\right\}  =\left\{\tfrac{47}{72},\tfrac{13}{24},\tfrac{23}{72},\tfrac{71}{72}\right\}\;, \; \nonumber
\\
&\textrm{Tor}[\rho_\alpha] = \tfrac{9}{8} \csc^2\left(\tfrac{4(2\alpha+1)\pi }{9}\right)\;. \nonumber
\end{align} 
Thus, according to \eqref{unitarity}, the associated modular structure is non-unitary. 

\paragraph{$(m007)_{p/q=0} =   S^2 (3,3, -3)$} There are 2 irreducible $SL(2,\mathbb{C})$ flat connections $\{\rho_\alpha\}_{\alpha=0}^1$ with
\begin{align}
&\left\{ CS[\rho_\alpha]\right\}  =\left\{\tfrac{2}{3},\tfrac{11}{12}\right\}\;, \quad \textrm{Tor}[\rho_\alpha] =1\;. \nonumber
\end{align} 
Thus, according to \eqref{unitarity}, the associated modular structure is unitary. 

\paragraph{$(m007)_{p/q=2} =   S^2 (2,\frac{3}2 , -3 )$} There is only one irreducible $SL(2,\mathbb{C})$ flat connection $\{\rho_\alpha\}_{\alpha=0}^2$ with
\begin{align}
&CS[\rho_{\alpha=0}] =\tfrac{19}{24}\;, \quad \textrm{Tor}[\rho_{\alpha=0}]  =\tfrac{1}2\;. \nonumber
\end{align} 
Thus, according to \eqref{unitarity}, the associated modular structure is unitary. 

\subsection{$\textrm{TFT}[M=S^2 ( \frac{p_1}{q_1}, \frac{p_2}{q_2},\frac{p_3}{q_3})]$ with trivial $H_1(M, \mathbb{Z}_2)$ and  $\textrm{rank}\leq 4$}  The  list  of unitary cases is given in Table~\ref{Table : full list} while non-unitary cases are given in Table~\ref{Table : full list-2}. Detailed analysis of modular structures of the $\textrm{TFT}[M]$ is given in Appendix \ref{app : modular structure of full list}. 

\begin{table}[h]
	\begin{center}
		\begin{tabular}{|c|c|c|c|c|}
			\hline
			$\quad \textrm{Rank}$ \quad &  3-manifold  & \textrm{TFT}[M]  
			\\
			\hline 
			2 & $S^2 (2,3,\frac{5}2)$ &  $\textrm{Gal}_{\sharp=2}\left[(A_1,3)_{1/2}\right]$
			\\
			\hline
			3 & $S^2 (2,3,\frac{7}2)$ & $\textrm{Gal}_{\sharp=2}\left[(A_1,5)_{1/2}\right]$
			\\
			& $S^2 (2,3,\frac{7}3)$ & $\textrm{Gal}_{\sharp=3}\left[(A_1,5)_{1/2}\right]$
			\\
			\hline
			& $S^2 (2,3,\frac{9}2) $ & $\textrm{Gal}_{\sharp=2}\left[(A_1,7)_{1/2}\right]$
			\\
			&  $S^2 (2,3,\frac{9}4) $  & $\textrm{Gal}_{\sharp=4}\left[(A_1,7)_{1/2}\right]$
			\\
			4 & $S^2 (\frac{3}{2},3,\frac{5}{2}) $ & $\textrm{Gal}_{\sharp=2}\left[(A_1,3)_{1/2}\right] \otimes (2^B_{\pm 1})$ 
			\\
			& $S^2 (2,5,\frac{5}{2})$ & $\textrm{Gal}_{\sharp=2}\left[(A_1,3)_{1/2}\right] \otimes (A_1,3)_{1/2} $ 
			\\
			& $S^2 (2,\frac{5}2,\frac{5}{2})$ & $\textrm{Gal}_{\sharp=2}\left[(A_1,3)_{1/2}\right] \otimes \textrm{Gal}_{\sharp=2} \left[(A_1,3)_{1/2}\right] $ 
			\\
			\hline
		\end{tabular}
	\end{center}
	\caption{List of non-unitary $\textrm{TFT}[M=S^2 ( \frac{p_1}{q_1}, \frac{p_2}{q_2},\frac{p_3}{q_3})]$ with trivial $H_1(M, \mathbb{Z}_2)$ up to rank 4. The $(A_1, k)_{1/2}$ theory admits its non-unitary Galois conjugations $\textrm{Gal}_{\sharp}(A_1, k)_{1/2}$ where $1<\sharp<\frac{k+2}2 $ is an positive integer relatively prime to $k+2$ \cite{Ardonne_2011}. $\textrm{Gal}_{\sharp=2}(A_1, 3)_{1/2}$ corresponds to the Lee-Yang model. }
	\label{Table : full list-2}
\end{table}

\section{Extension to  non-trivial $H_1(M, \mathbb{Z}_2)$}  \label{sec : non-trivial Z2 homology}
Here we generalize the analysis in the previous sections to non-hyperbolic 3-manifolds with non-trivial $H_1(M, \mathbb{Z}_2)$.

\subsection{$\textrm{TFT}[M]$ and $\widetilde{\textrm{TFT}}[M]$}  
When $H_1(M, \mathbb{Z}_2)$ is non-trivial, there are some additional discrete choices in the  geometrical construction in  \eqref{wrapped M5-branes} as recently studied in \cite{Eckhard:2019jgg}. The choice is in the one-to-one correspondence with the choice of a subgroup $H$ of the cohomology group $H^1(M, \mathbb{Z}_2)$. See Appendix \ref{appendix : T[M]} for a brief review  on the choice. Correspondingly, there are discrete variants of $\textrm{TFT}[M]$ depending on the choices. For example, we denote the topological theory associated to the choice $H = \emptyset$ by $\widetilde{\textrm{TFT}}[M]$ while the theory associated to $H=H^1(M, \mathbb{Z}_2)$ by $\textrm{TFT}[M]$. 

From the geometrical analysis in \cite{Eckhard:2019jgg}, $\widetilde{\textrm{TFT}}[M]$ is expected to have the cohomology group $H^1 (M, \mathbb{Z}_2)$ as 1-form flavor symmetry \cite{Gaiotto:2014kfa}.\footnote{$\widetilde{\textrm{TFT}}[M]$ can have bigger 1-form symmetry than the geometrically expected symmetry. 
	It is also possible that some subgroup of $H^1(M, \mathbb{Z}_2)$ trivially acts on all anyons and thus is effectively absent in the topological theory.  } 
Generally, 1-form symmetries in a topological theory are   generated by anyons.  We define such anyons as
\begin{align}
A_{\eta} \;:\; \textrm{Anyon generating 1-form symmetry $\eta \in H^1 (M, \mathbb{Z}_2)$}\;. \label{A-eta}
\end{align}
Since $\eta^2 = 1$ for $\eta \in H^1 (M, \mathbb{Z}_2)$, the corresponding anyon satisfies the fusion rule \footnote{For a $\mathbb{Z}_N$ symmetry, the symmetry generating anyon $A$ shows $A^N=1$. }
\begin{align}
A_\eta \times A_\eta  =1\;.
\end{align}
On spin manifolds, a 1-form symmetry $\eta$ is anomaly free only if the generating anyon $A_{\eta}$ has integer or half-integer spin \cite{Hsin:2018vcg}. In fact, the global 1-form symmetry $H^1 (M, \mathbb{Z}_2)$ is anomaly free \cite{Eckhard:2019jgg}. Thus the anyons $A_\eta$ have topological spins $h_\eta=0$ or $\frac{1}{2}$ mod 1.

We can gauge a 1-form symmetry if the symmetry is anomaly free. In the context of anyon theories, gauging a 1-form symmetry is also called as {\it anyon condensation} of the symmetry generating anyon. Gauging a subgroup $H\subset H^1(M,\mathbb{Z}_2)$ leads to another topological phase which we call TFT$[M;H]$. The topological phase TFT$[M]$ is in fact TFT$[M;H]$ with $H=H^1(M, \mathbb{Z}_2)$ obtained by gauging all the anomaly free 1-form symmetries in $\widetilde{\textrm{TFT}}[M]$.

In the subsequent sections, we give an algorithm for determining (partial) modular structures of $\widetilde{\textrm{TFT}}[M]$ and $\textrm{TFT}[M]$ from topological information of the 3-manifold $M$. 
 
\subsection{Modular structures of $\widetilde{\textrm{TFT}}[M]$}  
Some basic modular data of the topological theory is summarized in \eqref{Table : MTC[M]-2}.  The table can  be derived following  the similar logic  used in deriving Table \ref{Table : MTC[M]} combined with analysis on Bethe-vacua of $\widetilde{\mathcal{T}}[M] := \mathcal{T}[M;H=\emptyset]$ theory in \cite{Eckhard:2019jgg}.
\begin{table}[h]
	\begin{center}
		\begin{tabular}{|c|c|}
			\hline
			$\widetilde{\textrm{TFT}}[M] $ \quad &  3-manifold $M$
			\\
			\hline
			Anyons &$\{\rho^{PSL} \otimes  \eta \}$
			\\
			\hline
			&i) $ \pm CS [\rho^{PSL} \otimes \eta ] $ (\textrm{mod 1})
			\\
			$h_\alpha - \frac{c_{2d}}{24}$ \;  & (\textrm{for anyons with $w_2(\rho^{PSL}) =0$})\;,
			\\
			&ii) \textrm{Equation \eqref{top spin for non-zero w2}}
			\\
			& (\textrm{for anyons with non-trivial $w_2(\rho^{PSL})$})
			\\
			\hline
			$(S_{0 \alpha})^2$&  $(2 |H^1 (M, \mathbb{Z}_2)|  \textrm{Tor}[\rho^{PSL}])^{-1}$
			\\
			\hline	
			\textrm{$\mathcal{W}_{\rho^{PSL} \otimes  \eta'} (A_\eta) $}&  $(-1)^{\int_M \eta \cup w_2 (\rho^{PSL})} \in \{\pm 1\}$
			\\
			\hline	
		\end{tabular}
	\end{center}
	\caption{Some modular data of $\widetilde{\textrm{TFT}}[M]$, the topological phase associated to a non-hyperbolic 3-manifold $M$ satisfying \eqref{3-manifolds for top} with general $H_1(M, \mathbb{Z}_2)$.  Here $w_2$ denotes the 2nd  Stiefel-Whitney class.  For a $PSL(2,\mathbb{C})$ flat connection $\rho^{PSL}_\alpha$ with trivial $w_2$, the $\rho^{PSL}_\alpha \otimes \eta$  can be regarded as $SL(2,\mathbb{C})$ flat connections uplifted from the $PSL(2,\mathbb{C})$ flat connection and the Chern-Simons action $CS[\rho^{PSL}_\alpha \otimes  \eta]$ is well-defined modulo 1. $|H^1(M, \mathbb{Z}_2)|$ denotes the order of the cohomology group. }
	\label{Table : MTC[M]-2}
\end{table}
In the table,  $\rho^{PSL}$ is an irreducible $PSL(2,\mathbb{C})$ flat connection on $M$ and $\eta\in  H^1 (M, \mathbb{Z}_2)$ is the $\mathbb{Z}_2$ flat connection. The holonomy matrix of the $PSL(2,\mathbb{C})$ flat connection can be  written as
\begin{align}
\rho^{PSL} (a) = [\rho^{SL}(a)]\;,
\end{align}
where $\rho^{SL}(a) \in SL(2,\mathbb{C})$ is a representative uplift from $PSL(2,\mathbb{C})$ to $SL(2,\mathbb{C})$.  Here, $[\ldots]$ denotes the equivalence class under the $\mathbb{Z}_2$ in $PSL(2,\mathbb{C})= SL(2,\mathbb{C})/\mathbb{Z}_2$,
\begin{align}
[g] = [-g ] \in PSL(2,\mathbb{C})\;, \quad g\in SL(2,\mathbb{C})\;.
\end{align} 
Generally, $\rho^{SL}$ does not  give a homomorphism in $\textrm{Hom}\left[\pi_1 M \rightarrow SL(2,\mathbb{C})\right]$. Only for $ \rho^{PSL}$ with trivial 2nd Stiefel-Whitney class, i.e. $w_2 (\rho^{PSL})=0$, we can choose $\rho^{SL}$ to be a homomorphism and give a $SL(2,\mathbb{C})$ flat connection. 
For $ \rho^{PSL}$ with trivial $w_2$, the set $\{ \rho^{PSL}\otimes \eta \}_{\eta \in H^1(M, \mathbb{Z}_2)}$ can be regarded as the set of $SL(2,\mathbb{C})$ flat connections, where $\otimes$ represents a formal tensor product of $\rho^{PSL}$ and the $\mathbb{Z}_2$ flat connection  $\eta$ under the equivalence relation
\begin{align}
\begin{split}
&\rho^{PSL}   \otimes \eta_1=  \rho^{PSL}  \otimes \eta_2 \;\; \textrm{ if \;$\exists\; g \in SL(2,\mathbb{C})$ such that}
\\
&\eta_1(a)  \rho^{SL} (a)  = g\cdot \left( \eta_2(a)  \rho^{SL} (a) \right)  \cdot g^{-1} \;, \quad \forall a\in \pi_1 (M)\;. \label{PSLotimesZ2}
\end{split}
\end{align}
For the cases with trivial $H_1(M,\mathbb{Z}_2)$, the dictionary in  Table \ref{Table : MTC[M]-2} is just a subset of that in Table \ref{Table : MTC[M]}. In this case, $PSL(2,\mathbb{C})$ flat connections are in one-to-one correspondence with $SL(2,\mathbb{C})$ flat connections since their $w_2 \in H^2(M, \mathbb{Z}_2) = H_1(M, \mathbb{Z}_2)$ is automatically trivial. 
 
The 1-form symmetry generating anyon   $A_\eta$ in \eqref{A-eta}  can be identified with 
\begin{align}
A_\eta = \rho^{PSL}_{\alpha=0} \otimes \eta\;,
\end{align} 
where $\rho_{\alpha=0}$ is the flat-connection whose corresponding  $\rho_{\alpha=0}^{PSL}\otimes 1$ is the trivial anyon. 
Hence the fusion rules between the   $A_\eta$  and other anyons have the following simple structure,
\begin{align}
A_\eta  \times  (\rho^{PSL}\otimes \tilde{\eta}) = \rho^{PSL}\otimes (\eta \cdot \tilde{\eta}) \;. \label{fusion with A-eta}
\end{align}
The fusion rule determines  how the anyon $A_\eta$ along $B$-cycle in two-torus acts on the ground states in the Hibert-space  as illustrated in \eqref{Action of 1-form on ground states}. The action of the anyon along $A$-cycle is determined by $\mathcal{W}_{\rho^{PSL} \otimes  \eta'} (A_\eta) $ in the Table \ref{Table : MTC[M]-2}, which is   called  {\it monodromy charge}. This action is particularly important in the gauging procedure.
Upon gauging a 1-form symmetry  generated by $A_\eta$, the anyons $\rho^{PSL} \otimes  \tilde{\eta}$ with monodromy charge $\mathcal{W}_{\rho^{PSL} \otimes  \tilde{\eta}} (A_\eta)=-1$ are all projected out. 
 
The cohomology group  $H^1(M,\mathbb{Z}_2)$ admits a $\mathbb{Z}_2$ grading defined as
\begin{align}
\begin{split}
&\eta \in H^1 (M, \mathbb{Z}_2) \textrm{ is called } 
\\
&\begin{cases}
\textrm{bosonic }, \; \textrm{if $CS[\eta \otimes \rho] = CS[\rho]$ for all $\rho$}\\
\textrm{fermionic }, \; \textrm{otherwise}\;.\\
\end{cases} 
\end{split}\label{bosonic/fermionc of Z2-flat}
\end{align}
Here $\rho$ are $SL(2,\mathbb{C})$ flat connections on $M$ and $\eta \otimes \rho$ denotes the tensor product of $\rho$ and $\eta$ whose holonomy matrix is given by
\begin{align}
(\eta \otimes \rho)(a) = \eta (a) \rho(a)\;, \quad \eta (a) \in \{\pm 1\} \;.
\end{align}

The anyon $A_\eta$ for $\eta\in H^1 (M, \mathbb{Z}_2)$ has the topological spin
\begin{align}
h_\eta = \begin{cases}
0 \; , \qquad \textrm{if $\eta$ is bosonic}\\
\frac{1}2, \;\qquad \textrm{if $\eta$ is fermionic}\\
\end{cases} 
\end{align}
The quantum number $\check{h}:=h- \frac{c_{2d}}{24}$ (mod 1) for anyons $\rho^{PSL}\otimes \eta$ with non-trivial $w_2 (\rho^{PSL})$ satisfies following conditions
\begin{align}
\begin{split}
i) \;\; &(\check{h}  \textrm{ of }\rho^{PSL}\otimes \eta) = CS[\rho^{PSL}] \;\; \left(\textrm{mod $\frac{1}2$}\right) \;,
\\
ii) \;\; &\left(\check{h}  \textrm{ of }\rho^{PSL}\otimes (\eta \cdot \tilde{\eta})\right) -  \left(\check{h}  \textrm{ of }\rho^{PSL}\otimes  \tilde{\eta}\right)  \;\textrm{for all $\tilde{\eta}$}
\\
&= \begin{cases}
0\; , \;  \quad \textrm{if $\epsilon(\eta) \times (-1)^{\eta \cup w_2 (\rho^{PSL})}  = 1$ }\\
\frac{1}2\;, \; \quad \textrm{if $\epsilon(\eta) \times (-1)^{\eta \cup w_2 (\rho^{PSL})} = -1$}\ ,
\end{cases}  
\end{split}\label{top spin for non-zero w2}
\end{align}
where we defined
\begin{align}
\begin{split}
&\epsilon (\eta) :=  \begin{cases}
1\;, \; \textrm{if $\eta$ if bosonic}\\
-1\; , \; \textrm{if $\eta$ if fermionic}\\
\end{cases}
\label{bosonic/fermionc of Z2-flat-2}
\end{split}
\end{align}
 Note that  $CS[\rho^{PSL}]$ for  a flat connection $\rho^{PSL} $ with non-trivial $w_2$  is  defined only modulo $1/2$. The condition {\it ii}) follows from the universal property of bosonic topological phases, e.g.  proposition 2.10 in \cite{Bruillard_2017}, combined with the fusion rule in  \eqref{fusion with A-eta} and the dictionary for monodromy charge in Table \ref{Table : MTC[M]-2}. The above conditions uniquely determine the spectrum of $\check{h}$ when all $\eta \in H^1(M, \mathbb{Z}_2)$ are bosonic. On the other hand, we leave the detailed analysis of the cases with the fermionic $\eta$ as a future work.

\subsection{Modular structures of $\textrm{TFT}[M]$}  

The  topological theory TFT$[M]$ after gauging all  $\eta \in H^1 (M, \mathbb{Z}_2)$ could be either a spin (fermionic) TQFT or a symmetry enriched topological phase with non-self-dual anyons \cite{Barkeshli:2014cna,Gaiotto:2015zta}.   Some basic modular data of the topological theory is summarized in Table \eqref{Table : MTC[M]-3}.   Detailed derivation of the dictionary in Table  \ref{Table : MTC[M]-3} will be reported in \cite{To-appear}. 
\begin{table}[h]
	\begin{center}
		\begin{tabular}{|c|c|}
			\hline
			$ \textrm{TFT}[M]$ \quad &  3-manifold $M$
			\\
			\hline
			& \textrm{ Irreducible $PSL(2,\mathbb{C})$ flat connections   }
			\\
			\textrm{Anyons } 	&\textrm{$\rho^{PSL}$ on $M$ with $w_2 (\rho^{PSL}) =0$. }
			\\
			&\textrm{(Count with multiplicity $|\textrm{Inv}(\rho^{PSL})|$)}
			\\
			\hline
			\textrm{Bosonic/Fermionic } 	& Equation \eqref{bosonic/fermionic}
			\\
			\hline
			$h_\alpha - \frac{c_{2d}}{24}$& $ \pm CS [\rho^{PSL} ]\; \textrm{(mod 1) for bosonic } $
			\\
			& $\pm (CS [\rho^{PSL} ]\; \textrm{(mod $\frac{1}2$) for fermionic } $
			\\
			\hline
			$(S_{0 \alpha})^2$&  $|H^1 (M, \mathbb{Z}_2)|(2 \textrm{Tor}[\rho] |\textrm{Inv}(\rho^{PSL})|^2)^{-1}$
			\\
			\hline	
		\end{tabular}
	\end{center}
	\caption{Some modular data of $\textrm{TFT}[M]$, topological phase associated to non-hyperbolic 3-manifold $M$ satisfying \eqref{3-manifolds for top} with general $H_1(M, \mathbb{Z}_2)$. }
	\label{Table : MTC[M]-3}
\end{table}

We propose the following criterion:
\begin{align}
\begin{split}
&\textrm{TFT}[M]\textrm{ is }
\\
&\begin{cases}
\textrm{bosonic }, \; \textrm{if all $\eta \in H^1 (M, \mathbb{Z}_2)$ are bosonic}\\
\textrm{fermionic }, \; \textrm{otherwise}\\
\end{cases}  \label{bosonic/fermionic}
\end{split}
\end{align}
In the table, $\textrm{Inv}(\rho^{PSL}) $ denotes the  subgroup of $H^1(M, \mathbb{Z}_2)$ 
invariant under tensoring with a flat connection $\rho^{PSL}$. 
\begin{align}
\textrm{Inv}(\rho^{PSL}) : = \{ \eta \in H^1 (M, \mathbb{Z}_2)\;  : \; \rho^{PSL}  \otimes \eta =  \rho^{PSL} \otimes 1  \}  \;.\label{Inv}
\end{align}
We remark that for a bosonic $\eta$, each anyon $\rho^{PSL}\otimes \eta$ with non-trivial $\textrm{Inv}(\rho^{PSL})$ in the mother theory $\widetilde{\textrm{TFT}}[M]$ becomes multiple anyons in TFT$[M]$ with multiplicity  $|\textrm{Inv}(\rho^{PSL})|$, the order of symmetry generated by $A_\eta$ \cite{Hsin:2018vcg}. 
For a fermionic case, the topological spin $h$ is  defined only up to modulo $\frac{1}2$  since two anyons, $\rho^{PSL}\otimes 1$  and $\rho^{PSL}\otimes \eta$ for a fermionic $\eta$ whose topological spins differ by $1/2$, are regarded as  a single anyon $\rho^{PSL}$ after the gauging.

\subsection{$M=S^2 (2,3,8)$ : fermionic topological phase}
Let us consider the $\textrm{TFT}[M]$ with $M= S^2 (2,3,8)$, which is one of the simplest fermionic TFTs. The 3-manifold $M$ has non-trivial $H_1(M, \mathbb{Z}_2 ) = H^1(M, \mathbb{Z}_2 ) = \mathbb{Z}_2$ and  allows  4 irreducible $PSL(2,\mathbb{C})$ flat connections. The adjoint torsions for the flat connections are
\begin{align}
\{\textrm{Tor}[\rho^{PSL}_\alpha] \}_{\alpha=0}^3 =\left\{\csc ^2\left(\frac{\pi (\alpha+1)}{8}\right)\right\}_{\alpha=0}^3  \label{Tor for S2(2,3,8)}
\end{align}
Among them,  $\rho^{PSL}_{\alpha}$ for $\alpha=0$ and $\alpha=2$  have trivial 2nd Stiefel-Whitney class $w_2$ and they are not invariant under the tensoring with $H^1(M, \mathbb{Z}_2)$, i.e. $|\textrm{Inv}(\rho^{PSL}_{\alpha=0,2})|=1$. 
According to the dictionary  in Table \ref{Table : MTC[M]-3}, we have
\begin{align}
\begin{split}
&\{\textrm{GSD}_g [\textrm{TFT}[S^2 (2,3,8)]]  \}_{g=0}^\infty
\\
&=\{\sum_{\alpha=0,2} (S_{0\alpha})^{2-2g} \}= \{\sum_{\alpha=0,2} \csc^{2g-2} \left(\frac{\pi (\alpha+1)}8\right) \}
\\
&=  \{1,2,8,48,320,\ldots \}\;. \nonumber
\end{split}
\end{align}
From this result, one can confirm that $\textrm{TFT}[M=S^2 (2,3,8)]$ is not included in the  classification of  UMTCs, for example, in \cite{2007arXiv0712.1377R}. 
The reason is   that this theory is  actually a spin TQFT, not a bosonic topological theory. 

There are again 4 irreducible $SL(2,\mathbb{C})$ flat connections on $M=S^2 (2,3,8)$. These are uplifts of the two $PSL(2,\mathbb{C})$ flat connections $\rho^{PSL}_{\alpha=0}$ and $\rho^{PSL}_{\alpha=2}$.
\begin{align}
&\textrm{4 irreducible $SL(2,\mathbb{C})$ flat connections on $M= S^2 (2,3,8)$} \nonumber
\\
& =  \{\rho^{PSL}_{\alpha =0}, \rho^{PSL}_{\alpha =0}\otimes \eta , \rho^{PSL}_{\alpha =2}, \rho^{PSL}_{\alpha =2}\otimes \eta \}\;.
\end{align}
Their Chern-Simons invariants are
\begin{align}
\begin{split}
& CS[\rho^{PSL}_{\alpha =0}] = \frac{25}{96}\;,  \; CS[\rho^{PSL}_{\alpha =2}] =  \frac{49}{96}\;,
\\
& \; CS[\rho^{PSL}_{\alpha =0} \otimes \eta] =  \frac{73}{96}\;,\; CS[\rho^{PSL}_{\alpha =2}\otimes \eta ] = \frac{1}{96}\;.  \label{CS for S2(2,3,8)}
\end{split}
\end{align}
Since $CS[\rho^{PSL}_{\alpha =0,2} \otimes \eta] - CS[\rho^{PSL}_{\alpha =0,2}]  = \frac{1}2 $, $\textrm{TFT}[S^2 (2,3,8)] $ is fermionic according to the criterion in \eqref{bosonic/fermionic}.

This spin TQFT turns out to be obtained by condensating the fermionic anyon in the $SU(2)_6$ theory \cite{Bruillard_2017}, i.e.
\begin{align}
\begin{split}
&\textrm{TFT}[S^2 (2,3,8)] 
\\
&= \textrm{Condensating fermionic anyon  in $SU(2)_6$}\;.
\end{split}
\end{align}
This theory enjoys the following exotic fusion relation \cite{Bruillard_2017}
\begin{align}
\tau^2 = 1+2\tau\;,
\end{align}
which can not be realized in bosonic topological phases \cite{ostrik2003fusion}.

Moreover, one can confirm that the mother theory $\widetilde{\textrm{TFT}}[M=S^2 (2,3,8)]$ is actually the $SU(2)_6$ theory, which is a rank $7$ UMTC, using Table \ref{Table : MTC[M]-2}.  
Among 4 irreducible $PSL(2,\mathbb{C})$ flat connections, the last flat connection $\rho^{PSL}_{\alpha=3}$ is invariant under the tensoring with $H^1(M, \mathbb{Z}_2)  = \{1,\eta \}$, namely
\begin{align}
\eta \otimes \rho^{PSL}_{\alpha=3} =1\otimes  \rho^{PSL}_{\alpha=3}\;.
\end{align}
Thus there are in total $7$ anyons  in $\widetilde{\textrm{TFT}}[M=S^2 (2,3,8)]$:
\begin{align}
\{\rho^{PSL}_\alpha\otimes 1 \}_{\alpha=0}^3 \;, \quad \{\rho^{PSL}_\alpha\otimes \eta \}_{\alpha=0}^2\;. 
\end{align}
Using the table and the equation \eqref{Tor for S2(2,3,8)}, one can compute
\begin{align}
\begin{split}
(S_{0\alpha})^2 =\frac{1}2 \sin^2\left(\frac{\pi (\alpha+1)}{8}\right)\,
\end{split}
\end{align}
for the 7 anyons, which agrees with the   $S^2_{0\alpha}$ of 7 anyons in the $SU(2)_6$ theory. From the results in \eqref{CS for S2(2,3,8)}, one can further confirm that the topological spins $\rho^{PSL}\otimes \eta$   of 4 anyons with trivial $w_2(\rho^{PSL})$ in $\widetilde{\textrm{TFT}}[M=S^2 (2,3,8)]$ are
\begin{align}
\begin{split}
&h (\rho^{PSL}_{\alpha =0}) =0\;, \quad h (\rho^{PSL}_{\alpha =2}) =\frac{1}4\;,
\\
&h (\rho^{PSL}_{\alpha =0}\otimes \eta) =\frac{1}2\;, \quad h (\rho^{PSL}_{\alpha =2}\otimes \eta ) =\frac{3}4\;, 
\end{split}
\end{align}
This spectrum nicely matches the topological spins of $\mathbb{Z}_2$ invariant  4 anyons in the $SU(2)_6$ theory.

\section{Conclusions}\label{sec : Discussion}

In this paper, we have proposed a correspondence between a certain class of non-hyperbolic 3-manifolds  \eqref{3-manifolds for top} and $(2+1)d$ topological phases based on the physics of M-theory 5-branes compactified on the 3-manifolds. From this new correspondence, we developed a systematic program for generating and classifying the topological phases. Specifically, we provided an algorithm to read off the modular structure of the topological phases from topological data of the 3-manifold, and worked out infinitely many examples of Seifert fiber manifolds $S^2(\frac{p_1}{q_1},\frac{p_2}{q_2},\frac{p_3}{q_3})$. From this, we have successfully reproduced (and thus classified) all the known UMTCs up to rank 4. See the table \ref{Table : full list}. Not only this, we have also illustrated that our scheme can be universally applicable to the fermionic anyon theories, non-unitary MTCs, and symmetry-enriched models. 

This encourages us to conjecture that all (2+1)d topological phases of matter can be geometrically engineered by M5-branes wrapping non-hyperbolic 3-manifolds.
We have shown this for the theories with rank $\le4$. For higher rank TQFTs, in addition to the ingredients used in this paper one may need to scan over all the non-hyperbolic manifolds and/or to increase the number of M5-branes wrapped on them, though we've already produced in this paper two higher rank examples by using only 2 M5-branes: the $SU(2)_4$ (rank $5$) and the $SU(2)_6$ (rank $7$) UMTCs. Some more progress along this direction will be reported in \cite{To-appear}.
Our discussion in this paper merely opened a new door toward a bigger program to understand topological phases in terms of their concrete geometrical realizations, and we hope this program provides a more systematic classification of topological phases of matter. 

Going beyond the current results, there are a number of non-trivial future extensions. Let us list some of them below. 

\paragraph{Beyond modular structure} As recently noticed in \cite{Bonderson:2018ryx}, there are additional basic structures beyond modular structure, $S$- and $T$-matrices, that we studied in this paper. These additional structures are important because the modular data is not enough to fully characterize the different TQFTs at higher ranks \cite{MS}. It may be possible that this additional data is also encoded geometrically in some topological invariants of non-hyperbolic 3-manifolds. 
An interesting problem is to find and compute such topological invariants on generic  3-manifolds, which may allow us to achieve geometrically a complete classification of (2+1)d TQFTs.

\paragraph{Mathematics of Non-hyperbolic 3-Manifolds} our work connects two seemingly unrelated research fields, physics of  topological phases and mathematics of non-hyperbolic 3-manifolds, in an unexpected way. This connection will provide new implications to both fields. Via this connection, some universal properties on topological phases can be translated into non-trivial mathematical predictions on non-hyperbolic 3-manifolds satisfying \eqref{3-manifolds for top}.  For instance, the additional structures of TQFTs other than modular data can be related to new topological invariants on 3-manifolds.
 As concrete examples, we now propose the following non-trivial mathematical conjectures:
\begin{align}
\begin{split}
&\textrm{Let $\{\rho_\alpha\}$ be the set of irreducible flat connections on $M$}
\\
&\textrm{satisfying conditions in \eqref{3-manifolds for top} and having trivial $H_1(M, \mathbb{Z}_2)$.}
\\
&\textbf{Conjecture 1 :  }\sum_\alpha \frac{1}{2 \textrm{Tor}[\rho_\alpha]}  =1\;. 
\\
&\textbf{Conjecture 2 : } \textrm{There exists a special $\rho_{\alpha=0}$ satisfying }
\\
&\qquad \qquad \qquad  \frac{1}{\sqrt{2 \textrm{Tor}[\rho_{\alpha =0}]}} = \bigg{|}\sum_\delta \frac{\exp (-2\pi i CS[\rho_\delta])}{2 \textrm{Tor}[\rho_\delta]} \bigg{|}\;.
\end{split}
\end{align}
The first conjecture simply follows from the uniqueness of ground states on a two-sphere , i.e. $\textrm{GSD}_{g=0}=1$ in \eqref{Z[M-pg]}, and the 2nd follows from the explanations around the equation \eqref{trivial vacuum}.  It is also interesting to compare the 1st conjecture with its counterpart for hyperbolic 3-manifolds. For a hyperbolic $M$, the quantity in the 1st conjecture is claimed to vanish \cite{Benini:2019dyp,Gang:2019dbv,yoon2020vanishing}.

More recently, an interesting universal property of bosonic topological phases is proposed in \cite{Kong:2020bmb}
\begin{align}
c_{2d}  \times \textrm{GSD}_g \in 2 \mathbb{Z}\;, \; \textrm{when $g \geq 3$}\;.
\end{align}
The proposed property can be translated into following conjecture.
\begin{align}
\begin{split}
&\textbf{Conjecture 3 :  } CS[\rho_{\alpha=0}]\times \sum_\delta (2 \textrm{Tor}[\rho_\delta])^{g-1}  \in \frac{\mathbb{Z}}{48} \;, 
\\
& \qquad \qquad  \quad \qquad \textrm{for $g \geq 3$}\;.
\end{split}
\end{align}
Here  $\rho_{\alpha=0}$ is the special flat connection in the above 2nd conjecture. In the above we use the fact that $CS[\rho_{\alpha=0}] = \pm \frac{1}{24}c_{ 2d}$ where $c_{2d}$ is only well-defined modulo $\frac{1}2$. 
We checked that the above  conjectures hold for several examples of  $M=S^2(\frac{p_1}{q_1},\frac{p_2}{q_2},\frac{p_3}{q_3})$ with trivial $H_1 (M, \mathbb{Z}_2)$. It would be interesting to prove/improve or disprove those conjectures.

\section{Acknowledgments} 
We would like to thank Byungmin Kang, Seok Kim, Kimyeong Lee, Sungjay Lee for valuable discussions and comments. GYC is supported by the National Research Foundation of Korea(NRF) grant funded by the Korea government(MSIT) (No. 2020R1C1C1006048 and No. 2020R1A4A3079707). The research of DG is supported in part by the National Research Foundation of Korea under grant 2019R1A2C2004880. DG also acknowledges support by the appointment to the JRG program at the APCTP through the Science and Technology Promotion Fund and Lottery Fund of the Korean Government, as well as support by the Korean Local Governments, Gyeongsangbuk-do Province, and Pohang City. The research of HK is supported by the POSCO Science Fellowship of POSCO TJ Park Foundation and the National Research Foundation of Korea (NRF) Grant 2018R1D1A1B07042934.

\newpage 

\appendix

\section{Brief Review of UMTC}\label{UMTC-Suppl}
Here we present a brief summary of UMTC, which is based on \cite{Kitaev, Bonderson, Bernevig}. We will focus on listing the important facts. We recommend the interested readers to read \cite{Kitaev, Bernevig}. To begin with, UMTC is a mathematical framework describing the physics of anyons with the consistent implementation of fusion and braiding. Roughly speaking, it describes a theory of anyons with the fundamental particle being a boson. 

Hence, we consider a theory of a finite number of anyons, 
\begin{align}
\alpha, \beta, \gamma \cdots \in \mathcal{C}. 
\end{align}
For all $\alpha \in \mathcal{C}$, there is a unique anti-particle $\bar{\alpha}$. There is an unique vacuum (or trivial particle) $1 = \bar{1}$. The anyons satisfy the fusion rule, i.e., 
\begin{align}
\alpha \times \beta = \sum_{\gamma} N^{\gamma}_{\alpha \beta} \gamma.  
\end{align}
Here $N^{\gamma}_{\alpha \beta} =N^{\gamma}_{\beta\alpha}$ is a positive integer, which represents the multiplicity in the fusion. It satisfies certain algebraic equations. For instance, the associativity of the fusion requires $\sum_{\gamma}N^{\gamma}_{\alpha \beta}N_{\gamma \delta}^{\eta} = \sum_{\phi}N^{\eta}_{\alpha \phi}N_{\beta \delta}^{\phi}$. Also $N^{1}_{\alpha \bar{\alpha}} = 1$, i.e., when $\alpha$ and its antiparticle are fused, they generate a unique vacuum (among other anyons). 

Next we introduce $F$-symbol. For this, we first introduce the fusion $V_{\alpha \beta}^{\gamma}$ and splitting $V^{\alpha \beta}_{\gamma}$ spaces, which are dual to each other. A vector in $V^{\gamma}_{\alpha \beta}$ represents an amplitude of the particle fusion process. When multiple particles are split (or fused), there can be multiple different ways of decomposing the corresponding spaces into $V^{\alpha \beta}_{\gamma}$'s (or $V_{\alpha \beta}^{\gamma}$'s). For instance, when $V^{\alpha \beta \gamma}_{\delta}$ is considered, i.e., the anyon $\delta$ is decomposed into the three $\alpha$, $\beta$, $\gamma$, then there are two different ways of splitting it: $V^{\alpha \beta \gamma}_{\delta} = \sum_{\eta} V^{\alpha \beta}_{\eta} \otimes V^{\eta \gamma}_{\delta}$ or $V^{\alpha \beta \gamma}_{\delta} = \sum_{\phi} V^{\alpha \phi}_{\delta} \otimes V^{\beta \gamma}_{\phi}$. Since they all represent the same physical process, there must be an unitary mapping between them. This unitary matrix is the $F$-symbol, which must satisfy the pentagon equation. 

One important quantity, which can be derived from $F$, is the quantum dimension $d_\alpha$ for a given anyon $\alpha$,  
\begin{align}
\Big|F^{\alpha \bar{\alpha} \alpha}_{\alpha 11}\Big|  = \frac{1}{d_\alpha}. 
\end{align}
Physically, $d_\alpha$ represents the asymptotic scaling of the dimension of the Hilbert space under the repeated fusion of $\alpha$. That is, when we fuse $\alpha$ N-times, $\alpha \times \alpha \times \alpha \times \alpha \cdots \times \alpha$, the dimension of the fusion Hilbert space scales as $\sim d_{\alpha}^N$. For the Abelian particle, $d_\alpha = 1$. The consistency of the fusion requires $d_\alpha d_\beta = \sum_{\gamma} N^{\gamma}_{\alpha \beta} d_{\gamma}$.

Next we discuss the braiding of the anyons. The central object is $R^{\alpha \beta}_{\gamma}$ relating $V^{\alpha \beta}_{\gamma}$ and $V^{\beta\alpha}_{\gamma}$. Consistency between the fusion and the braiding requires so-called hexagon equation. What's important for us in relation with the main text is the topological spin $\theta_\alpha$ of anyon $\alpha \in \mathcal{C}$. This can be written in terms of $R$:    
\begin{align}
\theta_\alpha = \frac{1}{d_\alpha} \sum_\gamma d_\gamma \text{Tr}\Big(R^{\alpha \alpha}_\gamma \Big), 
\end{align}
in which the trace is taken over the multiplicities in the fusion channel $\alpha \times \alpha \to \gamma$. 

With these, we are now ready to write out the formula for the modular data. The first is the S-matrix:  
\begin{align}
S_{\alpha \beta} = \frac{1}{D} \sum_\gamma N^{\gamma}_{\alpha \beta} \frac{\theta_\gamma }{\theta_\alpha \theta_\beta} d_\gamma = S_{\beta \alpha}, 
\end{align}
in which $D = \sqrt{\sum_\alpha d_{\alpha}^2}$. This represents the quantum amplitude associated with the linking of the two anyons $\alpha$ and $\beta$. Note that the S-matrix is unitary and hence has an inverse. The theory with this property, i.e., unitary S-matrix, is called as UMTC. For instance, if the S-matrix is not unitary, we have a non-unitary MTC, which is studied in our main text. We can also relate certain part of the S-matrix with the quantum dimensions $d_\alpha$, i.e., $S_{1\alpha} = \frac{d_\alpha}{D}$. The second modular data is the T-matrix such that 
\begin{align}
T_{\alpha \beta} = \exp \left(-2\pi i \frac{c_{2d}}{24} \right) \times \delta_{\alpha \beta} \theta_\alpha. 
\end{align}
Remarkably, the chiral central charge $c_{2d}$ mod 8 can be determined solely from these modular data. 
\begin{align}
\frac{1}{D} \sum_{\alpha}d_{\alpha}^2 \theta_{\alpha} = \exp \Big(2\pi i \frac{c_{2d}}{8}\Big),   
\end{align}
from which we can actually determine the chiral central charge for a series of UMTC. In contrast to $R$- and $F$- symbols above, the modular data, S- and T-matrices, are gauge-independent though they do not completely characterize the MTC \cite{MS}. 

A side remark is that, in the main text, the charge conjugation $C$ plays an important role. The charge conjugation can be defined in terms of the modular data as $C = S^2  = \delta _{\bar{\alpha}\beta}$ and $C^2 = 1$. In the main text, we called the theory with $C= 1$ as ``self-dual". 

\section{Irreducible $SL(2,\mathbb{C})$ flat connections and topological invariants of 3-manifolds } \label{app : CS and Tor}
Here we explain the two  invariants, $CS[\rho_\alpha]$ and $\textrm{Tor}[\rho_\alpha]$ which play crucial role in the Table \ref{Table : MTC[M]}, using perturbative analysis of $SL(2,\mathbb{C})$ Chern-Simons theory on $M$.  The mathematical  invariants can be  understood  as the first two perturbative expansion coefficients (the leading and the next subleading) of the $SL(2,\mathbb{C})$ Chern-Simons theory. The classical action for the complex Chern-Simons  theory is 
\begin{align}
CS [\mathcal{A}]= \frac{1}{8\pi^2} \int_M \textrm{Tr} (\mathcal{A}\wedge d\mathcal{A} + \frac{2}3 \mathcal{A}\wedge \mathcal{A}\wedge \mathcal{A})\;.
\end{align}
The classical solutions are $SL(2,\mathbb{C})$ flat connections:
\begin{align}
\frac{\delta CS[\mathcal{A}]}{\delta \mathcal{A}}\bigg{|}_{\mathcal{A} = \mathcal{A}_\alpha} = 0\;  \Rightarrow  \; d\mathcal{A}_\alpha + \mathcal{A}_\alpha \wedge \mathcal{A}_\alpha=0\;.
\end{align} 
One can consider following  perturbative expansion of the complex Chern-Simons theory around a flat connection $\mathcal{A}_\alpha$:
\begin{align}
\begin{split}
& \int \frac{D(\delta \mathcal{A})}{(\textrm{gauge})} e^{-\frac{4\pi^2 }{\hbar} CS[\mathcal{A}_\alpha + \delta \mathcal{A}]}
\\
& \xrightarrow{\quad \hbar \rightarrow 0\quad }  \frac{1}{\textrm{vol}(H_\alpha)} \exp \left(\frac{1}{\hbar} F_0^\alpha + F_1^\alpha + \ldots \right) \label{CS theory perturbation} \ ,
\end{split}
\end{align}
where $H_\alpha$ is the unbroken gauge group by the flat connection $\mathcal{A}_\alpha$, i.e.
\begin{equation}
H_\alpha = \{ h \in SL(2,\mathbb{C}) :  [h, \rho_\alpha (\gamma)] =0 \; \forall \gamma \in \pi_1 (M)\} \ ,
\end{equation}
and $\textrm{vol}(H_\alpha)$ is the volume of the unbroken subgroup. Here, $\rho_\alpha(\gamma)$ is the holonomy matrix for $\mathcal{A}_\alpha$.  Note that only the flat connection $\mathcal{A}_\alpha$ with finite $\textrm{vol}(H_\alpha)$ has non-trivial perturbative expansion. This  is possible only  if $\textrm{dim}_{\mathbb{C}}H_\alpha =0 $.
\begin{align}
\begin{split}
\textrm{A flat connection $\rho_\alpha$ is  } 
&\begin{cases}
&\textrm{irreducible}, \quad \textrm{if $\textrm{dim}_{\mathbb{C}}H_\alpha =0$ }
\\
&\textrm{reducible}, \quad  \textrm{otherwise}\;.
\end{cases}
\end{split}
\end{align}
Note that only the irreducible flat connections appear in the Table \ref{Table : MTC[M]}. The classical part $F^\alpha_0$ is nothing but the classical Chern-Simons action
\begin{align}
F_0^\alpha  =-4\pi^2  CS[\rho_\alpha] \ .
\end{align}
The next leading part $F_1^\alpha$ is the 1-loop contribution. This term is related to the adjoint Reidemeister torsion as
\begin{align}
\textrm{Tor}[\rho_\alpha ] = \exp \left(-2F_1^\alpha \right)\;.
\end{align}

\section{3d theory $\mathcal{T}[M]$ and 3d-3d relations} \label{appendix : T[M]}
The theory $\mathcal{T}[M]$ in \eqref{6d def of T[M]} is defined as the (2+1)d field theory obtained from a  compactification of the (5+1)d world-volume theory of two M5-branes along a 3-manifold $M$. For a given 3-manifold $M$, the compactification  procedure is not unique. We need to specify some additional discrete choices \cite{Gang:2018wek,Eckhard:2019jgg}. We now explain   the discrete choices used in  the definition of $\mathcal{T}[M]$ in this paper.

First, we need to specify a component of vacua  in the compactification. There could be several discrete components of vacua on $\mathbb{R}^{1,2}$ in the compactification. The set of  vacua  is somehow related to the set of flat $SL(2,\mathbb{C})$ (or $PSL(2,\mathbb{C})$) connections on $M$. See \cite{Gang:2018wek} for details. In the relation, the subset of  irreducible flat connections corresponds to a component of vacua on $\mathbb{R}^{1,2}$ and we choose the `irreducible' component  in the definition of $\mathcal{T}[M]$. 

 Secondly, we need to specify a discrete  polarization choice   of the (5+1)d world-volume theory as studied in the context of 4d-2d correspondence \cite{Tachikawa:2013hya} and 3d-3d correspondence \cite{Gang:2018wek,Eckhard:2019jgg}. According to \cite{Eckhard:2019jgg}, the discrete choice can be characterized by the choice of a subgroup of $H \subset H^1 (M, \mathbb{Z}_2)$. Hence the (2+1)d theory in the compactification is labelled by a 3-manifold $M$ and a subgroup $H$,
\begin{align}
\mathcal{T}[M;H]\;, \quad H \subset H^1 (M, \mathbb{Z}_2)\;.
\end{align}
This theory has the subgroup $H$ as the 0-form flavor symmetry. 
In our convention, we denote by $\mathcal{T}[M]$ for the theory with  the choice $ H = H^1(M, \mathbb{Z}_2)$,
\begin{align}
\mathcal{T}[M] := \mathcal{T}[M;H= H^1 (M, \mathbb{Z}_2)] \ .
\end{align}
The theories in this class were studied in \cite{Gang:2018wek}. 
There is also a theory with the opposite polarization choice, where the subgroup $H$ is chosen to be trivial. We call this theory as
\begin{align}
\widetilde{\mathcal{T}}[M] := \mathcal{T}[M;H= \emptyset]
\end{align}
The theory $\widetilde{\mathcal{T}}[M]$ has the  cohomology group  $H^1(M, \mathbb{Z}_2)$ as the 1-form flavor symmetry. The theories $\mathcal{T}[M;H]$ with other generic choices of $H$  can be obtained from $\widetilde{\mathcal{T}}[M]$ by gauging the corresponding 1-form symmetries,
\begin{align}
\begin{split}
&\mathcal{T}[M;H] 
\\
&= \textrm{Gauging the 1-form symmetry $H\subset H^1(M, \mathbb{Z}_2)$ of $\widetilde{\mathcal{T}}[M]$ }\;. \nonumber
\end{split}
\end{align}
It then follows that
\begin{align}
\mathcal{T}[M] = \left(\textrm{Gauging $H=H^1(M, \mathbb{Z}_2) $ of $\widetilde{\mathcal{T}}[M]$ } \right)\;.
\end{align}
In condensed matter context, gauging a 1-form symmetry can be interpreted as condensation of the anyon generating the 1-form symmetry \cite{Gaiotto:2015zta}.

One nice feature of the $\mathcal{T}[M]$ is that its supersymmetric partition function can be related to invariants of $SL(2,\mathbb{C})$ Chern-Simons theories on $M$. The relation is called {\it 3d-3d relation}. For example \cite{Dimofte:2011py,Lee:2013ida,Yagi:2013fda,Benini:2019dyp},
\begin{align}
\begin{split}
&\mathcal{I}_{\rm sci}^{UV}(x) =\frac{1}2 \sum_\alpha B_M^\alpha (x)  B_M^{\overline{\alpha}}  (x^{-1}) \;,
\\
&\mathcal{I}_{\rm top}^{UV}(x) =\frac{1}2 \sum_\alpha B_M^\alpha (x)  B^\alpha_M(x^{-1})  \;. \label{3d-3d relation}
\end{split}
\end{align}
Here the summations run over all irreducible $SL(2,\mathbb{C})$ flat connections $\rho_\alpha$. Note the subtle but important difference between two lines, there is bar in $B^{\overline{\alpha}}_M(x^{-1})$ in the first line while no bar in the second line. $B^{\alpha}_M (x)$ is so-called {\it holomorphic block in $SL(2,\mathbb{C})$ Chern-Simons theory} associated to a flat connection $\rho_\alpha$ while $B^{\overline{\alpha}}_M (x)$ is associated to the flat connection $(\rho_\alpha)^*$, complex conjugation of $\rho_{\alpha}$. 
$\mathcal{I}^{UV}_{\rm sci} (x)$ and $\mathcal{I}^{UV}_{\rm top} (x)$ are the superconformal index \cite{Kim:2009wb,Imamura:2011su} and the refined  twisted index  at UV R-symmery choice respectively. 

The superconformal index $\mathcal{I}^{UV}_{\rm sci}(x)$ is a 3-manifold invariant called `3D index'  \cite{Garoufalidis:2016ckn} and can be systematically computed using a Dehn surgery description of $M$ along a knot $\mathcal{K}$ and an ideal triangulation of the knot complement \cite{Dimofte:2011py,Gang:2018wek}. 
The  holomorphic block  $B_M^\alpha (x)$ is defined by the following path-integral on $M$,
\begin{align}
B^\alpha_M (x):= \int_{\mathcal{C}^\alpha (\mathcal{A})} \frac{D \mathcal{A}}{(\textrm{gauge})} e^{- \frac{4\pi^2 }\hbar  CS[\mathcal{A}]}\bigg{|}_{\hbar  = \log x}\;.
\end{align}
$\mathcal{A}$ is the $SL(2,\mathbb{C})$ connections on $M$ and the path-integral contour $\mathcal{C}^\alpha (\mathcal{A})$ is the Leftschetz thimble associated to an irreducible flat connection $\rho_\alpha$ \cite{Witten:2010cx}. The $B^\alpha_M(x)$ shares  the same perturbative expansion with \eqref{CS theory perturbation} in the limit $x:=e^{\hbar} \rightarrow 1$,
\begin{align}
\log B^\alpha_M(x) \xrightarrow{\quad x=e^{\hbar};\;\hbar \rightarrow 0 \quad } \sum_{n} F_n^\alpha \hbar^{n-1}\;. \label{perturbative of hol block}
\end{align}
Thus, $B^\alpha_M(x)$ can be considered as a non-perturbative completion of the perturbative expansion \eqref{CS theory perturbation}. Via the 3d-3d relation \cite{Beem:2012mb}, the holomorphic block $B^\alpha_M$ of the $SL(2,\mathbb{C})$ theory is identical to the holomorphic block $\mathbb{B}^\alpha$ in  \eqref{factorization}  of the $\mathcal{T}[M]$ theory up to an overall factor
\begin{align}
(\mathbb{B}^\alpha (x) \textrm{ of }\mathcal{T}[M])  = \frac{1}{\sqrt{2}} \times B^\alpha_M (x)  \;.
\end{align}
The  overall factor $1/\sqrt{2}$ should be replaced by $1/\sqrt{K}$ when the number of M-branes in the geometrical  set-up  \eqref{6d def of T[M]} is  $K$ instead of 2. The  overall factor is important in  the study of large $K$ limit of supersymmetric partition functions \cite{Gang:2019uay,Benini:2019dyp}.

\section{IR phases of $\mathcal{T}[M]$} \label{appendix : IR phases}
Here we explain how to determine  basic properties of IR phases of  the 3d supersymmetric gauge theory  $\mathcal{T}[M]$  from basic topological properties of the 3-manifold $M$.
\\
\paragraph{(No irreducible $SL(2,\mathbb{C})$ flat connection on $M$) $\Rightarrow$ (Supersymmetry in $\mathcal{T}[M]$ is spontaneously broken)} Obviously from the 3d-3d relations in  \eqref{3d-3d for twisted parition functions} and \eqref{3d-3d relation}, all the supersymmetric partition functions on $\mathcal{M}_{g,p}$ (such as $\mathcal{I}_{\rm sci}(x), \mathcal{I}_{\rm top}(x),\ldots$) vanish when there is no irreducible $SL(2,\mathbb{C})$ flat connection on $M$. This implies that the supersymmetry of the theory $\mathcal{T}[M]$ is spontaneously broken. 
In the broken phase, the vacuum partition functions vanish due to the fermion zero modes coming from the broken supercharges. Among the supersymmetric partition functions, the Witten index (the partition function on $\mathcal{M}_{g=1,p=0}$) is independent of the choice of R-symmetry. Thus the above conclusion is still valid even if the  UV R-symmetry is different from the IR R-symmetry. 

One famous example of such a 3-manifold having no irreducible flat connection is the Lens space.  The fundamental groups of the Lens spaces are all Abelian and thus there is no irreducible flat connection. Accordingly, the supersymmetry of the associated (2+1)d theories $\mathcal{T}[M]$ will be spontaneously broken.
One can in fact use the property $\mathcal{I}^{UV}_{\rm sci}(x)=0$ of the $\mathcal{T}[M]$ from the Lens spaces $M$  to check if a 3-manifold given by a Dehn surgery description is a Lens space or not \cite{Gang:2018gyt}.
\\
\paragraph{(Hyperbolic 3-manifold $M$) $\Rightarrow$ (non-trivial superconformal field theory)} In the case, there is a canonical complex flat connection $\mathcal{A}_{\alpha={\rm hyp}}$ constructed from the unique hyperbolic structure on $M$. Locally, the flat connection can be written as
\begin{align}
\mathcal{A}_{\alpha={\rm hyp}} = \omega + i e\;.
\end{align}
Here $\omega$ and $e$ are the spin-connection and the dreibein,  respectively, constructed from the hyperbolic structure. Both can be considered as $SO(3)$-valued 1-forms on $M$. The above complex combination gives a  $SL(2,\mathbb{C})$  connection which is actually the flat connection satisfying  $d\mathcal{A}+\mathcal{A} \wedge\mathcal{A} =0$. 
The flat connection gives the most exponentially dominant contribution to the superconformal index $\mathcal{I}_{\rm sci}(x)$ in the limit $x\rightarrow 1$ \cite{Bobev:2019zmz,Benini:2019dyp}:
\begin{align}
\log \mathcal{I}^{UV}_{\rm sci}(x) \xrightarrow{\quad x=e^{-\nu};\;\nu \rightarrow i 0^- \quad } -  \frac{2 i}{\nu} \textrm{vol}(M)+ o (\nu^0)\;. \nonumber
\end{align}
The above follows from \eqref{3d-3d relation} and \eqref{perturbative of hol block} combined with following fact
\begin{align}
\textrm{Im}\left[CS[\mathcal{A}_{\alpha = \textrm{\rm hyp}}]\right] =- \frac{1}{4\pi^2} \textrm{vol}(M)\;.
\end{align}
Here $\textrm{vol}(M)$ is the hyperbolic volume of the hyperboilc 3-manifold. The non-triviality of the superconformal index guarantees that the  $\mathcal{T}[M]$ flows to a non-trivial superconformal field theory.
More generally,  from the same argument,  one expects that
\begin{align}
\begin{split}
&\textrm{If there is an irreducible $SL(2,\mathbb{C})$ flat connection $\rho$ on $M$}
\\
& \textrm{with $\textrm{Im}[CS[\rho]] \neq 0$}
\\
&\Rightarrow 
\\
&\textrm{$\mathcal{T}[M]$ flows to a non-trivial superconformal field theory}. \nonumber
\end{split}
\end{align}
\\
\paragraph{\textrm{$M$ satisfying   conditions in \eqref{3-manifolds for top}} $\Rightarrow\ B_M^{\overline{\alpha}} (x) = B_M^{\alpha} (x)$}

For a $SL(2,\mathbb{R})$ or $SU(2)$ flat connection $\mathcal{A}_\alpha$, its complex conjugation $(\mathcal{A}_\alpha)^*$ equals to itself up to a transpose or sign 
\begin{align}
&(\mathcal{A}_\alpha)^* =\mathcal{A}_\alpha \;, \quad \textrm{for $SL(2,\mathbb{R})$ flat connection}\;,\nonumber
\\
&(i \mathcal{A}_\alpha)^* =(i\mathcal{A}_\alpha)^T\;,  \quad \textrm{for $SU(2)$ flat connection}\;.\nonumber
\end{align}
This implies that two flat connections $\mathcal{A}_\alpha$ and $(\mathcal{A}_\alpha)^*$  have the  same  perturbative expansion in \eqref{CS theory perturbation}, i.e. $F_n^\alpha = F^{\overline{\alpha}}_n=(F_n^\alpha)^* $ for all $n\geq 0$.
Therefore, one finds that for a 3-manifold subject to the conditions \eqref{3-manifolds for top},
\begin{align}
\begin{split}
& B_M^{\overline{\alpha}} (x) = B_M^{\alpha} (x) \;\; \textrm{for all irreducible flat connections $\rho_\alpha$} \;,
\\
&\Rightarrow\ \mathcal{I}^{UV}_{\rm sci}(x) = \mathcal{I}^{UV}_{\rm top}(x)  \textrm{ from  \eqref{3d-3d relation}}  \;. \nonumber
\end{split}
\end{align}

\section{Irreducible flat connections on $M=S^2 (3,3,-4/3), S^2 (3,3,-5/3)$ and $S^2 (2,3,8)$} \label{app : examples of irreducible flat connections}
To obtain irreducible $SL(2,\mathbb{C})$ flat connections on $S^2 (\frac{p_1}{q_1},\frac{p_2}{q_2},\frac{p_3}{q_3})$, we start from following ansatz for their holonomy matrices
\begin{align}
&\rho(x_1) =   \left(
\begin{array}{cc}
 m & \epsilon \\
0 & m^{-1} \\
\end{array}
\right)\bigg{|}_{\epsilon \in \{0,1\}}\;, \quad \rho(x_2) =   \left(
\begin{array}{cc}
a & b\\
c & d \\
\end{array}
\right)\;,\nonumber
\\
&\rho(x_3) = \left(\rho(x_1) \cdot \rho(x_2) \right)^{-1}\;, \quad  \rho(h) \in  \{ + \mathbf{1},-\mathbf{1}\} \;.
\end{align}
Then, we solve the matrix equations for the holonomy matrices constrained from the relations \eqref{fund group} in the fundamental group. Among those solutions,  we need to discard reducible flat connections for which all holonomy matrices are mutually commuting, and choose gauge inequivalent sets of the irreducible ones. 
\\
\paragraph{$M=S^2 (3,3,-4/3)$  } There are 3 irreducible flat connections $\rho_{\alpha =0,1,2}$ on $M=S^2 (3,3,-4/3)$ whose $SL(2,\mathbb{C})$ holonomy matrices are
\begin{align}
\begin{split}
&\alpha =0:
\\
&\rho_{\alpha} (x_1) = \left(
\begin{array}{cc}
e^{ \frac{\pi i}{3}}& 0 \\
0 & e^{ -\frac{\pi i}{3}} \\
\end{array}
\right),  \; \rho_{\alpha=0 } (h) = \left(
\begin{array}{cc}
-1 & 0\\
0 & -1 \\
\end{array}
\right),  
\\
&\rho_{\alpha} (x_2) =\left(
\begin{array}{cc}
\frac{1}2 + \frac{i}{6} \left(\sqrt{3}-2 \sqrt{6}\right) & -\frac{\sqrt{2}}{3} \\
1 & \frac{1}2 - \frac{i}{6} \left(\sqrt{3}-2 \sqrt{6}\right)   \\
\end{array}
\right),
\\
&\rho_{\alpha} (x_3) =\left(
\begin{array}{cc}
\frac{1}{\sqrt{2}}-\frac{i}{\sqrt{3}}+\frac{i}{\sqrt{6}} & \frac{1+i \sqrt{3}}{3 \sqrt{2}} \\
\frac{1}{2} i \left(\sqrt{3}+i\right) & \frac{1}{\sqrt{2}}+\frac{i}{\sqrt{3}}-\frac{i}{\sqrt{6}} \\
\end{array}
\right) , 
\\
&\alpha=1:
\\
&\rho_{\alpha } (x_1) = \left(
\begin{array}{cc}
e^{- \frac{2\pi i}{3}}& 0 \\
0 & e^{ \frac{2\pi i}{3}} \\
\end{array}
\right), \;  \rho_{\alpha} (x_2) = \left(
\begin{array}{cc}
\frac{e^{-\frac{5 i \pi }{6} }}{\sqrt{3}} & -\frac{2}3 \\
1 & \frac{e^{\frac{5 i \pi }{6} }}{\sqrt{3}}  \\
\end{array}
\right),  
\\
&\rho_{\alpha } (x_3) = \left(
\begin{array}{cc}
-\frac{i}{\sqrt{3}}& \frac{2}3  e^{\frac{4i \pi}3 }  \\
e^{-\frac{i \pi}3 } & \frac{i}{\sqrt{3}}\\
\end{array}
\right), \;  \rho_{\alpha} (h) = \left(
\begin{array}{cc}
1 & 0\\
0 & 1 \\
\end{array}
\right),  
\\
&\alpha =2:
\\
&\rho_{\alpha} (x_1) = \left(
\begin{array}{cc}
e^{ \frac{\pi i}{3}}& 0 \\
0 & e^{ -\frac{\pi i}{3}} \\
\end{array}
\right),  \; \rho_{\alpha } (h) = \left(
\begin{array}{cc}
-1 & 0\\
0 & -1 \\
\end{array}
\right),  
\\
&\rho_{\alpha} (x_2) =\left(
\begin{array}{cc}
\frac{1}2 + \frac{i}{6} \left(\sqrt{3}+2 \sqrt{6}\right) & \frac{\sqrt{2}}{3} \\
1 & \frac{1}2 - \frac{i}{6} \left(\sqrt{3}+2 \sqrt{6}\right)   \\
\end{array}
\right),
\\
&\rho_{\alpha } (x_3) =\left(
\begin{array}{cc}
-\frac{1}{\sqrt{2}}-\frac{i}{\sqrt{3}}-\frac{i}{\sqrt{6}} & -\frac{i \left(\sqrt{3}-i\right)}{3 \sqrt{2}} \\
\frac{1}{2} i \left(\sqrt{3}+i\right) & -\frac{1}{\sqrt{2}}+\frac{i}{\sqrt{3}}+\frac{i}{\sqrt{6}} \\
\end{array}
\right) . \label{holonomy matrices}
\end{split}
\end{align}
One can confirm that all of them are  conjugate to either $SU(2)$ or $SL(2,\mathbb{R})$ flat connections by checking that $\textrm{Tr}(\rho(a)) \in \mathbb{R}$ for all $a \in \pi_1 (M)$. $\rho_{\alpha=2}$ is conjugate to $SL(2,\mathbb{R})$ flat connection while the other twos are conjugate to $SU(2)$ flat connections. 
Thus the 3-manifold $M$ satisfies the topological conditions in \eqref{3-manifolds for top}.  The eigenvalues of the holonomy matrices are
\begin{align}
\begin{split}
&\alpha=0 \; : \; n_{\alpha 1} = \frac{1}2\;,  n_{\alpha 2} = \frac{1}2\;,  n_{\alpha 3} = \frac{1}2\;, \lambda_{\alpha} = \frac{1}2 \;,
\\
&\alpha=1 \; : \; n_{\alpha 1} = 1\;,  n_{\alpha 2} =1\;,  n_{\alpha 3} = \frac{1}2\;, \lambda_{\alpha} =0 \;,
\\
&\alpha=2 \; : \; n_{\alpha 1} = \frac{1}2\;,  n_{\alpha 2} = \frac{1}2\;,  n_{\alpha 3} = \frac{3}2\;, \lambda_{\alpha} = \frac{1}2\;.
\end{split}
\end{align}
See eq.~\eqref{n-lambda} for the definition of the above $n_{\alpha i}$ and $\lambda_\alpha$. From this result, we obtain  $CS[\rho_\alpha]$ and $\textrm{Tor}[\rho_\alpha]$ given in \eqref{CS-Tor for M[3,3,-4/3]} using the formula in  \eqref{CS-and-torsion-Seifert}

\paragraph{$M=S^2 (3,3,-5/3)$  } There are 4 irreducible flat connections $\rho_{\alpha =0,1,2,3}$ on $M=S^2 (3,3,-5/3)$ whose $SL(2,\mathbb{C})$ holonomy matrices are
\begin{align}
\begin{split}
&\alpha =0:
\\
&\rho_{\alpha} (x_1) = \left(
\begin{array}{cc}
e^{- \frac{2\pi i}{3}}& 0 \\
0 & e^{ \frac{2 \pi i}{3}} \\
\end{array}
\right),  \; \rho_{\alpha=0 } (h) = \left(
\begin{array}{cc}
1 & 0\\
0 & 1 \\
\end{array}
\right),  
\\
&\rho_{\alpha} (x_2) =\left(
\begin{array}{cc}
-\frac{1}{2}-\frac{i \left(\sqrt{5}+2\right)}{2 \sqrt{3}}  & \frac{\sqrt{5}}{3} \\
1 &-\frac{1}{2}+i \sqrt{\frac{\sqrt{5}}{3}+\frac{3}{4}}   \\
\end{array}
\right),
\\
&\rho_{\alpha} (x_3) =\left(
\begin{array}{cc}
u & -\frac{i \left(\sqrt{3}-i\right)}{3 \sqrt{2}} \\
\frac{1}{2} i \left(\sqrt{3}+i\right) & u^* \\
\end{array}
\right), 
\\
&\alpha=1:
\\
&\rho_{\alpha } (x_1) = \left(
\begin{array}{cc}
e^{ \frac{\pi i}{3}}& 0 \\
0 & e^{ -\frac{\pi i}{3}} \\
\end{array}
\right), \;    \rho_{\alpha} (h) = \left(
\begin{array}{cc}
-1 & 0\\
0 & -1 \\
\end{array}
\right),  
\\
&\rho_{\alpha} (x_2) =\left(
\begin{array}{cc}
\frac{1}{6} \left(3-i \sqrt{15}\right) & -\frac{1}{3} \\
1 & \frac{1}{6} \left(3+i \sqrt{15}\right) \\
\end{array}
\right) ,  
\\
&\rho_{\alpha } (x_3) =\left(
\begin{array}{cc}
\frac{1}{6} \sqrt[6]{-1} \left(\sqrt{15}-3 i\right) & \frac{\sqrt[3]{-1}}{3} \\
(-1)^{2/3} & \frac{1}{6} \sqrt[3]{-1} \left(3-i \sqrt{15}\right) \\
\end{array}
\right) 
\\
&\alpha =2:
\\
&\rho_{\alpha} (x_1) = \left(
\begin{array}{cc}
e^{ -\frac{2\pi i}{3}}& 0 \\
0 & e^{ \frac{2\pi i}{3}} \\
\end{array}
\right),  \; \rho_{\alpha } (h) = \left(
\begin{array}{cc}
1 & 0\\
0 & 1 \\
\end{array}
\right),  
\\
&\rho_{\alpha} (x_2) =\left(
\begin{array}{cc}
-\frac{1}{2}-\frac{i \left(-\sqrt{5}+2\right)}{2 \sqrt{3}}  &- \frac{\sqrt{5}}{3} \\
1 &-\frac{1}{2}+i \sqrt{\frac{-\sqrt{5}}{3}+\frac{3}{4}}   \\
\end{array}
\right)\;,
\\
&\rho_{\alpha } (x_3) =\left(
\begin{array}{cc}
v  & -\frac{1}{3} \sqrt[3]{-1} \sqrt{5} \\
-(-1)^{2/3} & v^* \\
\end{array}
\right),
\\
&\alpha=3:
\\
&\rho_{\alpha } (x_1) = \left(
\begin{array}{cc}
e^{ \frac{\pi i}{3}}& 0 \\
0 & e^{ -\frac{\pi i}{3}} \\
\end{array}
\right), \;    \rho_{\alpha} (h) = \left(
\begin{array}{cc}
-1 & 0\\
0 & -1 \\
\end{array}
\right),  
\\
&\rho_{\alpha} (x_2) =\left(
\begin{array}{cc}
\frac{1}{6} \left(3+i \sqrt{15}\right) & -\frac{1}{3} \\
1 & \frac{1}{6} \left(3-i \sqrt{15}\right) \\
\end{array}
\right) ,  
\\
&\rho_{\alpha } (x_3) =\left(
\begin{array}{cc}
\frac{1}{6} \sqrt[6]{-1} \left(-\sqrt{15}-3 i\right) & \frac{\sqrt[3]{-1}}{3} \\
(-1)^{2/3} & \frac{1}{6} \sqrt[3]{-1} \left(3+i \sqrt{15}\right) \\
\end{array}
\right) \;.
\label{holonomy matrices2}
\end{split}
\end{align}
Here $\{u \simeq  -0.809017-1.04444 i ,\;v \simeq 0.309017-0.398939 i ,\; u^*,\;v^*\}$ are the solutions to an  algebraic equation $9 x^4+9 x^3+9x^2-6 x+4=0$.  $\rho_{\alpha=0}$ is conjugate to an $SL(2,\mathbb{R})$ flat connection while the other threes are conjugate to $SU(2)$ flat connections.

\paragraph{$M=S^2 (2,3,8)$ } This manifold has a non-trivial $H_1(M, \mathbb{Z}_2) = \mathbb{Z}_2$. Thus, the set of $PSL(2,\mathbb{C})$ flat connections is different from the set of  $SL(2,\mathbb{C})$ flat connections. There are four irreducible $PSL(2,\mathbb{C})$ flat connections $\{\rho^{PSL}_{\alpha} \}_{\alpha =0}^3$ whose holonomy matrices are $\rho_\alpha^{PSL}(a)= [\rho^{SL}(a)]$ with
\begin{align}
\begin{split}
&\rho^{SL}_{\alpha } (x_1) = \left(
\begin{array}{cc}
-i & 0 \\
0 & i \\
\end{array}
\right)\;,  \quad \rho^{SL}_{\alpha } (h) =  \left(
\begin{array}{cc}
-1 & 0 \\
0 & -1 \\
\end{array}
\right)\;,
\\
&\rho^{SL}_{\alpha =0} (x_2)  = \left(
\begin{array}{cc}
\frac{1}{2} \left(1-i \sqrt{\sqrt{2}+2}\right) & \frac{1}{4} \left(\sqrt{2}-1\right) \\
1 & \frac{1}{2} \left(1+i \sqrt{\sqrt{2}+2}\right) \\
\end{array}
\right)\;,
\\
&\rho^{SL}_{\alpha =0} (x_3)  = \left(
\begin{array}{cc}
\frac{1}{2} \left(-\sqrt{\sqrt{2}+2}+i\right) & \frac{1}{4} i \left(\sqrt{2}-1\right) \\
-i & \frac{1}{2} \left(-\sqrt{\sqrt{2}+2}-i\right) \\
\end{array}
\right)\;,
\\
&\rho^{SL}_{\alpha =1} (x_2)  =\left(
\begin{array}{cc}
\frac{1}{2}-\frac{i}{\sqrt{2}} & -\frac{1}{4} \\
1 & \frac{1}{2}+\frac{i}{\sqrt{2}} \\
\end{array}
\right) \;,
\\
&\rho^{SL}_{\alpha =1} (x_3)  =\left(
\begin{array}{cc}
-\frac{1}{\sqrt{2}}+\frac{i}{2} & -\frac{i}{4} \\
-i & -\frac{1}{\sqrt{2}}-\frac{i}{2} \\
\end{array}
\right)\;,
\\
&\rho^{SL}_{\alpha =2} (x_2)  = \left(
\begin{array}{cc}
\frac{1}{2} \left(1-i \sqrt{2-\sqrt{2}}\right) & \frac{1}{4} \left(-\sqrt{2}-1\right) \\
1 & \frac{1}{2} \left(1+i \sqrt{2-\sqrt{2}}\right) \\
\end{array}
\right)\;,
\\
&\rho^{SL}_{\alpha =2} (x_3)  =\left(
\begin{array}{cc}
\frac{1}{2} \left(-\sqrt{2-\sqrt{2}}+i\right) & -\frac{1}{4} i \left(\sqrt{2}+1\right) \\
-i & \frac{1}{2} \left(-\sqrt{2-\sqrt{2}}-i\right) \\
\end{array}
\right) \;,
\\
&\rho^{SL}_{\alpha =3} (x_2)  =\left(
\begin{array}{cc}
\frac{1}{2} & -\frac{3}{4} \\
1 & \frac{1}{2} \\
\end{array}
\right) \;, \quad \rho^{SL}_{\alpha =3} (x_3)  =\left(
\begin{array}{cc}
\frac{i}{2} & -\frac{3i}{4}  \\
-i & -\frac{i}{2} \\
\end{array}
\right) \;.
\end{split}
\end{align}
There are two $\mathbb{Z}_2$ flat connections, $H^1 (M, \mathbb{Z}_2) = \{1, \eta\}$, where
\begin{align}
\eta (x_1) = \eta (x_3) =-1\;, \quad \eta (x_2)  = \eta(h)=0\;.
\end{align}
Since the $\tr (\rho^{PSL}_{\alpha=0,1,2} (x_3)) \neq 0$, the first three $PSL(2,\mathbb{C})$ flat connections, $\rho_{\alpha=0,1,2}$, are not invariant under the tensoring with $\eta$. On the other hand,  the last flat connection $\rho_{\alpha=3}$ is invariant under the tensoring. Among the 4 $PSL(2,\mathbb{C})$ flat connections, $\rho^{PSL}_{\alpha=0,2}$ can be uplifted to $SL(2,\mathbb{C})$ flat connections $\rho_{\alpha=0,2}$ whose holonomy matrices are  $\rho^{SL}_{\alpha=0,2}$ given above. Taking into account of tensoring with the $\mathbb{Z}_2$ flat connections, there are 4 irreducible $SL(2,\mathbb{C})$ flat connections, $\rho_{\alpha=0,2} $ and $\eta\otimes \rho_{\alpha=0,2}$

\section{Modular structure of TFT$[M]$ in Table \ref{Table : full list}  and \ref{Table : full list-2}} \label{app : modular structure of full list}

Here we give modular structures of $\textrm{TFT}[M]$ in Table \ref{Table : full list} and Table \ref{Table : full list-2}. For simplicity of  presentation, we only give explicit modular data for a specific choice of true vacuum if there are several possible choices. See the related discussion around the equation \eqref{True vacuum choice}. The modular data of $\textrm{TFT}[M]$ with other choices can be easily obtained from the Lorentz symmetry fractionalization procedure or the parity operation. 

\subsection{Trivial $H_1(M, \mathbb{Z}_2)$ }
\paragraph{$S^2 \left(2,3,3\right)$}\label{para:trivial} There is only one irreducible $SL(2,\mathbb{C})$ flat connection $\rho$ with
\begin{align}
\begin{split}
&\textrm{Tor}[\rho]= \tfrac{1}2 \;, \quad CS[\rho] = -\tfrac{7}{24}    \;. \nonumber
\end{split}
\end{align}
Note that the central charge  $c_{2d} = \pm 24 CS[\rho] = 0\; (\textrm{mod } \frac{1}2)$ and $\textrm{GSD}_{g}=(2 \textrm{Tor}[\rho])^{g-1}=1$ for all $g\geq 0$.  The modular structure is identical to that of the trivial bosonic topological phase $1^B_1$.

\paragraph{$S^2 (2,3,5)$} \label{para:Fibonacci}
There are two irreducible $SL(2,\mathbb{C})$ flat connections $\rho_{\alpha=0,1}$ with
\begin{align}
\begin{split}
&\{\textrm{Tor}[\rho_{\alpha}]\}_{\alpha=0}^1 = \left\{\tfrac{\left(5+\sqrt{5}\right)}{4} ,\tfrac{\left(5-\sqrt{5}\right)}{4} \right\} \;,
\\
&\{CS[\rho_{\alpha}]\}_{\alpha=0}^1  = \left\{\tfrac{41}{120},\tfrac{89}{120}  \right\} \;. \nonumber
\end{split}
\end{align}
According to \eqref{unitarity}, the corresponding topological phase is unitary. The flat-connection-to-loop operator map is
\begin{align}
\rho_{\alpha} \rightarrow \left(a= x_3, R = \textrm{Sym}^\alpha\Box \right) \;. \nonumber
\end{align}
The $S$-matrix and the topological spins $h_\alpha= \pm (CS[\rho_{\alpha}]-CS[\rho_{\alpha=0}])$  are
\begin{align}
\begin{split}
&S = \left(
\begin{array}{cc}
\sqrt{\frac{1}{10} \left(5-\sqrt{5}\right)} & \sqrt{\frac{1}{10} \left(5+\sqrt{5}\right)} \\
\sqrt{\frac{1}{10} \left(\sqrt{5}+5\right)} & -\sqrt{\frac{1}{10} \left(5-\sqrt{5}\right)} \\
\end{array}
\right) \;,
\\
&\{h_\alpha\}_{\alpha=0}^1 =\pm  \left\{ 0,   \tfrac{2}5 \right\}\;.\nonumber
\end{split}
\end{align}
The ground state degeneracy is
\begin{align}
\{\textrm{GSD}_g\}_{g=0,1,\ldots} = \left\{ 1,2,5,15 ,50, 175,\ldots \right\} \;. \nonumber
\end{align}
The modular structure is identical to that of $2^B_{\pm 14/5}$.

\paragraph{$S^2\left(2,3,\frac{5}{2} \right)$}
There are two irreducible $SL(2,\mathbb{C})$ flat connections $\rho_{\alpha=0,1}$ with
\begin{align}
\begin{split}
&\{\textrm{Tor}[\rho_{\alpha}]\}_{\alpha=0}^1 = \left\{\tfrac{\left(5-\sqrt{5}\right)}{4} , \tfrac{\left(\sqrt{5}+5\right)}{4}  \right\} \;,
\\
&\{CS[\rho_{\alpha}]\}_{\alpha=0}^1  = \left\{\tfrac{83}{120} ,\tfrac{107}{120}  \right\} \;. \nonumber
\end{split}
\end{align}
According to \eqref{unitarity}, the corresponding topological phase is non-unitary. The flat-connection-to-loop operator map is
\begin{align}
\rho_{\alpha} \rightarrow \left(a= x_3^2, R = \textrm{Sym}^\alpha\Box \right)\;. \nonumber
\end{align}
The $S$-matrix and the topological spins $h_\alpha= \pm (CS[\rho_{\alpha}]-CS[\rho_{\alpha=0}])$  are
\begin{align}
\begin{split}
&S = \left(
\begin{array}{cc}
\sqrt{\frac{1}{10} \left(5+\sqrt{5}\right)} & -\sqrt{\frac{1}{10} \left(\sqrt{5}-5\right)} \\
-\sqrt{\frac{1}{10} \left(\sqrt{5}-5\right)} & -\sqrt{\frac{1}{10} \left(5+\sqrt{5}\right)} \\
\end{array}
\right) \;,
\\
&\{h_\alpha\}_{\alpha=0}^1  = \pm \left\{0, \tfrac{1}5 \right\} \nonumber
\end{split}
\end{align}
The ground state degeneracy is
\begin{align}
\{\textrm{GSD}_g\}_{g=0,1,\ldots} = \left\{ 1,2,5,15 ,50, 175,\ldots \right\} \;. \nonumber
\end{align}
The modular structure is identical to that of $\textrm{Gal}_{\sharp =2}[(A_1, 3)_{1/2}]  = (\textrm{Lee-Yang model})$. 
\paragraph{$S^2 (3,3,3)$} \label{para:Semion}
There are two irreducible flat connections $\rho_{\alpha=0,1}$ with
\begin{align}
\begin{split}
&\{\textrm{Tor}[\rho_{\alpha}]\}_{\alpha=0}^1 = \left\{1,1 \right\} \;,
\\
&\{CS[\rho_{\alpha}]\}_{\alpha=0}^1  = \left\{0 , \tfrac{3}4 \right\} \;. \nonumber
\end{split}
\end{align}
 According to \eqref{unitarity}, the corresponding topological phase is unitary. The flat-connection-to-loop operator map is
\begin{align}
\rho_{\alpha} \rightarrow \left(a= x_3, R = \textrm{Sym}^\alpha\Box \right)\;.  \nonumber
\end{align}
The $S$-matrix and the topological spin $\{h_\alpha\}_{\alpha=0}^1$  are
\begin{align}
\begin{split}
&S = \frac{1}{\sqrt{2}}\left(
\begin{array}{cc}
1& 1 \\
1 & -1 \\
\end{array}
\right) \;,
\\
&\{h_\alpha\}_{\alpha=0}^1 = \pm \left\{0,   \tfrac{3}4\right\}\;. \nonumber
\end{split}
\end{align}
The ground state degeneracy is
\begin{align}
\textrm{GSD}_g =2^g \;. \nonumber
\end{align}
The modular structure is identical to that of $2^B_{\pm 1}$.

\paragraph{$S^2 (2,3,7)$} \label{para:A1-5} There are three irreducible flat $SL(2,\mathbb{C})$ connections $\rho_{\alpha=0,1,2}$ with
\begin{align}
\begin{split}
&\{\textrm{Tor}[\rho_{\alpha}]\}_{\alpha=0}^2 =\tfrac{7}{8} \left\{ \csc ^2\left(\tfrac{\pi }{7}\right), \csc ^2\left(\tfrac{2\pi }{7}\right), \csc ^2\left(\tfrac{3\pi }{7}\right)\right\} \;,
\\
&\{CS[\rho_{\alpha}]\}_{\alpha=0}^2  =  \left\{\tfrac{127}{168} ,\tfrac{151}{168},\tfrac{79}{168}\right\} \;. \nonumber
\end{split}
\end{align}
According to \eqref{unitarity}, the corresponding topological phase is unitary. The flat-connection-to-loop operator map is
\begin{align}
\rho_{\alpha} \rightarrow  \left(a= x_3, R = \textrm{Sym}^\alpha\Box \right)  \nonumber
\end{align}
The $S$-matrix and the topological spin $\{h_\alpha\}_{\alpha=0}^2$  are
\begin{align}
\begin{split}
&S = \left(
\begin{array}{ccc}
\frac{2 \sin \left(\frac{\pi }{7}\right)}{\sqrt{7}} & \frac{2 \cos \left(\frac{3 \pi }{14}\right)}{\sqrt{7}} & \frac{2 \cos \left(\frac{\pi }{14}\right)}{\sqrt{7}} \\
\frac{2 \cos \left(\frac{3 \pi }{14}\right)}{\sqrt{7}} & -\frac{2 \cos \left(\frac{\pi }{14}\right)}{\sqrt{7}} & \frac{2 \sin \left(\frac{\pi }{7}\right)}{\sqrt{7}} \\
\frac{2 \cos \left(\frac{\pi }{14}\right)}{\sqrt{7}} & \frac{2 \sin \left(\frac{\pi }{7}\right)}{\sqrt{7}} & -\frac{2 \cos \left(\frac{3 \pi }{14}\right)}{\sqrt{7}} \\
\end{array}
\right)\;,
\\
&\{h_\alpha\}_{\alpha=0}^2  =\pm  \{0, \tfrac{1}7 ,\tfrac{5}7 \}\;. \nonumber
\end{split}
\end{align}
The ground state degeneracy is
\begin{align}
\left\{\textrm{GSD}_g   \right\}_{g=0,1,\ldots}=\left\{1,3,14,98,833,7546,\ldots   \right\}\;. \nonumber
\end{align}
The modular structure is identical to that of $3^B_{\pm 8/7}$.

\paragraph{$S^2 \left(2, 3, \frac{7}{2} \right)$} There are three irreducible $SL(2,\mathbb{C})$ flat  connections $\rho_{\alpha=0,1,2}$ with
\begin{align}
\begin{split}
&\{\textrm{Tor}[\rho_{\alpha}]\}_{\alpha=0}^2 = \tfrac{7}{8} \left\{ \sec ^2\left(\tfrac{\pi }{14}\right), \csc ^2\left(\tfrac{\pi }{7}\right), \sec ^2\left(\tfrac{3 \pi }{14}\right)\right\}\;,
\\
&\{CS[\rho_{\alpha}]\}_{\alpha=0}^2  =  \left\{\tfrac{25}{168},\tfrac{121}{168} ,\tfrac{1}{168} \right\}\;. \nonumber
\end{split}
\end{align}
 According to \eqref{unitarity}, the corresponding topological phase is non-unitary. The flat-connection-to-loop operator map is
\begin{align}
\rho_{\alpha} \rightarrow \left(a= x_3^3, R = \textrm{Sym}^\alpha\Box \right)   \nonumber
\end{align}
The $S$-matrix and the topological spins $h_\alpha$  are
\begin{align}
\begin{split}
&S = \left(
\begin{array}{ccc}
\frac{2 \cos \left(\frac{\pi }{14}\right)}{\sqrt{7}} & \frac{2 \sin \left(\frac{\pi }{7}\right)}{\sqrt{7}} & -\frac{2 \cos \left(\frac{3 \pi }{14}\right)}{\sqrt{7}} \\
\frac{2 \sin \left(\frac{\pi }{7}\right)}{\sqrt{7}} & \frac{2 \cos \left(\frac{3 \pi }{14}\right)}{\sqrt{7}} & \frac{2 \cos \left(\frac{\pi }{14}\right)}{\sqrt{7}} \\
-\frac{2 \cos \left(\frac{3 \pi }{14}\right)}{\sqrt{7}} & \frac{2 \cos \left(\frac{\pi }{14}\right)}{\sqrt{7}} & -\frac{2 \sin \left(\frac{\pi }{7}\right)}{\sqrt{7}} \\
\end{array}
\right)\;,
\\
&\{h_\alpha\}_{\alpha=0}^2  =\pm  \{0, \tfrac{4}7 ,\tfrac{6}7 \}\;. \nonumber
\end{split}
\end{align}
The ground state degeneracy is
\begin{align}
\left\{\textrm{GSD}_g   \right\}_{g=0,1,\ldots}=\left\{1,3,14,98,833,7546,\ldots   \right\}\;. \nonumber
\end{align}
The modular structure is identical to that of $\textrm{Gal}_{\sharp=2}[(A_1,5)_{1/2}]$.

\paragraph{$S^2 \left(2,3, \frac{7}{3} \right)$} There are three $SL(2,\mathbb{C})$ irreducible flat connections $\rho_{\alpha=0,1,2}$ with
\begin{align}
\begin{split}
&\{\textrm{Tor}[\rho_{\alpha}]\}_{\alpha=0}^2 =\tfrac{7}{8}  \left\{ \sec ^2\left(\tfrac{3 \pi }{14}\right), \sec ^2\left(\tfrac{\pi }{14}\right), \csc ^2\left(\tfrac{\pi }{7}\right)\right\}\;,
\\
&\{CS[\rho_{\alpha}]\}_{\alpha=0}^2  =  \left\{\tfrac{19}{168},\tfrac{139}{168},\tfrac{115}{168}\right\}  \;. \nonumber
\end{split}
\end{align}
According to \eqref{unitarity}, the corresponding topological phase is non-unitary. The flat-connection-to-loop operator map is
\begin{align}
\rho_{\alpha} \rightarrow \left(a= x_3^2, R = \textrm{Sym}^\alpha\Box \right)\;. \nonumber
\end{align}
The $S$-matrix and the topological spins $h_\alpha$  are
\begin{align}
\begin{split}
&S = \left(
\begin{array}{ccc}
\frac{2 \cos \left(\frac{3 \pi }{14}\right)}{\sqrt{7}} & -\frac{2 \cos \left(\frac{\pi }{14}\right)}{\sqrt{7}} & \frac{2 \sin \left(\frac{\pi }{7}\right)}{\sqrt{7}} \\
-\frac{2 \cos \left(\frac{\pi }{14}\right)}{\sqrt{7}} & -\frac{2 \sin \left(\frac{\pi }{7}\right)}{\sqrt{7}} & \frac{2 \cos \left(\frac{3 \pi }{14}\right)}{\sqrt{7}} \\
\frac{2 \sin \left(\frac{\pi }{7}\right)}{\sqrt{7}} & \frac{2 \cos \left(\frac{3 \pi }{14}\right)}{\sqrt{7}} & \frac{2 \cos \left(\frac{\pi }{14}\right)}{\sqrt{7}} \\
\end{array}
\right)\;,
\\
&\{h_\alpha\}_{\alpha=0}^2  =\pm  \{0, \tfrac{5}7 ,\tfrac{4}7 \}\;. \nonumber
\end{split}
\end{align}
The ground state degeneracy is
\begin{align}
\left\{\textrm{GSD}_g   \right\}_{g=0,1,\ldots}=\left\{1,3,14,98,833,7546,\ldots   \right\}\;. \nonumber
\end{align}
The modular structure is identical to that of $\textrm{Gal}_{\sharp=3}[(A_1,5)_{1/2}]$.

\paragraph{$S^2 \left(3, \frac{3}{2}, 4 \right)$} \label{para:Ising}
There are three irreducible flat connections $\rho_{\alpha=0,1,2}$ with
\begin{align}
\begin{split}
&\{\textrm{Tor}[\rho_{\alpha}]\}_{\alpha=0}^3 = \left\{2,1,2\right\},
\\
&\{CS[\rho_{\alpha}]\}_{\alpha=0}^3  =  \left\{\tfrac{11}{16},\tfrac{3}{4}  ,\tfrac{3}{16} \right\} \;. \nonumber
\end{split}
\end{align}
According to \eqref{unitarity}, the corresponding topological phase is unitary. The flat-connection-to-loop operator map is
\begin{align}
\rho_{\alpha} \rightarrow  \left(a= x_3, R = \textrm{Sym}^\alpha\Box \right)  \;.\nonumber
\end{align}

The $S$-matrix and the topological spin $\{h_\alpha\}_{\alpha=0}^2$  are
\begin{align}
\begin{split}
&S  = \frac{2}3 \left(
\begin{array}{ccc}
\frac{1}{2} & \frac{1}{\sqrt{2}} & \frac{1}{2} \\
\frac{1}{\sqrt{2}} & 0 & -\frac{1}{\sqrt{2}} \\
\frac{1}{2} & -\frac{1}{\sqrt{2}} & \frac{1}{2} \\
\end{array}
\right)\;,
\\
&\{h_\alpha\}_{\alpha=0}^2= \pm \left\{0,\tfrac{1}{16},\tfrac{1}{2}\right\}  \;. \nonumber
\end{split}
\end{align}
The ground state degeneracy is
\begin{align}
\left\{\textrm{GSD}_g   \right\}_{g=0,1,\ldots}=\left\{1,3,10,36,136,528,\ldots   \right\}\;. \nonumber
\end{align}
The modular structure is identical to that of $3^B_{\pm 1/2}$.

\paragraph{$S^2 \left(3,3,4\right)$ } \label{para:A1-2}
There are three irreducible flat connections $\rho_{\alpha=0,1,2}$ with
\begin{align}
\begin{split}
&\{\textrm{Tor}[\rho_{\alpha}]\}_{\alpha=0}^3 = \left\{2,1,2\right\},
\\
&\{CS[\rho_{\alpha}]\}_{\alpha=0}^3  =  \left\{\tfrac{37}{48},\tfrac{1}{12}  ,\tfrac{13}{48} \right\} \;. \nonumber
\end{split}
\end{align}
 According to \eqref{unitarity}, the corresponding topological phase is unitary. The flat-connection-to-loop operator map is
\begin{align}
\rho_{\alpha} \rightarrow \left(a= x_3, R = \textrm{Sym}^\alpha\Box \right)  \nonumber
\end{align}
The $S$-matrix and the topological spins $\{h_\alpha\}_{\alpha=0}^2$  are
\begin{align}
\begin{split}
&S  = \frac{2}3 \left(
\begin{array}{ccc}
\frac{1}{2} & \frac{1}{\sqrt{2}} & \frac{1}{2} \\
\frac{1}{\sqrt{2}} & 0 & -\frac{1}{\sqrt{2}} \\
\frac{1}{2} & -\frac{1}{\sqrt{2}} & \frac{1}{2} \\
\end{array}
\right)\;,
\\
&\{h_\alpha\}_{\alpha=0}^2= \pm \left\{0,\tfrac{5}{16},\tfrac{1}{2}\right\}  \;. \nonumber
\end{split}
\end{align}
The ground state degeneracy is
\begin{align}
\left\{\textrm{GSD}_g   \right\}_{g=0,1,\ldots}=\left\{1,3,10,36,136,528,\ldots   \right\}\;. \nonumber
\end{align}
The modular structure is identical to that of $3^B_{\pm 5/2}$.

\paragraph{$S^2 (2,3,9)$ } \label{para:239} There are four irreducible flat connections $\rho_{\alpha=0,1,2,3}$ with
\begin{align}
\begin{split}
&\{\textrm{Tor}[\rho_{\alpha}]\}_{\alpha=0}^3 = \tfrac{9}{8} \left\{ \csc ^2\left(\tfrac{\pi }{9}\right), \csc ^2\left(\tfrac{2 \pi }{9}\right),\tfrac{4}{3},\sec ^2\left(\tfrac{\pi }{18}\right)\right\},
\\
&\{CS[\rho_{\alpha}]\}_{\alpha=0}^3  = \left\{\tfrac{55}{72} ,\tfrac{31}{72},\tfrac{13}{24},\tfrac{7}{72}\right\} \;. \nonumber
\end{split}
\end{align}
 According to \eqref{unitarity}, the corresponding topological phase is unitary. The flat-connection-to-loop operator map is
\begin{align}
\rho_{\alpha} \rightarrow \left(a= x_3, R = \textrm{Sym}^\alpha\Box \right)  \nonumber
\end{align}
The $S$-matrix and the topological spins $\{h_\alpha\}_{\alpha=0}^3$  are
\begin{align}
&S = \left(
\begin{array}{cccc}
\frac{2}{3} \sin \left(\frac{\pi }{9}\right) & \frac{2}{3} \sin \left(\frac{2 \pi }{9}\right) & \frac{1}{\sqrt{3}} & \frac{2}{3} \cos \left(\frac{\pi }{18}\right) \\
\frac{2}{3} \sin \left(\frac{2 \pi }{9}\right) & -\frac{2}{3} \cos \left(\frac{\pi }{18}\right) & \frac{1}{\sqrt{3}} & -\frac{2}{3}  \sin \left(\frac{\pi }{9}\right) \\
\frac{1}{\sqrt{3}} & \frac{1}{\sqrt{3}} & 0 & -\frac{1}{\sqrt{3}} \\
\frac{2}{3} \cos \left(\frac{\pi }{18}\right) & -\frac{2}{3}  \sin \left(\frac{\pi }{9}\right) & -\frac{1}{\sqrt{3}} & \frac{2}{3} \sin \left(\frac{2 \pi }{9}\right) \\
\end{array}
\right),\nonumber
\\
&\{h_\alpha\}_{\alpha=0}^3 = \pm \left\{0,\tfrac{2}{3},\tfrac{7}{9},\tfrac{1}{3}\right\}\;. \nonumber
\end{align}
The ground state degeneracy is
\begin{align}
\left\{\textrm{GSD}_g   \right\}_{g=0,1,\ldots}=\left\{1,4,30,414,7317,137862,\ldots   \right\}\;. \nonumber
\end{align}
The modular structure is identical to that of $4^B_{\pm 10/3}$.

\paragraph{$S^2 (2,3,\frac{9}{2}) $} There are four irreducible flat connections $\rho_{\alpha=0,1,2,3}$ with
\begin{align}
\begin{split}
&\{\textrm{Tor}[\rho_{\alpha}]\}_{\alpha=0}^3 = \tfrac{9}{8}  \left\{\sec^2\left(\tfrac{\pi }{18}\right), \csc ^2\left(\tfrac{\pi }{9}\right),\tfrac{4}{3}, \csc ^2\left(\tfrac{2 \pi }{9}\right)\right\},
\\
&\{CS[\rho_{\alpha}]\}_{\alpha=0}^3  = \left\{\tfrac{29}{72},\tfrac{53}{72},\tfrac{7}{24},\tfrac{5}{72}\right\}  \;. \nonumber
\end{split}
\end{align}
  According to \eqref{unitarity}, the corresponding topological phase is non-unitary. The flat-connection-to-loop operator map is
\begin{align}
\rho_{\alpha} \rightarrow  \left(a= x_3^5, R = \textrm{Sym}^\alpha\Box \right) \;. \nonumber
\end{align}
The $S$-matrix and the topological spins $\{h_\alpha\}_{\alpha=0}^3$  are
\begin{align}
&S  = \frac{2}3 \left(
\begin{array}{cccc}
\cos \left(\frac{\pi }{18}\right) & -\sin \left(\frac{\pi }{9}\right) & -\frac{\sqrt{3}}{2} & \sin \left(\frac{2 \pi }{9}\right) \\
-\sin \left(\frac{\pi }{9}\right) & -\sin \left(\frac{2 \pi }{9}\right) & -\frac{\sqrt{3}}{2} & -\cos \left(\frac{\pi }{18}\right) \\
-\frac{\sqrt{3}}{2} & -\frac{\sqrt{3}}{2} & 0 & \frac{\sqrt{3}}{2} \\
\sin \left(\frac{2 \pi }{9}\right) & -\cos \left(\frac{\pi }{18}\right) & \frac{\sqrt{3}}{2} & -\sin \left(\frac{\pi }{9}\right) \\
\end{array}
\right) \;, \nonumber
\\
&\{h_\alpha\}_{\alpha=0}^3  = \pm  \left\{0,\tfrac{1}{3},\tfrac{8}{9},\tfrac{2}{3}\right\}\;.  \nonumber
\end{align}
The ground state degeneracy is
\begin{align}
\left\{\textrm{GSD}_g   \right\}_{g=0,1,\ldots}=\left\{1,4,30,414,7317,137862,\ldots   \right\}\;. \nonumber
\end{align}
The modular structure is identical to that of $\textrm{Gal}_{\sharp =2}[(A_1, 7)_{1/2}]$.

\paragraph{$S^2 (2,3,\frac{9}{4})$ } 
There are four irreducible  $SL(2,\mathbb{C})$ flat connections $\rho_{\alpha=0,1,2,3}$ with
\begin{align}
\begin{split}
&\{\textrm{Tor}[\rho_{\alpha}]\}_{\alpha=0}^3 = \tfrac{9}{8} \left\{\csc ^2\left(\tfrac{2 \pi }{9}\right),\sec ^2\left(\tfrac{\pi }{18}\right),\tfrac{4}{3},\csc ^2\left(\tfrac{\pi }{9}\right)\right\} ,
\\
&\{CS[\rho_{\alpha}]\}_{\alpha=0}^3  =  \left\{\tfrac{25}{72},\tfrac{1}{72},\tfrac{19}{24},\tfrac{49}{72}\right\}   \;. \nonumber
\end{split}
\end{align}
According to \eqref{unitarity}, the corresponding topological phase is non-unitary. The flat-connection-to-loop operator map is
\begin{align}
\rho_{\alpha} \rightarrow \left(a= x_3^2, R = \textrm{Sym}^\alpha\Box \right)\;. \nonumber
\end{align}
The $S$-matrix and the topological spins $\{h_\alpha\}_{\alpha=0}^3$  are
\begin{align}
&S  = \frac{2}3 \left(
\begin{array}{cccc}
\sin \left(\frac{2 \pi }{9}\right) & -\cos \left(\frac{\pi }{18}\right) & \frac{\sqrt{3}}{2} & -\sin \left(\frac{\pi }{9}\right) \\
-\cos \left(\frac{\pi }{18}\right) & \sin \left(\frac{\pi }{9}\right) & \frac{\sqrt{3}}{2} & -\sin \left(\frac{2 \pi }{9}\right) \\
\frac{\sqrt{3}}{2} & \frac{\sqrt{3}}{2} & 0 & -\frac{\sqrt{3}}{2} \\
-\sin \left(\frac{\pi }{9}\right) & -\sin \left(\frac{2 \pi }{9}\right) & -\frac{\sqrt{3}}{2} & -\cos \left(\frac{\pi }{18}\right) \\
\end{array}
\right)\;, \nonumber
\\
&\{h_\alpha\}_{\alpha=0}^3  =  \pm \left\{0,\tfrac{2}{3},\tfrac{4}{9},\tfrac{1}{3}\right\} \;. \nonumber
\end{align}
The ground state degeneracy is
\begin{align}
\left\{\textrm{GSD}_g   \right\}_{g=0,1,\ldots}=\left\{1,4,30,414,7317,137862,\ldots   \right\}\;. \nonumber
\end{align}
The modular structure is identical to that of $\textrm{Gal}_{\sharp =4}[(A_1, 7)_{1/2}]$.

\paragraph{$S^2 \left(3,3,5\right)$ }  \label{para:335}
There are four irreducible flat connections $\rho_{\alpha=0,1,2,3}$ with
\begin{align}
\begin{split}
&\{\textrm{Tor}[\rho_{\alpha}]\}_{\alpha=0}^3 =\ \left\{\tfrac{\sqrt{5}+5}{2},\tfrac{\sqrt{5}-5}{2},\tfrac{\sqrt{5}-5}{2},\tfrac{\sqrt{5}+5}{2} \right\}\; ,
\\
&\{CS[\rho_{\alpha}]\}_{\alpha=0}^3  =  \left\{\tfrac{47}{60} ,\tfrac{2}{15},\tfrac{23}{60},\tfrac{8}{15}\right\}   \;. \nonumber
\end{split}
\end{align}
According to \eqref{unitarity}, the corresponding topological phase is unitary. The flat-connection-to-loop operator map is
\begin{align}
\rho_{\alpha} \rightarrow \left(a= x_3, R = \textrm{Sym}^\alpha\Box \right)\;. \nonumber
\end{align}
The $S$-matrix and the topological spins $\{h_\alpha\}_{\alpha=0}^2$  are
\begin{align}
\begin{split}
&S  = \frac{1}2\left(
\begin{array}{cccc}
\sqrt{1-\tfrac{1}{\sqrt{5}}} & \sqrt{1+\tfrac{1}{\sqrt{5}}} & \sqrt{1+\frac{1}{\sqrt{5}}} & \sqrt{1-\frac{1}{\sqrt{5}}} \\
\sqrt{1+\frac{1}{\sqrt{5}}} & \sqrt{1-\frac{1}{\sqrt{5}}} & -\sqrt{1-\frac{1}{\sqrt{5}}} & -\sqrt{1+\frac{1}{\sqrt{5}}} \\
\sqrt{1+\frac{1}{\sqrt{5}}} & -\sqrt{1-\frac{1}{\sqrt{5}}} & -\sqrt{1-\frac{1}{\sqrt{5}}} & \sqrt{1+\frac{1}{\sqrt{5}}} \\
\sqrt{1-\frac{1}{\sqrt{5}}} & -\sqrt{1+\frac{1}{\sqrt{5}}} & \sqrt{1+\frac{1}{\sqrt{5}}} & -\sqrt{1-\frac{1}{\sqrt{5}}} \\
\end{array}
\right)\;,
\\
&\{h_\alpha\}_{\alpha=0}^2 =\pm  \left\{0,\tfrac{7}{20},\tfrac{3}{5},\tfrac{3}{4}\right\}  \;. \nonumber
\end{split}
\end{align}
The ground state degeneracy is
\begin{align}
\left\{\textrm{GSD}_g   \right\}_{g=0,1,\ldots}=\left\{1,4,20,120,800,5600,\ldots   \right\}\;. \nonumber
\end{align}
The modular structure is identical to that of $4^B_{\pm 19/5}$.

\paragraph{$S^2 \left(\frac{3}2, 3, \frac{5}2 \right)$}

There are four irreducible flat connections $\rho_{\alpha=0,1,2,3}$ with
\begin{align}
\begin{split}
&\{\textrm{Tor}[\rho_{\alpha}]\}_{\alpha=0}^3 =\ \left\{\tfrac{\sqrt{5}-5}{2},\tfrac{\sqrt{5}+5}{2},\tfrac{\sqrt{5}+5}{2},\tfrac{\sqrt{5}-5}{2} \right\}\; ,
\\
&\{CS[\rho_{\alpha}]\}_{\alpha=0}^3  =\left\{\tfrac{3}{5},\tfrac{2}{5},\tfrac{17}{20},\tfrac{13}{20}\right\}  \;. \nonumber
\end{split}
\end{align}
According to \eqref{unitarity}, the corresponding topological phase is non-unitary. The flat-connection-to-loop operator map is
\begin{align}
&\rho_{\alpha=1} \rightarrow \left(a= x_3^2, R = \Box \right)\;, \quad \rho_{\alpha=2} \rightarrow \left(a= x_2, R = \Box \right)\;, \nonumber
\\
&\rho_{\alpha=3} \rightarrow \left(a= x_3^2, R = \Box \right) \otimes  \left(a= x_2, R = \Box \right)\;. \nonumber
\end{align}
The $S$-matrix and the topological spins $\{h_\alpha\}_{\alpha=0}^3$  are
\begin{align}
\begin{split}
&S  = \frac{1}2\left(
\begin{array}{cccc}
\sqrt{1+\frac{1}{\sqrt{5}}} & -\sqrt{1-\frac{1}{\sqrt{5}}} & \sqrt{1+\frac{1}{\sqrt{5}}} & -\sqrt{1-\frac{1}{\sqrt{5}}} \\
-\sqrt{1-\frac{1}{\sqrt{5}}} & -\sqrt{1+\frac{1}{\sqrt{5}}} & -\sqrt{1-\frac{1}{\sqrt{5}}} & -\sqrt{1+\frac{1}{\sqrt{5}}} \\
\sqrt{1+\frac{1}{\sqrt{5}}} & -\sqrt{1-\frac{1}{\sqrt{5}}} & -\sqrt{1+\frac{1}{\sqrt{5}}} & \sqrt{1-\frac{1}{\sqrt{5}}} \\
-\sqrt{1-\frac{1}{\sqrt{5}}} & -\sqrt{1+\frac{1}{\sqrt{5}}} & \sqrt{1-\frac{1}{\sqrt{5}}} & \sqrt{1+\frac{1}{\sqrt{5}}} \\
\end{array}
\right) \;,
\\
&\{h_\alpha\}_{\alpha=0}^3 =\pm \left\{0,\tfrac{4}{5},\tfrac{1}{4},\tfrac{1}{20}\right\}   \;. \nonumber
\end{split}
\end{align}
The ground state degeneracy is
\begin{align}
\left\{\textrm{GSD}_g   \right\}_{g=0,1,\ldots}=\left\{1,4,20,120,800,5600,\ldots   \right\}\;. \nonumber
\end{align}
The modular structure is identical to that of $\left(\textrm{Gal}_{\sharp 2}[(A_1,3)_{1/2}]\right)\otimes (2^B_{\pm 1})$. 
\paragraph{$S^2 \left(2,5,5\right)$}  \label{para:255}
There are four irreducible flat connections $\rho_{\alpha=0,1,2,3}$ with
\begin{align}
\begin{split}
&\{\textrm{Tor}[\rho_{\alpha}]\}_{\alpha=0}^3 =\left\{\tfrac{5(3+\sqrt{5})}{4} ,\tfrac{5}{2},\tfrac{5}{2},  \tfrac{5(3-\sqrt{5})}{4} \right\} ,
\\
&\{CS[\rho_{\alpha}]\}_{\alpha=0}^3  = \left\{\tfrac{31}{40} ,\tfrac{3}{8} ,\tfrac{3}{8}  ,\tfrac{39}{40} \right\}  \;. \nonumber
\end{split}
\end{align}
  According to \eqref{unitarity}, the corresponding topological phase is unitary. The flat-connection-to-loop operator map is
\begin{align}
&\rho_{\alpha=1} \rightarrow \left(a= x_2, R = \Box \right)\;, \quad \rho_{\alpha=2} \rightarrow \left(a= x_3, R = \Box \right)\;, \nonumber
\\
&\rho_{\alpha=3} \rightarrow \left(a= x_2, R = \Box \right) \otimes  \left(a= x_3, R = \Box \right)\;.\nonumber
\end{align}
The $S$-matrix and the topological spins $\{h_\alpha\}_{\alpha=0}^2$  are 
\begin{align}
&S= \frac{1}{\sqrt{5}}\left(
\begin{array}{cccc}
\tfrac{\left(\sqrt{5}-1\right) }{2} & 1 & 1 & \frac{ \left(\sqrt{5}+1\right)}{2} \\
1 & \frac{ \left(1-\sqrt{5}\right) }{2}& \frac{\left(\sqrt{5}+1\right)}{2}  & -1 \\
1 & \frac{\left(\sqrt{5}+1\right) }{2} & \frac{ \left(1-\sqrt{5}\right) }{2}& -1 \\
\frac{ \left(\sqrt{5}+1\right) }{2}& -1 & -1 & \frac{ \left(\sqrt{5}-1\right)}{2} \\
\end{array}
\right) \; \;, \nonumber
\\
&\{h_\alpha\}_{\alpha=0}^3 =  \pm  \left\{0,\tfrac{3}{5},\tfrac{3}{5},\tfrac{1}{5}\right\} \;.\nonumber
\end{align}
The ground state degeneracy is
\begin{align}
\left\{\textrm{GSD}_g   \right\}_{g=0,1,\ldots}=\left\{1,4,25,225,2500,30625,\ldots   \right\}\;. \nonumber
\end{align}
The modular structure is identical to that of $4^B_{\pm 12/5}$.

\paragraph{$S^2 \left(2,4, \frac{5}4\right)$ }  \label{para:245/4} 
There are four irreducible flat connections $\rho_{\alpha=0,1,2,3}$ with
\begin{align}
\begin{split}
&\{\textrm{Tor}[\rho_{\alpha}]\}_{\alpha=0}^3 =\left\{\tfrac{5(3+\sqrt{5})}{4} ,\tfrac{5}{2},\tfrac{5}{2},  \tfrac{5(3-\sqrt{5})}{4}  \right\} ,
\\
&\{CS[\rho_{\alpha}]\}_{\alpha=0}^3  = \left\{\tfrac{5}{8} ,\tfrac{9}{40}  ,\tfrac{1}{40}  ,\tfrac{5}{8}  \right\}  \;. \nonumber
\end{split}
\end{align}
 According to \eqref{unitarity}, the corresponding topological phase is unitary. The flat-connection-to-loop operator map is
\begin{align}
\begin{split}
&\rho_{\alpha=1} \rightarrow \left(a= x_2, R = \Box \right)\;, \quad \rho_{\alpha=2} \rightarrow \left(a= x_3, R = \Box \right)\;,
\\
&\rho_{\alpha=3} \rightarrow \left(a= x_2, R = \Box \right) \otimes  \left(a= x_3, R = \Box \right)\;. \nonumber \end{split}
\end{align}
The $S$-matrix and the topological spins $\{h_\alpha\}_{\alpha=0}^3$  are 
\begin{align}
\begin{split}
&S= \frac{1}{\sqrt{5}}\left(
\begin{array}{cccc}
\frac{\left(\sqrt{5}-1\right) }{2} & 1 & 1 & \frac{ \left(\sqrt{5}+1\right)}{2} \\
1 & \frac{ \left(1-\sqrt{5}\right) }{2}& \frac{\left(\sqrt{5}+1\right)}{2}  & -1 \\
1 & \frac{\left(\sqrt{5}+1\right) }{2} & \frac{ \left(1-\sqrt{5}\right) }{2}& -1 \\
\frac{ \left(\sqrt{5}+1\right) }{2}& -1 & -1 & \frac{ \left(\sqrt{5}-1\right)}{2} \\
\end{array}
\right) \; \;,
\\
&\{h_\alpha\}_{\alpha=0}^3 =  \pm  \left\{0,\tfrac{3}{5},\tfrac{2}{5}, 0 \right\} \;. \nonumber
\end{split}
\end{align}
The ground state degeneracy is
\begin{align}
\left\{\textrm{GSD}_g   \right\}_{g=0,1,\ldots}=\left\{1,4,25,225,2500,30625,\ldots   \right\}\;. \nonumber
\end{align}
The modular structure is identical to that of $4_0^{B,c}$. 

\paragraph{$S^2 \left(2, 5,\frac{5}2\right)$}  There are four irreducible flat connections $\rho_{\alpha=0,1,2,3}$ with
\begin{align}
\begin{split}
&\{\textrm{Tor}[\rho_{\alpha}]\}_{\alpha=0}^3 =\left\{\tfrac{5}{2}, \tfrac{5\sqrt{5}+3}{4} , \tfrac{5(3-\sqrt{5})}{4},\tfrac{5}{2}\right\} ,
\\
&\{CS[\rho_{\alpha}]\}_{\alpha=0}^3  =\left\{\tfrac{37}{40},\tfrac{29}{40},\tfrac{21}{40},\tfrac{13}{40}\right\}  \;. \nonumber
\end{split}
\end{align}
According to \eqref{unitarity}, the corresponding topological phase is non-unitary. The flat-connection-to-loop operator map is
\begin{align}
&\rho_{\alpha=1} \rightarrow \left(a= x_3^2, R = \Box \right)\;, \quad \rho_{\alpha=2} \rightarrow \left(a= x_2, R = \Box \right)\;, \nonumber
\\
&\rho_{\alpha=3} \rightarrow \left(a= x_3^2, R = \Box \right) \otimes  \left(a= x_2, R = \Box \right)\;.  \nonumber
\end{align}
The $S$-matrix and the topological spins $\{h_\alpha\}_{\alpha=0}^3$  are 
\begin{align}
&S=\frac{1}{\sqrt{5}} \left(
\begin{array}{cccc}
1 & \frac{\left(1-\sqrt{5}\right)}{2}  & -\frac{\left(1+\sqrt{5}\right) }{2} & 1 \\
\frac{\left(1-\sqrt{5}\right)}{2}  & -1 & 1 & \frac{\left(\sqrt{5}+1\right)}{2}  \\
-\frac{ \left(1+\sqrt{5}\right)}{2} & 1 & -1 & \frac{ \left(\sqrt{5}-1\right)}{2} \\
1 & \frac{ \left(\sqrt{5}+1\right)}{2} & \frac{\left(\sqrt{5}-1\right)}{2}  & 1 \\
\end{array}
\right) \; \;,\nonumber
\\
&\{h_\alpha\}_{\alpha=0}^3  =\textrm{diag} \left\{ 0,\tfrac{4}{5},\tfrac{3}{5},\tfrac{2}{5} \right\}   \;. \nonumber
\end{align}
The ground state degeneracy is
\begin{align}
\left\{\textrm{GSD}_g   \right\}_{g=0,1,\ldots}=\left\{1,4,25,225,2500,30625,\ldots   \right\}\;. \nonumber
\end{align}
The modular structure is identical to that of $ \left( \textrm{Gal}_{\sharp=2}[(A_1,3)_{1/2}] \right) \otimes (A_1,3)_{1/2} $.

\paragraph{$S^2 \left(2, \frac{5}2,\frac{5}2\right)$}  There are four irreducible flat connections $\rho_{\alpha=0,1,2,3}$ with
\begin{align}
\begin{split}
&\{\textrm{Tor}[\rho_{\alpha}]\}_{\alpha=0}^3 =\left\{ \tfrac{5(3-\sqrt{5})}{4} , \tfrac{5}{2}, \tfrac{5}{2},\tfrac{5(3+\sqrt{5})}{4}  \right\} ,
\\
&\{CS[\rho_{\alpha}]\}_{\alpha=0}^3  = \left\{\tfrac{3}{40},\tfrac{7}{8},\tfrac{7}{8},\tfrac{27}{40}\right\}\;. \nonumber
\end{split}
\end{align}
According to \eqref{unitarity}, the corresponding topological phase is non-unitary. The flat-connection-to-loop operator map is
\begin{align}
&\rho_{\alpha=1} \rightarrow \left(a= x_3^2, R = \Box \right)\;, \quad \rho_{\alpha=2} \rightarrow \left(a= x_2^2, R = \Box \right)\;, \nonumber
\\
&\rho_{\alpha=3} \rightarrow \left(a= x_3^2, R = \Box \right) \otimes  \left(a= x_2^2, R = \Box \right)\;.  \nonumber
\end{align}
The $S$-matrix and the topological spins $\{h_\alpha\}_{\alpha=0}^3$  are 
\begin{align}
&S=\frac{1}{\sqrt{5}} \left(
\begin{array}{cccc}
\frac{\left(\sqrt{5}+1\right)}{2}  & -1 & -1 & \frac{\left(\sqrt{5}-1\right)}{2}  \\
-1 & \frac{\left(\sqrt{5}-1\right) }{2} & \frac{\left(-\sqrt{5}-1\right) }{2} & 1 \\
-1 & \frac{\left(-\sqrt{5}-1\right) }{2} & \frac{\left(\sqrt{5}-1\right)}{2}  & 1 \\
\frac{\left(\sqrt{5}-1\right)}{2}  & 1 & 1 & \frac{\left(\sqrt{5}+1\right)}{2}  \\
\end{array}
\right) \; \;, \nonumber
\\
&\{h_\alpha\}_{\alpha=0}^3  =\pm \left\{ 0,\tfrac{4}{5},\tfrac{4}{5},\tfrac{3}{5} \right\}   \;. \nonumber
\end{align}

The ground state degeneracy is
\begin{align}
\left\{\textrm{GSD}_g   \right\}_{g=0,1,\ldots}=\left\{1,4,25,225,2500,30625,\ldots   \right\}\;. \nonumber
\end{align}
The modular structure is identical to that of $ \left( \textrm{Gal}_{\sharp=2}[(A_1,3)_{1/2}] \right) \otimes \left( \textrm{Gal}_{\sharp=2}[(A_1,3)_{1/2}] \right) $.

\subsection{Non-trivial $H_1(M, \mathbb{Z}_2)$}

\paragraph{$S^2(2,3,6)$} \label{App : Z3}
There are two irreducible $PSL(2,\mathbb{C})$ connections, $\{\rho^{PSL}_\alpha\}_{\alpha =0}^1$, with trivial 2nd Stiefel-Whitney class.  There are  two $\mathbb{Z}_2$ flat connections $\{1,\eta\} \in H^1 (M, \mathbb{Z}_2)$. Among the two $PSL(2,\mathbb{C})$ connections, the $ \rho^{PSL}_{\alpha =1} $ is invariant under tensoring with $\eta$, i.e. 
\begin{align}
\rho^{PSL}_{\alpha=1} \otimes \eta = \rho^{PSL}_{\alpha=1} \otimes 1\;\Rightarrow\; \textrm{Inv}(\rho_{\alpha=1}) = \mathbb{Z}_2\;. \nonumber
\end{align}
Accoring to Table~\ref{Table : MTC[M]-3}, there are $1+2*1=3$ anyons  in  TFT$[S^2(2,3,6)]$, which are
\begin{align}
\rho^{PSL}_{\alpha=0} \;  \textrm{ and }\; \rho^{PSL}_{\alpha=1} \; (\textrm{with multiplicity 2})\;. \nonumber
\end{align}
We choose the flat connection which is not invariant under tensoring with $\eta$ as the trivial anyon. 
Torsions for the flat connections are
\begin{align}
\textrm{Tor}[\rho_{\alpha=0}^{PSL}] =3\;, \quad \textrm{Tor}[\rho_{\alpha=1}^{PSL}] =\frac{3}4\;. \nonumber
\end{align}
From the dictionary in  Table~\ref{Table : MTC[M]-3}, we have
\begin{align}
(S_{0\alpha}^2 \textrm{ for all three anyons}) = \frac{1}3\;. \nonumber
\end{align} 
There are 3 $SL(2,\mathbb{C})$ flat connections which are
\begin{align}
\rho^{PSL}_{\alpha=0} \otimes 1\;, \; \rho^{PSL}_{\alpha=0} \otimes \eta\;, \; \rho^{PSL}_{\alpha=1}\otimes 1\;. \nonumber
\end{align}
Since the $\rho^{PSL}_{\alpha=0,1}$ have  trivial 2nd Stiefel-Whitney class, $\rho^{PSL}_{\alpha=0,1} \otimes \delta$ with $\delta \in H^1 (M, \mathbb{Z}_2)$ can be regarded as $SL(2,\mathbb{C})$ flat connections on $M$. Their Chern-Simons actions are
\begin{align}
\begin{split}
&CS[\rho_{\alpha=0}\otimes 1] = CS[\rho_{\alpha=0}\otimes \eta] =\frac{3}4\;,
\\
&CS[\rho_{\alpha=1}\otimes 1]  =\frac{5}{12}\;. \nonumber
\end{split}
\end{align}
From the 1st line, we can confirm that $\eta$ is bosonic and thus TFT$[S^2(2,3,6)]$ is a bosonic topological theory, see \eqref{bosonic/fermionc of Z2-flat} and \eqref{bosonic/fermionic}. From the computation of the Chern-Simons invariants, we have 
\begin{align}
h(\rho^{PSL}_{\alpha=0} ) =0\;, \quad h(\rho^{PSL}_{\alpha=1} )  = \pm \frac{1}3\;. \nonumber
\end{align}
Recall that there are two anyons associated to $\rho^{PSL}_{\alpha=1} $. The spectrum of $\{S_{0\alpha}^2\}$ and $\{h_\alpha\}$ are identical to the that of $3^B_{\pm 2}$.

This also implies that the mother theory $\widetilde{\textrm{TFT}}[M=S^2 (2,3,6)]$ is the $SU(2)_4$ theory. This theory is a UMTC with 5 anyons including an anyon generating the 1-form symmetry $H_1(M,\mathbb{Z}_2)$. By condensing this symmetry generating anyon, this theory reduces to the TFT$[S^2(2,3,6)]$.
\\
\paragraph{$S^2(\frac{4}{q_1},\frac{4}{q_2},\frac{3}{2})$} \label{App : Z4}

There are three irreducible $PSL(2,\mathbb{C})$ connections, $\{\rho^{PSL}_\alpha\}_{\alpha =0}^2$, with trivial 2nd Stiefel-Whitney class.  There are  two $\mathbb{Z}_2$ flat connections $\{1,\eta\} \in H^1 (M, \mathbb{Z}_2)$. Among the three $PSL(2,\mathbb{C})$ connections,  the $ \rho^{PSL}_{\alpha =2} $ is invariant under tensoring with $\eta$, i.e. 
\begin{align}
\rho^{PSL}_{\alpha=2} \otimes \eta = \rho^{PSL}_{\alpha=2} \otimes 1\;\Rightarrow\; \textrm{Inv}(\rho_{\alpha=2}) = \mathbb{Z}_2\;. \nonumber
\end{align}
According to Table~\ref{Table : MTC[M]-3}, there are $2+2*1=4$ anyons  in  TFT$[S^2(2,3,6)]$, which are
\begin{align}
\rho^{PSL}_{\alpha=0} \;,  \; \rho^{PSL}_{\alpha=1} \; \textrm{ and } \;  \rho^{PSL}_{\alpha=2} \; (\textrm{with multiplicity 2})\;. \nonumber
\end{align}
Torsions for the flat connections are
\begin{align}
\textrm{Tor}[\rho_{\alpha=0}^{PSL}]  = \textrm{Tor}[\rho_{\alpha=1}^{PSL}] =4\;, \;\; \textrm{Tor}[\rho_{\alpha=2}^{PSL}] =1\;. \nonumber
\end{align}
From the dictionary in  Table~\ref{Table : MTC[M]-3}, we have
\begin{align}
(S_{0\alpha}^2 \textrm{ for all four anyons}) = \frac{1}4\;. \nonumber
\end{align} 
There are 5 $SL(2,\mathbb{C})$ flat connections which are
\begin{align}
\rho^{PSL}_{\alpha=0,1} \otimes 1\;, \; \rho^{PSL}_{\alpha=0,1} \otimes \eta\;, \; \rho^{PSL}_{\alpha=2}\otimes 1\;. \nonumber
\end{align}
Since the $\rho^{PSL}_{\alpha=0,1}$ have  trivial 2nd Stiefel-Whitney class, $\rho^{PSL}_{\alpha=0,1} \otimes \delta$ with $\delta \in H^1 (M, \mathbb{Z}_2)$ can be regarded as $SL(2,\mathbb{C})$ flat connections on $M$. 
Their Chern-Simons actions are
\begin{align}
\begin{split}
&CS[\rho_{\alpha=0}\otimes 1] = CS[\rho_{\alpha=0}\otimes \eta] =\pm   \frac{\ell}{48}\;,
\\
&CS[\rho_{\alpha=1}\otimes 1] = CS[\rho_{\alpha=1}\otimes \eta] = \pm \left(\frac{\ell}{48}+\frac{1}2 \right)\;,
\\
&CS[\rho_{\alpha=2}\otimes 1] = \pm \left( \frac{\ell}{48} + \frac{(\frac{q_1+q_2}2 \textrm{ mod 4})}{8} \right) \;. \nonumber
\end{split}
\end{align}
Here $\ell \in 2 \mathbb{Z}$ depends on $\{q_i\}_{i=1}^3$. From the 1st and 2nd line, we can confirm that $\eta$ is bosonic and thus TFT$[S^2(\frac{4}{q_1},\frac{4}{q_2},\frac{3}{2})]$  is a bosonic topological theory, see \eqref{bosonic/fermionc of Z2-flat} and \eqref{bosonic/fermionic}. From the computation of the Chern-Simons invariants, we have 
\begin{align}
\begin{split}
&h(\rho^{PSL}_{\alpha=0} ) =0\;,  \; h(\rho^{PSL}_{\alpha=2} )  = \pm \frac{1}2 \;, 
\\
& h(\rho^{PSL}_{\alpha=2} )  =\pm  \frac{(\frac{q_1+q_2}2 \textrm{ mod 4})}{8} \;. \nonumber
\end{split}
\end{align}
Recall that there are two anyons associated to $\rho^{PSL}_{\alpha=2} $. The spectrum of $\{S_{0\alpha}^2\}$ and $\{h_\alpha\}$ are identical to the that of $ 4_{\pm (\frac{q_1+q_2}2 \textrm{ mod 4})}^{B} $.

\bibliographystyle{ytphys}
\bibliography{ref}

\end{document}